\documentclass{article}

\usepackage{iclr_aims,times}

\usepackage{hyperref}       % hyperlinks
\usepackage{url}            % simple URL 

\usepackage{natbib}
 \bibpunct[, ]{(}{)}{,}{a}{}{,}%
 \def\bibfont{\small}%
 \def\bibsep{\smallskipamount}%
 \def\bibhang{24pt}%

\usepackage{bm}
\usepackage{bbm}
\usepackage{psfrag,amssymb,thmtools}
\usepackage{mathrsfs}
\usepackage{hyperref}

\usepackage{xspace}
\usepackage{url}
\usepackage{dsfont}
\usepackage{natbib}
\usepackage{graphicx,graphics}
\usepackage{subfigure}
\usepackage{wrapfig}
\usepackage{xcolor}
\usepackage{enumitem}
\usepackage{booktabs} 
\usepackage{nicefrac} 
\usepackage{yfonts}
\usepackage[euler]{textgreek}
\usepackage[normalem]{ulem}
\usepackage[most]{tcolorbox}

\usepackage{amsmath}
\newtheorem{theorem}{Theorem}
\newtheorem{lemma}{Lemma}
\newtheorem{proof}{Proof}[section]
\newtheorem{proposition}{Proposition}
\newtheorem{definition}{Definition}
\usepackage[linesnumbered,ruled]{algorithm2e}

\title{A Robust Multi-Item Auction Design with Statistical Learning}
\author{%
  Jiale Han\\
  UCLA\\
  \texttt{jialehan@ucla.edu} 
   \And
  Xiaowu Dai \thanks{Corresponding author.}\\
  UCLA \\
  \texttt{daix@ucla.edu} 
}

\iclrfinalcopy % Uncomment for camera-ready version, but NOT for submission.

\begin{document}

\maketitle

\begin{abstract}
We propose a novel statistical learning method for multi-item auctions that incorporates credible intervals. Our approach employs nonparametric density estimation to estimate credible intervals for bidder types based on historical data. We introduce two new strategies that leverage these credible intervals to reduce the time cost of implementing auctions. The first strategy screens potential winners' value regions within the credible intervals, while the second strategy simplifies the type distribution when the length of the interval is below a threshold value. These strategies are easy to implement and ensure fairness, dominant-strategy incentive compatibility, and dominant-strategy individual rationality with a high probability, while simultaneously reducing implementation costs. We demonstrate the effectiveness of our strategies using the Vickrey-Clarke-Groves mechanism and evaluate their performance through simulation experiments. Our results show that the proposed strategies consistently outperform alternative methods, achieving both revenue maximization and cost reduction objectives.
\end{abstract}

\section{Introduction}
Designing optimal-revenue auctions is a fundamental problem in economics and computer science, with applications in a wide range of fields such as e-commerce, online advertising, and spectrum auctions. Although Myerson's celebrated work \citep{myerson1981optimal} provided an optimal design for single-item auctions by assuming knowing bidders' value distributions, the problem becomes increasingly complex in multi-item scenarios.
Additionally, in practical settings, bidders' value distributions are private, leaving the seller with limited information. Typically, sellers only have access to historical bidding data before an auction.  Moreover, the process of querying bidders before an auction can be time-consuming, especially when the number of bidders is large.

To address the challenges in multi-item mechanism design, various methods have been proposed, including \citet{6108212}, \citet{6108213}, \citet{10.1145/2229012.2229017}, and \citet{DBLP:journals/corr/abs-1210-3560}. However, these approaches often assume knowing bidder types  distributions, which may not be available in practice. Recently, there has been some progress in developing mechanisms that are robust to errors in the distribution of bidder types, such as \citet{8104086}, \citet{10.1145/3391403.3399541}, and \citet{10.1145/3490486.3538354}. Nevertheless, there remains a lack of practical algorithms that effectively utilize data for estimating bidders' types and reducing the time cost of implementing auctions, without relying on prior knowledge of bidders' types. In this paper, we propose a novel statistical learning method for optimal multi-item auctions that leverages historical data and credible intervals. Our approach aims to minimize the computational complexity and time cost of implementing auctions, providing a more efficient and practical solution for multi-item auctions.

Our proposed method leverages nonparametric density estimation to estimate credible intervals for bidder types  using their historical bidding data. We introduce two new strategies that leverage these credible intervals to reduce the time cost of implementing auctions. First, we propose a filtering method that screens potential winners by considering their value regions within the credible intervals generated by the estimated distribution. This approach ensures fairness while selecting winners. Second, we classify intervals based on their length and simplify the distribution of the type when the interval length of its credible interval falls below a threshold value.

To demonstrate the efficacy of our proposed strategies, we implement these two strategies within the Vickrey-Clarke-Groves (VCG) mechanism, with items being sold separately.  This has resulted in a new mechanism design that guarantees fairness, dominant-strategy incentive compatibility (DSIC), and dominant-strategy individual rationality (DSIR) with a high probability.
We conducted simulations using both small-scale and large-scale data to evaluate the effectiveness of our proposed strategies within the VCG mechanism. Our simulation results demonstrate that our strategies can efficiently reduce the implementation costs, as evidenced by a significant reduction in the number of queries required compared to the original VCG mechanism, while maintaining a comparable level of revenue. Overall, our proposed strategies offer an efficient data preprocessing approach for implementing multi-item auctions.

{\bf{{Related works.}}} 
The seminal work of Myerson \citep{myerson1981optimal} laid the foundation for optimal auction design in single-item settings, and the revelation principle in the paper suggests that achieving optimal design requires a round of direct communication in which bidders simultaneously and confidentially announce their value estimates to the seller. Therefore, to design a mechanism that maximizes revenue, the seller must make some queries to the bidders. However, the extension of this framework to multi-item scenarios poses significant challenges, as counterintuitive properties may arise that do not present in single-item settings \citep{10.1145/2845926.2845928}. The complexity can become intractable as the number of items increases, as demonstrated in \citet{DBLP:journals/corr/DughmiHN14} and \citet{10.1145/3055399.3055426}. In recent years, various strategies have been proposed to alleviate these analytical and computational difficulties for multi-item auctions, including \citet{6108212, 6108213, 10.1145/2229012.2229017, DBLP:journals/corr/abs-1210-3560, 10.1145/2213977.2214021, 6375290, chawla2007algorithmic, 10.1145/1806689.1806733, cai2013understanding, DBLP:journals/corr/MorgensternR16, 10.1007/978-3-662-54110-4_12}. However, the majority of these methods require knowing the true distribution of bidder types, which may not be available in practice. 
%There are studies showing that the use of Myerson's optimal auction or VCG with entry fees, as proposed in \citet{yao2014n, cai2016duality}, can achieve revenue at least $1/8$ of the optimal revenue when bidders have additive valuations. In this paper, we adopt the VCG mechanism and selling items separately, while providing a lower bound on the revenue. 
The presence of private and unknown bidder types has motivated research on robust mechanism design. Studies in this area focus on robustifying mechanisms to errors in the distribution of bidder types, without relying on Bayesian assumptions. Notable examples include \citet{8104086}, \citet{10.1145/3490486.3538354}, \citet{10.1145/3391403.3399541} and \citet{bergemann2011robust}, which have all aimed to design mechanisms that achieve optimal revenue with only approximated distributions of bidder types.
In contrast to these existing works, our approach does not rely on the approximated distribution of bidder types. Instead, we utilize estimated credible intervals for bidder types to design mechanisms that effectively reduce the time cost of implementing auctions and maximize revenue. \cite{han2025online} also studies mechanism design without prior knowledge of the values, but focuses on the single-item setting. Our method is designed to be efficient and easy to implement, providing a practical solution for multi-item auction applications.

%The rest of the paper is organized as follows. In Section \ref{sec:problem}, we provide the problem formulation. In Section \ref{sec:method}, we present the proposed mechanism design and the theoretical analysis of its properties. In Section \ref{Algorithm}, we provide the algorithms for implementing our proposed mechanism. In Section \ref{sec:experiment}, we provide numerical simulations to demonstrate the performance of our proposed mechanism. In Section \ref{sec:conclusion}, we conclude the paper and discuss future research directions. All proofs and additional simulation results are provided in the Supplementary Appendix.

\section{Preliminaries}
%\label{sec:problem}
%In this section, we introduce key definitions and notations. Specifically, we aim to formulate a mechanism design strategy for multi-item auctions with additive valuations that maximizes the expected revenue while ensuring incentive compatibility.
%\subsection{}
\label{Preliminaries}
In our auction setting, a seller aims to sell $N\geq 1$ items to a group of $m>1$ bidders, denoted as $\mathcal{B}=\{B_1, B_2, \dots, B_m\}$. Each bidder $B_i$ has a private valuation for the $N$ items, which is captured by their type $t_i=(t_{i,1},t_{i,2},\dots,t_{i,N}) \in \mathbb{R}^{1\times N}$, and $i\in\{1,\dots m\}\equiv[m]$. To accommodate the different preferences of bidders, we assume that each element $t_{i,j}$ is drawn from a private distribution $D_{i}^j$. We denote the set of all distributions as $D=\otimes_{i,j}D_{i}^j$, where the $mN$ type distributions are independent, but not necessarily identical.

\vspace{0.1in}
\noindent
{\bf{Valuation.}}  For the set of items $S\subset [N]$, we assume a bidder $\textcolor{black}{B_i}$'s value for $S$ is given by the additive valuation function as \cite{10.1007/978-3-662-54110-4_12}, 
\[v_i(t_i, S)=\sum_{j\in S}t_{i,j}.\]

\noindent
{\bf{Auction Mechanism.}} 
An auction mechanism $M$ is a mapping from the space of all the bidder types  $t = (t_1, t_2, \dots, t_m)^T\in\mathbb{R}^{m\times N}$ to the space of allocations $Q(t)=(q^j_i(t))$ and payment $P(t) = (p_1(t), p_2(t), \dots, p_m(t))$, where $q^j_i(t)\in[0,1]$ denotes the probability the items $j$ is allocated to the bidder $B_i$ and the $p_i(t)\in (-\infty,\infty)$ denotes the payment for bidder $B_i$. Let $p(t)=\sum_{j\in [m]}p_j(t)$ be the total payments received by the seller.
Since the bidder types  follow  the distribution $D$, the goal is to design a mechanism for the seller that maximum her expected revenue, 
\begin{equation*}
    \text{REV}(M,D)\equiv\mathbb{E}_{t\sim D}\left[p(t)\right].
\end{equation*}
Given a multi-item auction mechanism $M$ and all bidders' reporting bids $b\in\mathbb{R}^{m\times N}$, the utility of a bidder $B_i$ with private type $t_i$ is formulated as 
\begin{equation*}
    u_i(t_i,M(b))=\sum_{j=1}^N t_{i,j}q^j_i(b)-p_i(b).
\end{equation*}

\vspace{0.1in}
\noindent
{\bf{DSIC.}} 
To ensure truthfulness, we impose an incentive compatibility constraint. We adopt the definition of dominant-strategy incentive compatibility (DSIC); see, e.g., \citet{cai2016duality}.
%\citep[e.g.,][]{yao2014n}.
\begin{definition}[DSIC]
An auction mechanism $M$ is DSIC if, for every bidder ${\textcolor{black}{B_i}}$, $i\in [m]$, 
the utility of truthfully reporting type $t_i$ is no less than the  utility of reporting an alternative type $t'_i$. That is,
\[u_i(t_i,M(t_i,t_{-i}))\geq u_i(t_i,M(t'_i,t_{-i})).\]
\end{definition}

\vspace{0.1in}
\noindent
{\bf{DSIR.}} 
Next, we introduce a generalization of the dominant-strategy individual rationality (DSIR), denoted as $\delta$-DSIR, which is different from the traditional definitions in the literature \citep[e.g.,][]{yao2014n}. The introduction of $\delta$-DSIR is motivated by our proposed mechanism in Section \ref{sec:method}, which utilizes a method of simplifying the distribution of bidder types  based on the lower bound of their corresponding credible intervals.
 
\begin{definition}[$\delta$-DSIR] A mechanism $M$ is $\delta$-individual rationality ($\delta$-DSIR) if for all types $t=(t_1,t_2,\dots,t_m)$ and for all bidder $i\in[m]$,
\[\mathbb{P}\left\{u_i(t_i,M({\color{black}{t}}))\geq 0\right\} \geq 1-\delta.\]
A mechanism is DSIR if $\delta=0$, i.e. for all types $t=(t_1,t_2,\dots,t_m)$ and for all bidder ${\textcolor{black}{B_i}}$, $i\in[m]$, we have
$u_i(t_i,M({\color{black}{t}}))\geq 0$ almost surely.
\end{definition}

\vspace{0.1in}
\noindent
{\bf{VCG Mechanism.}} 
In this paper, we will implement our methods upon the foundation of the Single-Stage Vickrey-Clarke-Groves (VCG) mechanism or the 2nd-price Vickrey mechanism in our situation. The VCG mechanism operates by allocating the bundle of items in accordance with the bidder who expresses the highest valuation for the bundle and charges them the value of the second highest bid. Mathematically, let $b_{i,j}$ be the reporting bid of bidder $B_j$ to item $j$. 
The winner of the bundle $S$ is determined as
\begin{equation*}
\label{def:winner}
i_{*} :=\arg\max\limits_{i}\sum_{j\in[S]}{\textcolor{black}{b}}_{i,j},
\end{equation*}
and the winner is subsequently charged with 
$\max_{i,i\neq i_{*}}\sum_{j\in[S]}{\textcolor{black}{b}}_{i,j}$.
It is well-known that truthful reporting is a dominant strategy for each bidder in a VCG mechanism \citep[e.g.,][]{ausubel2006lovely} and VCG mechanism has also been widely recognized for its desirable properties of DSIR and DSIC \citep{vickrey1961counterspeculation, clarke1971multipart, che2001optimal}. This implies that, when using a VCG mechanism and soliciting bids from bidders, it can be expected that they will rationally provide truthful valuations, i.e. $b_{i,j}=t_{i,j}$ for any bidder $B_i$ and item $j$. 

The single-stage VCG mechanism simplifies cost measurement in implementation, requiring only that the seller make queries to bidders. Cost reduction can be achieved by strategically minimizing the number of queries.

\section{Main Results}
\label{sec:method}
%%%%%%%%%%%%%%%%%%%%%%%%%%%%%%%%%%%%%%%%%%%%%%%%%%%%%%%%%%%%%%%
In this section, we present the main results of this paper. 
We begin by introducing a method for estimating the distributions and constructing credible intervals for bidder types  in Section \ref{subsubsection3.2.1}. Next, we propose a novel strategy for reducing the number of bids considered in an auction in Section \ref{Winnow down the possible winners}. We then provide a mechanism that simplifies the distributions of the bidder types  in Section \ref{Designate the value}. Finally, in Section \ref{subsubsection3.2.5}, we implement the VCG mechanism based on the strategy in Section \ref{Designate the value} and give the lower bound of its revenue.

We consider the most challenging scenario where we lack any prior knowledge about the distributions of bidder types, and the bidder types distributions can be different for a certain item. Additionally, we assume that the distributions of the bidder types  are fixed throughout the bidding process.
%%%%%%%%%%%%%%%%%%%%%%%%%%%%%%%%%%%%%%%%%%%%%%%%%%%%%%%%%%%%%%%

\subsection{Estimation of Distribution and Credible Intervals for bidder types }\label{subsubsection3.2.1}

\noindent 
{{\color{black}  In this subsection, we make the assumption that each bidder has participated in numerous bidding instances with independent bids coming from her true value's distribution for every item, generating a certain amount of historical data available for analysis. 

Let the dataset of previous bids for bidder $B_i$ on item $j$ be denoted as $\Gamma_{i,j}$, which consists of $n_{i,j}$ independent and identically distributed samples, $\Gamma_{i,j}\equiv\{x_{i,j}^1,x_{i,j}^2,\ldots, x_{i,j}^{n_{i,j}}\}$.

%In order to estimate each individual bidder's private valuations, we collect historical bid data for each item from each bidder $B_i$. 

\vspace{0.1in}
\noindent
\textbf{Kernel Density Estimate.} With the data $\Gamma_{i,j}$, we can perform kernel density estimation of $D_{i}^j$ using the function $\widehat{D}_{i}^j$, as defined by:
\begin{equation*}
\widehat{D}_{i}^j(x) =\frac{1}{{n_{i,j}}h}\sum^{{n_{i,j}}}_{s=1}K\left(\frac{x-x_{i,j}^s}{h}\right).
\end{equation*}
Here, $K$ represents a density function known as the kernel, and $h$ is the bandwidth. This method, which dates back to the works of Rosenblatt (1956) and Parzen (1962), commonly utilizes a Gaussian kernel, $K(t)=\frac{1}{\sqrt{2\pi}}e^{-t^2/2}$. A choice of the bandwidth is $h=1.06 \cdot\min{\bigl \{\hat{\sigma},\frac{IQR}{1.34}}\bigr\} \cdot {n_{i,j}^{-1/5}}$, where $\hat{\sigma}$ is the sample standard deviation, $n_{i,j}$ denotes the sample size, and $IQR$ represents the sample interquartile range. This choice of $h$ is known to be  robust against outliers \citep{scott1992multivariate}, and is minimax optimal for estimation \citep{zambom2013review} with the mean integrated square error (MISE) of $\mathbb{E}\int[\widehat{D}_{i}^j(x)-D_{i}^j(x)]^2dx=O(n_{i,j}^{-4/5})$.

\vspace{0.1in}
\noindent
{\bf{Rejection Sampling}}: 
In order to sample from the estimated density, we can utilize the rejection sampling method. This involves selecting random samples from a given distribution, $g(x)$, and choosing a constant $c$ such that \begin{equation*}
    c\cdot g(x)>\widehat{D}_{i}^j(x), \quad\forall x. 
\end{equation*}  Then, we accept the sample $x$ with a probability of $\widehat{D}_{i}^j(x)/cg(x)$, resulting in a sampling process equivalent to that from $\widehat{D}_{i}^j(x)$ as described in \cite{casella2004generalized}. In our scenario, for each item $j\in[N]$, we can define the support of the density $\widehat{D}_{i}^j$ to be within the known minimum and maximum values, $a_j$ and $b_j$, respectively, which may be set by the seller or obtained from the market. Thus, we can set $g(x)$ as a uniform distribution over the interval $[a_j, b_j]$.

We can repeat this sampling process $\mathcal{N}$ times, and compute the $(\alpha/2)$th and  ($1-\alpha/2$)th quantiles of the samples as the lower and upper bounds of the $(1-\alpha)$ {{\color{black}credible }}interval of $t_{ij}$. As the MISE of $\hat{D}_{i}^j$ is $O(n_{i,j}^{-4/5})$, we can assert that when $n_{i,j}$ is large enough, the difference between $\widehat{D}_{i}^j$ and $D_{i}^j$ is negligible, resulting in high precision of the credible interval. We provide a detailed outline of this procedure in Algorithm \ref{alg:one}, which is deferred to Section  \ref{A1}.

\subsection{Winnow Down Potential Winners}\label{Winnow down the possible winners}

\noindent 
{{\color{black}We present a novel strategy that effectively reduces the number of bids considered in an auction while maintaining fairness in the game based on credible intervals. This approach narrows down the pool of potential winners, leading to improved efficiency in the auction process.}} We begin by defining the fairness of the mechanism.
\begin{definition}[$\delta$-fairness]\label{beta-fair}
Given a set of items $S\subset [N]$ and the winner of the bundle $S$, $B_{i*}$, a mechanism is $\delta$-fair if for any item $j$ in $S$, with probability at least $1-\delta$, for all $i \in [m], i \neq i_*$, it holds true that $t_{i,j}< t_{i_* j}$. A mechanism is fair if $\delta=0$.

%$\sum_{j\in S}t_{i,j}<\sum_{j\in S}t_{{i_*},j}$.
\end{definition}
\noindent
This definition draws inspiration from \citet{NIPS2016_eb163727}, in which fairness is defined through pairwise comparison. In other words, a fair mechanism assigns higher probabilities of winning to bidders with higher values. In our scenario, we only need to consider the winner, as the outcome is unique.
Based on the definition of $\delta$-fairness, it is clear that a smaller $\delta$ value corresponds to a fairer game.

Assume that we have already obtained the $(1-\alpha)$ credible interval $T_{i,j} = [{t}_{i,j}^L,{t}_{i,j}^U]$ for each bidder $B_i$'s type for item $j$, where ${t}_{i,j}^L$ and ${t}_{i,j}^U$ is the $(\alpha/2)$th and $(1-\alpha/2)$th quantiles of $D_{i}^{j}$ . {{\color{black}This can be achieved by employing the methodology outlined in Section \ref{subsubsection3.2.1}, which leverages historical data to gain the interval.}} Then with probability more than $(1-\alpha)$, the true value $t_{i,j}$ is within the interval $[{}t_{i,j}^L,{}t_{i,j}^U]$.
For each item $j$, we identify the bidder $B_{i_{*}^j}$ with the largest \emph{upper credible interval} among all bidders,
\begin{equation}
\label{eqn:defofwinneruci}
i_{*}^j=\arg\max\limits_{i\in[m]}{t}_{i,j}^U.
\end{equation}
This bidder has the potential to have the highest bid and is therefore more likely to win the item. However, other bidders whose estimated intervals overlap with $B_{i_{}^j}$'s may also have a chance to win. To account for this scenario, we define a new interval relationship. For each item $j$, we consider bidders $B_i$ and $B_{i^{'}}$ to be \emph{linked} if
\begin{equation}
\label{eqn:linked}
    [{t}_{i,j}^L, {t}_{i,j}^U ) \cap ({t}_{i^{'},j}^L,{t}_{i^{'},j}^U] \neq \emptyset.
\end{equation} 

We note that \citet{NIPS2016_eb163727} introduces the concept of "chained relation" in addition to the "linked relation" defined above. Two bidders are considered chained if they belong to the same component of the transitive closure of the linked relation. The authors then focus on all the arms chained to the arm with the highest upper confidence bound and disregard the others. Our approach is similar, but we utilize the linked relation instead of the chained relation, as only one selection is made in our setting and only one arm needs to be pulled.

In order to mitigate the computational complexity associated with determining the winner of a multi-item auction, we propose a mechanism that selectively considers bids from a subset of bidders for each item. Specifically, for each item $j\in[N]$, we only consider the bids from bidders that are linked to the bidder $B_{i_{*}^j}$, as defined by 
\begin{equation}\label{Bb}
    \mathcal{B}_j=\{B_i|i\in[m], B_i \text{ is linked to } B_{i_{*}^j} \text{ for item } j\}.
\end{equation}

Here $i_*^j$ and the linked relation are defined in Eqs.~\eqref{eqn:defofwinneruci} and \eqref{eqn:linked}, respectively. The algorithm of constructing the set $\mathcal{B}_j$ is given in Section \ref{A2}. 
We show that if just considering the bidders in set $\mathcal{B}_j$ in an original fair mechanism, it is still fair enough.
\begin{proposition}\label{fairness}
    Let $\mathcal{B}_j$ denote the set of bidders linked to the bidder with the highest upper confidence bound for item $j$, as defined by \eqref{Bb}. Then for each item $j \in [N]$, considering only bidders in the set $\mathcal{B}_j$ to implement a fair mechanism results in a new mechanism that is $\alpha$-fair. 
\end{proposition}
\noindent 
The proof of this proposition is given in Appendix \ref{fair}. {{\color{black}The reduction in the number of bidders that need to be considered results in decreased computational complexity and time cost. This is because there is no longer a need to calculate the allocation and payment for these bidders when implementing any mechanism.}}

%Note that it becomes difficult to sell items in a bundle, as the total value of a bundle would be the sum of the bidder types  for each item, which cannot be calculated with ignored bids. 
%Moreover, the allocation space $A(t)$ that the mechanism needs to search for becomes increasingly large as the number of items increases. As an example, we calculate the lower bound of the complexity of allocation when each bundle has at most 2 items. 
%\begin{proposition}\label{prop1}
%When each bundle has at most 2 items, the complexity of the allocation grows exponentially as a function of the number of items. Specifically, the number of potential allocations exceeds $(2N/3)^{\lfloor N/3\rfloor}$.
%\end{proposition}
%\noindent
%The proof of Proposition \ref{prop1} is provided in Appendix \ref{A}. Hence the number of allocations increases exponentially as a function of the number of items. 
%In this paper, we consider that the seller sells each item separately, which gives a practical solution to alleviate the computational complexity associated with determining the allocation of items in a bundle. 

%%%%%%%%%%%%%%%%%%%%%%%%%%%%%%%%%%%%%%%%%%%%%%%%%%%%%
{{\color{black}
\subsection{Lower Credible Bound Method}\label{Designate the value}

In this section, we present another strategy that utilizes credible intervals to reduce the cost of the seller to implement any mechanism with good incentive properties, such as the VCG mechanism that satisfies the DSIC and DSIR.
%As previously established in Section \ref{Preliminaries}, the use of  guarantees truthful reporting of bidders' values for each item. However, in practice, obtaining these values through a large number of queries can be costly.  The strategy outlined in Section \ref{Winnow down the possible winners} provides a solution to this problem. We now propose an additional approach to further reduce the number of queries. 

Regarding robust auction mechanism design, since the bidder types  are private, obtaining the true set of distributions $D$ is not possible, the goal is to ensure that the distance between the estimated distribution $\widehat{D}$ and the true distribution ${D}$ is minimized so that implement any mechanism on $\widehat{D}$ will also gain comparable revenue. Mathematically, as motivated by the characterization of the Prokhorov Metric \citep[e.g.,][]{strassen1965existence}, we want to ensure that there exists a coupling $\gamma$ of ${D}$ and $\widehat{D}$, such that
\begin{equation}
\label{eqn:Prokhorov}
\mathbb{P}_{(\hat{t}, t)\sim \gamma}[\|\hat{t}-t\|_{\infty}>\epsilon]\leq \delta_0,
\end{equation}
for some small values of $\epsilon,\delta_0>0$. 
%To achieve this, appropriate sampling methods must be employed to generate accurate estimates $\hat{t}$, and strategies must be devised to control the distance between the estimated and true distributions while also maximizing revenue. This requires the mechanism to be robust to estimation errors.

For any bidder $B_i,i\in[m]$ and item $j\in[N]$, $T_{i,j} = [{t}_{i,j}^L,{t}_{i,j}^U]$ is the $(1-\alpha)$ credible interval for $t_{i,j}$, as described in Sections \ref{Winnow down the possible winners}. Thus $\mathbb{P}({t}_{i,j}^L \leq t_{i,j} \leq{t}_{i,j}^U)=1-\alpha.$
%\begin{equation*}
%    \mathbb{P}({t}_{ij}^L-t_{ij}>0)\leq\alpha/2.
%\end{equation*} 
Additionally, we define $d_{i,j}={t}_{i,j}^U-{t}_{i,j}^L$ as the length of the interval. Then, we have 
\begin{equation*}
\begin{aligned}
\mathbb{P}({t}_{i,j}^L- d_{i,j} \leq t_{i,j} \leq{t}_{i,j}^L+d_{i,j}) & =\mathbb{P}(- d_{i,j} \leq t_{i,j}-{t}_{i,j}^L \leq d_{i,j})\\
& =\mathbb{P}(|t_{i,j}^L-{t}_{i,j}| \leq d_{i,j})\geq 1-\alpha,
\end{aligned}
\end{equation*}
where $t_{i,j}$ is a random variable and ${t}_{i,j}^L$ is a fixed value.

%Above all, we have $\mathbb{P}(|{t}_{i,j}^L-t_{i,j}|>d_{i,j})\leq \alpha$.
If we choose $\widehat{D}_i^j$ to be the one-point distribution, denoted by $\mathbb{P}(\hat{t}_{i,j}={t}_{i,j}^L)=1$ for all $i$ and $j$, and define the norm $||t||_{\infty}\equiv\max_{i,j}t_{i,j}$, then due to the independence of all $mN$ type distributions, we can deduce that:
\begin{equation*}
    \begin{aligned}
        \mathbb{P}_{(\hat{t}, t)\sim (\widehat{D},D)}\Bigl[\|\hat{t}-t\|_{\infty}\leq \max_{i,j} d_{i,j}\Bigr]&=\prod_{i\in[m],j\in[N]}\mathbb{P}(|t_{i,j}^L-{t}_{i,j}|\leq \max_{i,j} d_{i,j})\\
        &\geq \prod_{i\in[m],j\in[N]}\mathbb{P}(|t_{i,j}^L-{t}_{i,j}|\leq  d_{i,j})\geq (1-\alpha)^{mN},
    \end{aligned}
\end{equation*}
so it satisfies Eq.~\eqref{eqn:Prokhorov} with $\epsilon=\max_{i,j} d_{i,j}$ and $\delta_0=1-(1-\alpha)^{mN}\leq mN\alpha$. 
%\begin{equation*}
%    \mathbb{P}(t_{ij}-{t}_{ij}^L> d_{ij})\leq\alpha/2.
%\end{equation*}
Thus, the length of the intervals $d_{i,j}$ becomes a crucial factor in determining the accuracy of the degenerated $\widehat{D}_i^j$ as one point ${t}_{i,j}^L$.

By simplifying all $\widehat{D}_i^j$ to one-point distributions, we can significantly reduce the computational complexity and time cost involved. This is because we eliminate the need to deal with complex probability distributions when computing the reserve price \citep[e.g. implementing the optimal auction design for a single item,][]{myerson1981optimal} or designing a mechanism. However, it is important to acknowledge that this approach may lead to a considerable reduction in revenue, particularly if certain credible intervals have a large bandwidth. Therefore, it becomes necessary to carefully manage the set of bids utilizing the one-point distribution in order to minimize complexity while still maintaining revenue reductions within acceptable tolerances.

To deal with this trade-off dilemma, for every bidder $B_i,i\in[m]$, we first define two types of credible intervals based on a specified threshold value $d > 0$. Let 
\begin{equation}
\label{eqn:defofli}
Q_i = \{j|{t}_{i,j}^U-{t}_{i,j}^L > d, j\in[N]\},\quad L_i = \{j|{t}_{i,j}^U-{t}_{i,j}^L \leq d, j\in[N]\}.
\end{equation}
We let the distribution of the types for items in $L_i$ degenerate to one-point distribution, i.e. the distribution of the types in $L^{'}\equiv\{{t}_{i,j}|i\in[m], j\in L_i\}$ will be replaced by a one-point distribution. Then let 
\begin{equation}
\label{L}
Q=\{{t}_{i,j}|i\in[m], j\in Q_i\},\quad L =\{\hat{t}_{i,j}|i\in[m], j\in L_i\}.
\end{equation}
be the sets of types that the seller will use to do analysis, where ${t}_{i,j}\sim D^j_i$ and $\hat{t}_{i,j}\sim [\mathbb{P}(\hat{t}_{i,j}={t}_{i,j}^L)=1]$ are all random variables. We can let the degenerated type distribution be $\widehat{D}$ and $\hat{t}=Q\cup L$.
%We call the values in  $Q$ \emph{type I} types and values in $L$ \emph{type II} types. 

Let $|L_i| \equiv n_i$ so that $|Q_i| = N - n_i$, and  \begin{equation}
\label{eqn:defofn}
n \equiv \sum_{i\in [m]}n_i=|L|, \quad |Q|= mN-n
\end{equation}

The following theorem demonstrates the effectiveness of our strategy in maintaining both the DSIC and $\delta$-DSIR properties for any DSIC and DSIR mechanism. However, our mechanism is specifically trained using a simpler distribution $\widehat{D}$ and has query access limited to the types in $Q$. This approach reduces the computational complexity and time cost involved.

\begin{theorem}\label{New1}

Suppose we implement a DSIR mechanism $\widehat{M}$  w.r.t. $\widehat{D}$ and get the allocation $(\hat{q}^j_i(\hat{t}))$ and payment rule $(\hat{p}_i(\hat{t}))$. We can construct a mechanism $M$ using only query access to $\widehat{M}$ (which equals to only use query access to the types in $Q$), such that the mechanism ${M}$ is $\delta$-DSIR with $\delta = \alpha n/2$ w.r.t. $D$. 
If $\widehat{M}$ is also DSIC, and we assume that a change in type for a specific item only affects the allocation and payment of that item, then $M$ is DSIC w.r.t. $D$ as well.
%In addition, if we use the same way to train a DSIR and DSIC mechanism $\widetilde{M}$ w.r.t. ${D}$ and get the allocation $(\tilde{q}^j_i(t))$ and payment rule $(\tilde{p}_i(\hat{t}))$. 
%Let $L_i^*:=\{j|0 < {t}_{i,j}^U-{t}_{i,j}^L\leq d,j\in[N]\}$, 
%and
%\begin{equation}\label{K}
%k=\sum_{i\in[m],j\in L_i^*}\tilde{q}_i^j,
%\end{equation}
%if we also assume that then with probability $(1-\alpha/2)^{n}$,
%\[\text{REV}({M},{D}) \geq \text{REV}(\widetilde{M},{D})-kd\]

%Suppose we are given query access to a mechanism $\hat{M}$ that is DSIC and DSIR w.r.t. to the types defined as \eqref{L}, we can construct mechanism $M$ using only query access to $\hat{M}$ and obviously with respect to $D$, such that $M$ is $DSIC$ and the expected revenue of $M$ is at least $\text{REV}(\hat{M},\widehat{D})

%With probability $(1-\alpha/2)^{n}$, 
%employing any DSIC mechanism $M$ on  the expected revenue will be reduced by at most $kd$ compared to using $M$ with the true distribution $D$. 
\end{theorem}
The proof of this theorem is given in Appendix \ref{3.5}. Theorem \ref{New1} implies that our proposed mechanism is practically viable. The only challenge is to keep $\alpha n$ as small as possible to ensure that bidders are willing to participate in the auction. To make $\alpha$ small, we can use historical data or sample more extensively from the estimated distribution to obtain more precise credible intervals for $t$. Additionally, we can control $n$ by carefully choosing the value of $d$.

}}

\subsection{Degenerated VCG Mechanism}\label{subsubsection3.2.5}
%%%%%%%%%%%%%%%%%%%%%%%%%%%%

The VCG mechanism is both a DSIC and DSIR mechanism, and it meets the definition of fairness as outlined in Definition \ref{beta-fair}. Moreover, when there is a change in type for a specific item, it only affects the allocation and payment for that particular item. Therefore, the VCG mechanism satisfies all the assumptions stated in Proposition \ref{fairness} and Theorem \ref{New1}. Consequently, we can utilize the VCG mechanism to validate the properties of our strategies in Section \ref{Winnow down the possible winners} and Section \ref{Designate the value}. In this section, our attention will be directed towards the method discussed in Section \ref{Designate the value}. 

To make the validation process easier, we consider that the seller sells each item separately, which gives a practical solution to alleviate the computational complexity associated with determining the allocation of items in a bundle. As an example, we calculate the lower bound of the complexity of allocation when each bundle has at most 2 items in Appendix \ref{Complexity} and show the complexity of the allocation grows exponentially as a function of the number of items.

%\subsection{Analysis of Revenue} \label{subsubsection3.2.5}
\noindent 
Based lower credible bound method in Section \ref{Designate the value}, we can give a degenerated VCG mechanism,  this approach involves keeping the types in set $L$ fixed while only querying the types in $Q$. For any bidder $B_i, i\in[m]$, we have the set of  estimated types $L_i$, which is defined in Eq.~\eqref{eqn:defofli}. Let 
$L_i^*:=\{j|{t}_{i,j}^U-{t}_{i,j}^L>0,j\in[N]\}$, 
and
\begin{equation}\label{K}
K= \cup_{i\in[m]}(L_i\cap L_i^*).
\end{equation}
Let $k=|K|$ be the number of items in set $K$.
The revenue reduction is limited to the items in $K$ since only the distributions of bidder $B_i$'s valuation for items in $K$ are altered and degenerate into one-point distributions. We have the following result.
%types in set $Q$ are equal to true types and will not impact the final revenue. Additionally, for any $i \in[m]$ and $j\in [N]$, when ${t}_{i,j}^U-{t}_{i,j}^L = 0$, with probability $(1-\alpha/2)$, the true bid is smaller than the estimated bid.
\begin{theorem}\label{T1}
    Let $\widehat{M}$ be the VCG mechanism implemented on $\widehat{D}$ constructed in Section \ref{Designate the value}, and ${M}_0$ be the VCG mechanism implemented directly on ${D}$. $M$ is the degenerated VCG mechanism  using only query access to $\widehat{M}$. Then with probability $(1-\alpha/2)^{n}$, 
    \[\text{REV}({M},{D}) \geq \text{REV}({M}_0,{D})-kd.\]
\end{theorem}
The proof of this theorem is given in Appendix \ref{Theorem}.
We observe that when the number of items, $N$, is limited, it is likely that the number of items, $k$, in the set $K$ in Eq.~\eqref{K} will also be limited due to the constraint $k<N$. Consequently, $kd$ will also be relatively small, indicating the potential effectiveness of the proposed mechanism when $N$ is small. 
Moreover, our proposed mechanism holds a distinct advantage over the robust mechanism design described in \citet{10.1145/3490486.3538354} in situations where the number of bidders, $m$, is exceptionally large. 
While their revenue lower bound demonstrates a negative correlation with $m$, our revenue lower bound remains largely unaffected by $m$. This is due to the fact that $k$ is strictly smaller than $N$, and $d$ is a hyper-parameter that is typically reduced to control the number of $n$ as $m$ increases.
Therefore, our mechanism is particularly well-suited for environments with a high number of bidders and a limited number of items.

Theorem \ref{T1} shows that the performance of the revised VCG mechanism based on Section \ref{Designate the value} depends on $kd$. The value of $d$ is a hyper-parameter that is chosen based on the specific characteristics of the data set at hand. The density function, and subsequently the length of the confidence intervals, can vary from dataset to dataset, thus requiring the selection of an appropriate value of $d$. In Appendix \ref{A3}, we give the algorithms of the implemented VCG mechanism and the details of how to explore the appropriate range of values for $d$.

\section{Numerical Experiments}
\label{sec:experiment}
%In order to implement our proposed mechanism, it is necessary to acquire historical data from the bidders. 
In this section, we present our experimental results. More details and results are presented in the supplementary materials appendix \ref{ER}. We utilize a truncated re-normalized Gaussian distribution to model the private valuations of each bidder, employing distinct means and variances. Once the distribution $D$ is obtained, bidder types $t$ can be generated from $D$, enabling the simulation of historical data sets $\Gamma_{i,j}$ as described in Section \ref{subsubsection3.2.1}. To validate the effectiveness of our proposed methods, we compare three revised VCG Mechanisms with the original VCG Mechanism.

Mechanism 1 (M1) employs the method outlined in Section \ref{subsubsection3.2.1} to obtain credible intervals, followed by the utilization of the approach described in Section \ref{Winnow down the possible winners} to narrow down the dataset. Additionally, the method explained in Section \ref{Designate the value} is used to exclude queries to types whose distributions have degenerated into one-point distributions. 
In comparison to M1, Mechanism 2 (M2) does not employ the winnow-down method, while Mechanism 3 (M3) constructs a credible interval using the minimum and maximum values from the original data, without utilizing the distribution estimation method.
The benchmark is the revenue generated by implementing the original VCG mechanism without employing any methods from our paper.

The \emph{regret} is defined as the discrepancy in revenue between the original VCG mechanism and the modified VCG mechanism.

%{{\bf{Small-Scale Data}}}
\begin{figure}[htbp]
\centering
\subfigure[Comparison of Revenue]{
\begin{minipage}[t]{0.3\linewidth}
\centering
\label{1a}
\includegraphics[width=1.8in]{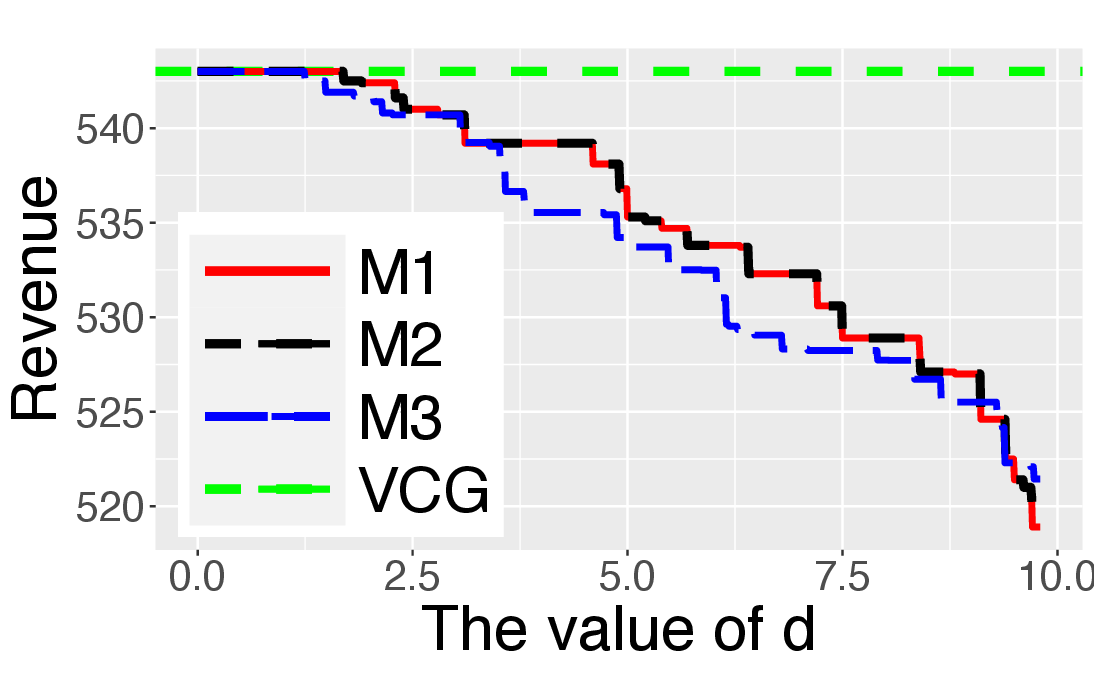}
%\caption{fig1}
\end{minipage}%
}%
\subfigure[Comparison of Regret]{
\begin{minipage}[t]{0.3\linewidth}
\centering
\label{1b}
\includegraphics[width=1.8in]{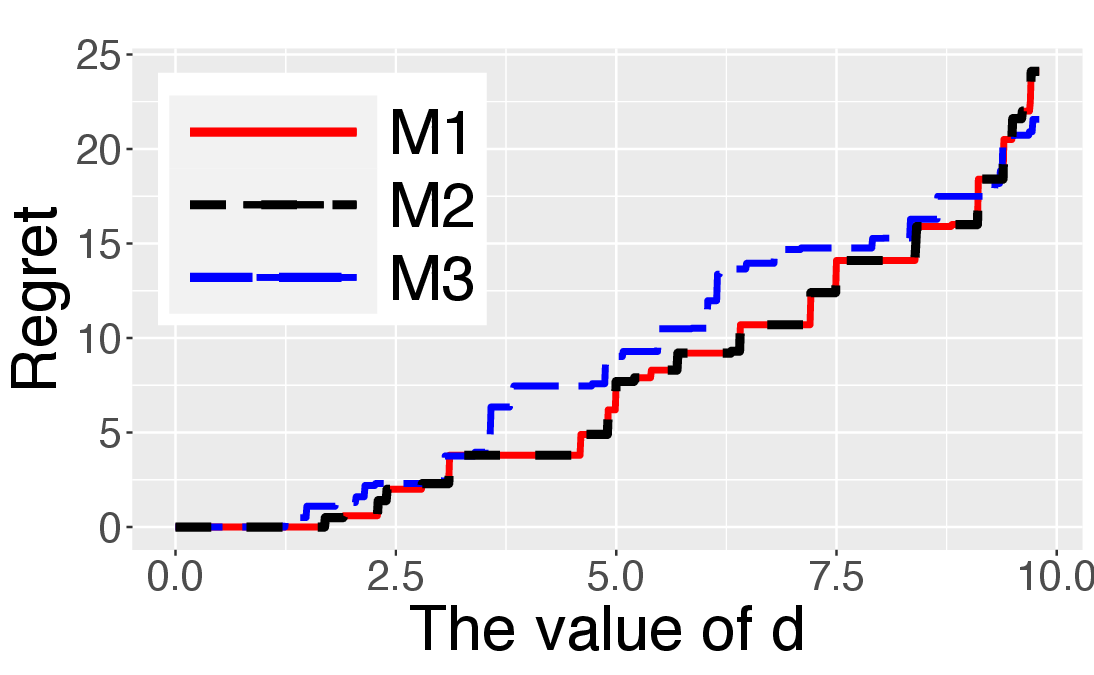}
%\caption{fig2}
\end{minipage}%
}%
\subfigure[Comparison of $kd$]{
\begin{minipage}[t]{0.3\linewidth}
\centering
\label{1c}
\includegraphics[width=1.8in]{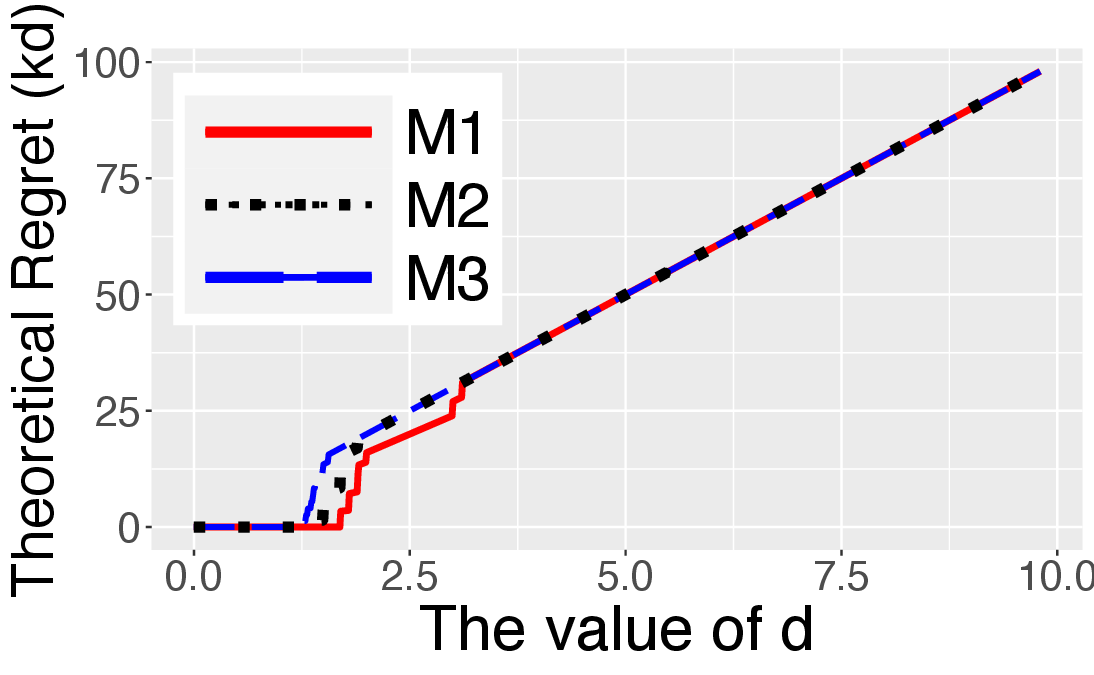}
%\caption{fig2}
\end{minipage}
}%
\centering
\caption{Comparison of the revenue and regret when $m=30$, $N=10$.}
\label{fig:one1}
\end{figure}

Figure \ref{fig:one1} illustrates a comparison of revenue and regret when the number of items $m$ is 30 and when the number of items $N$ is 10. The revenue of the original VCG mechanism is represented by a green dashed line at the top of Figure \ref{1a} and serves as a benchmark. Mechanism 1 is depicted by a red solid line, Mechanism 2 by a black dashed line, and Mechanism 3 by a blue long dash line. To further analyze the results, we present a modified version of Figure \ref{1a} in Figure \ref{1b}, focusing on regret. In Figure \ref{1b}, each line's position represents the discrepancy with the green line, allowing for a clearer understanding of the regret associated with each mechanism. 

It can be observed that the revenue of M1 and M2 overlap across all values of $d$, indicating that the winnow-down method described in Section \ref{Winnow down the possible winners} does not have an impact on the revenue. This finding validates Proposition \ref{fairness}, which asserts that the winnow-down method preserves fairness in the auction, as none of the winners are omitted. 
Moreover, it is worth noting that the revenue generated by M3 consistently remains lower than that of M1 and M2. This demonstrates the advantage of utilizing the method described in Section \ref{subsubsection3.2.1} to obtain credible intervals.
Additionally, we conducted a verification of the interval coverage and found that all of the bidders' bids fall within our constructed credible intervals. This implies that utilizing the lower bound as the estimated type still ensures the mechanism is DSIR, which aligns with the findings of Theorem \ref{New1} in Section \ref{Designate the value} and further affirms the feasibility of employing the method outlined in Section \ref{subsubsection3.2.1}. 
Furthermore, upon comparing Figure \ref{1b} and Figure \ref{1c}, it can be observed that the practical regret consistently remains smaller than the upper bound of the theoretical regret, which is $kd$. This observation serves to validate the result presented in Theorem \ref{T1}. Further refinement is possible. Currently, when calculating the upper bound of regret theoretically, the revenue for items in set $K$ defined in Eq.~\eqref{K} is reduced. In practice, a more precise method would be to sort the intervals' bandwidths and deduce the maximum bandwidth for each item. Specifically, we can define 
\[d_j=\max\limits_{i}\{{t}_{i,j}^U-{t}_{i,j}^L|i\in[m],0<{t}_{i,j}^U-{t}_{i,j}^L\leq d\}.\] 
By utilizing this approach, the expected revenue derived from the degenerated distribution $\widehat{D}$ would be lower than that based on the true distribution $D$ by a maximum of $\sum_{j\in K}d_j$, which is actually smaller than $kd$. This would yield a more accurate result than Theorem \ref{T1}.

%{{\bf{Large-Scale Data}}}

\begin{figure}[htbp]
\centering
\subfigure[Comparison of Type's Proportion Without Query]{
\begin{minipage}[t]{0.5\linewidth}
\centering
\label{2a}
\includegraphics[width=2.2in]{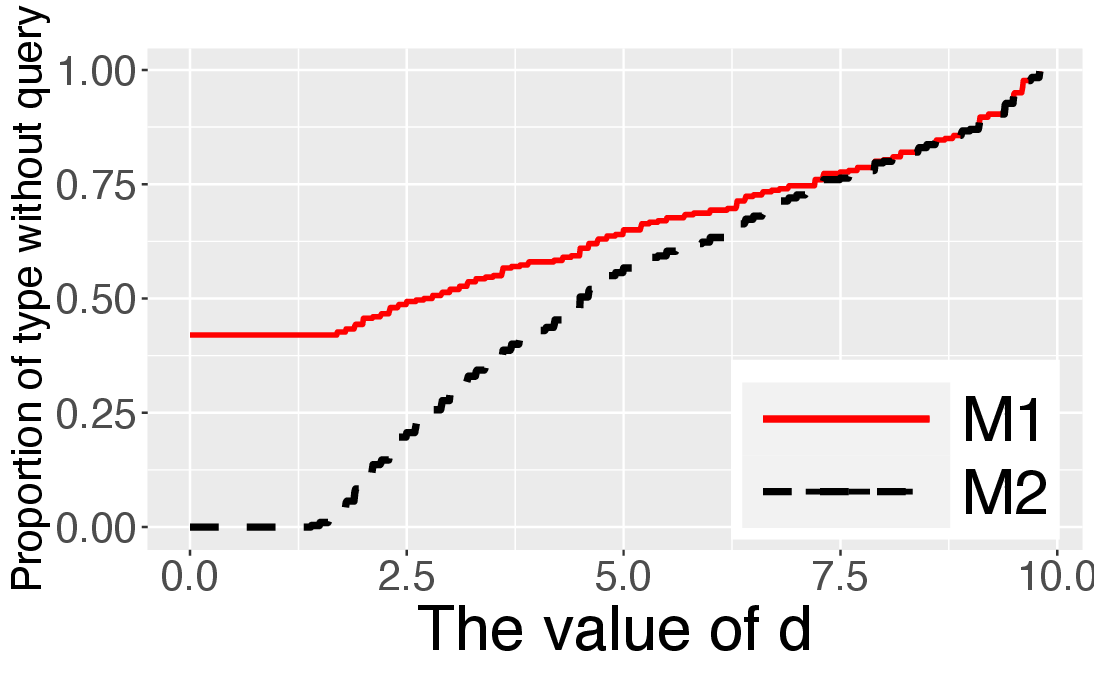}
%\caption{fig2}
\end{minipage}%
}%
\subfigure[Comparison of Confidence Rate]{
\begin{minipage}[t]{0.5\linewidth}
\centering
\label{2b}
\includegraphics[width=2.2in]{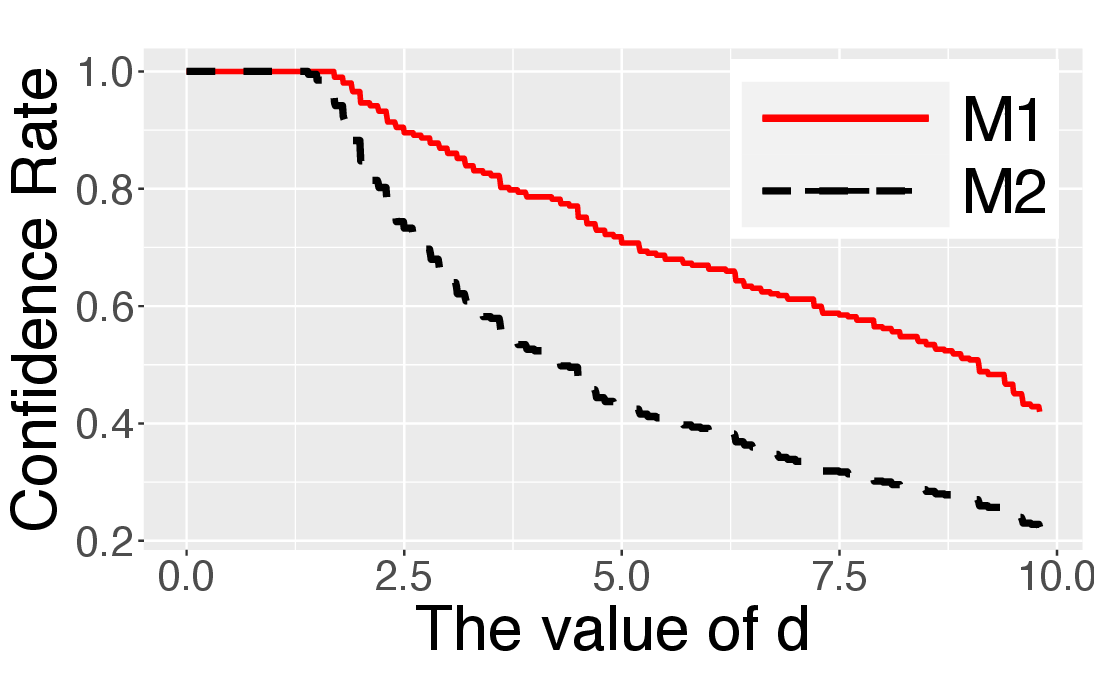}
%\caption{fig2}
\end{minipage}
}%
\centering
\caption{Comparison of the proportion of bidder types  without queries and comparison of the confidence rate when $m=30$, $N=10$.}
\label{fig:two1}
\end{figure}

To further compare M1 and M2, Figure \ref{2a} presents a comparison of the proportion of bidder types  without queries. It can be observed that after implementing the winnow-down method, approximately half of the queries are eliminated. When examined alongside Figure \ref{fig:one1}, it becomes apparent that these eliminated queries are redundant, underscoring the significant advantage of the winnow-down method in Section \ref{Winnow down the possible winners}.

Figure \ref{2b} illustrates the comparison of the confidence rate $(1-\alpha/2)^n$, where $n$ represents the total number of degenerated types distributions.
According to Theorem \ref{T1}, we can only guarantee that the regret is smaller than $kd$ with a probability of $(1-\alpha/2)^n$. We demonstrate in Section \ref{subsubsection3.3.5} that, when the confidence rate $\eta=0.9$ and $\alpha=0.01$, at most ${\color{black}{n=}}
21$ types distributions can degenerate to achieve a high level of precision. Nevertheless, our experimental results demonstrate that in practice, a greater number of distributions can be degenerated without significantly affecting the revenue. As observed in Figure \ref{2b}, even with a low confidence rate, the obtained regret in Figure \ref{1b} is still smaller than the upper bound $kd$ shown in Figure \ref{1c}. This finding provides support for the effectiveness of our proposed method, which employs the lower bound to construct one-point distributions. Moreover, it underscores the robustness of our approach in handling variations in the confidence rate.

We also conducted experiments with different parameter settings, specifically when $m=50$ and $N=30$, as well as when $m=50$ and $N=100$. Remarkably, in all of the cases, we arrived at the same conclusions and observed consistent results. The analysis of each of them can be found in Appendix \ref{ER} of the supplementary material.

\section{Conclusion}
\label{sec:conclusion}
This paper proposes a statistical learning method that leverages credible intervals to effectively reduce computation complexity and time costs in auction implementations. Our method utilizes kernel density estimation with historical data to estimate credible intervals. We propose two easily implementable strategies that ensure fairness, dominant-strategy incentive compatibility, and dominant-strategy individual rationality, while also reducing implementation costs.
Through simulation experiments using the Vickrey-Clarke-Groves mechanism, we demonstrate the effectiveness of our proposed strategies. We provide the lower bound of the expected revenue when using the lower credible bound method, highlighting the advantages of our approach. Our suggested mechanism offers a highly effective solution for sellers in multi-item auctions.
There are several directions for future research. For example, it is of interest to explore distribution-free methods for obtaining credible intervals, and we are currently working on this direction.

\bibliography{reference}

\begin{thebibliography}{32}
\providecommand{\natexlab}[1]{#1}
\providecommand{\url}[1]{\texttt{#1}}
\expandafter\ifx\csname urlstyle\endcsname\relax
  \providecommand{\doi}[1]{doi: #1}\else
  \providecommand{\doi}{doi: \begingroup \urlstyle{rm}\Url}\fi

\bibitem[Alaei(2014)]{6108212}
Saeed Alaei.
\newblock Bayesian combinatorial auctions: Expanding single buyer mechanisms to
  many buyers.
\newblock \emph{SIAM Journal on Computing}, 43\penalty0 (2):\penalty0 930--972,
  2014.

\bibitem[Alaei et~al.(2012)Alaei, Fu, Haghpanah, Hartline, and
  Malekian]{10.1145/2229012.2229017}
Saeed Alaei, Hu~Fu, Nima Haghpanah, Jason~D Hartline, and Azarakhsh Malekian.
\newblock Bayesian optimal auctions via multi-to single-agent reduction.
\newblock In \emph{13th ACM Conference on Electronic Commerce, EC'12}, pp.\
  ~17, 2012.

\bibitem[Ausubel et~al.(2006)Ausubel, Milgrom, et~al.]{ausubel2006lovely}
Lawrence~M Ausubel, Paul Milgrom, et~al.
\newblock The lovely but lonely vickrey auction.
\newblock \emph{Combinatorial auctions}, 17:\penalty0 22--26, 2006.

\bibitem[Babaioff et~al.(2017)Babaioff, Gonczarowski, and
  Nisan]{10.1145/3055399.3055426}
Moshe Babaioff, Yannai~A Gonczarowski, and Noam Nisan.
\newblock The menu-size complexity of revenue approximation.
\newblock In \emph{Proceedings of the 49th Annual ACM SIGACT Symposium on
  Theory of Computing}, pp.\  869--877, 2017.

\bibitem[Bergemann \& Schlag(2011)Bergemann and Schlag]{bergemann2011robust}
Dirk Bergemann and Karl Schlag.
\newblock Robust monopoly pricing.
\newblock \emph{Journal of Economic Theory}, 146\penalty0 (6):\penalty0
  2527--2543, 2011.

\bibitem[Brannan(2006)]{brannan2006first}
David~Alexander Brannan.
\newblock \emph{A first course in mathematical analysis}.
\newblock Cambridge University Press, 2006.

\bibitem[Brustle et~al.(2020)Brustle, Cai, and
  Daskalakis]{10.1145/3391403.3399541}
Johannes Brustle, Yang Cai, and Constantinos Daskalakis.
\newblock Multi-item mechanisms without item-independence: Learnability via
  robustness.
\newblock In \emph{Proceedings of the 21st ACM Conference on Economics and
  Computation}, pp.\  715--761, 2020.

\bibitem[Cai \& Daskalakis(2011)Cai and Daskalakis]{6108213}
Yang Cai and Constantinos Daskalakis.
\newblock Extreme-value theorems for optimal multidimensional pricing.
\newblock In \emph{2011 IEEE 52nd Annual Symposium on Foundations of Computer
  Science}, pp.\  522--531. IEEE, 2011.

\bibitem[Cai \& Daskalakis(2017)Cai and Daskalakis]{8104086}
Yang Cai and Constantinos Daskalakis.
\newblock Learning multi-item auctions with (or without) samples.
\newblock In \emph{2017 IEEE 58th Annual Symposium on Foundations of Computer
  Science (FOCS)}, pp.\  516--527. IEEE, 2017.

\bibitem[Cai \& Daskalakis(2022)Cai and Daskalakis]{10.1145/3490486.3538354}
Yang Cai and Constantinos Daskalakis.
\newblock Recommender systems meet mechanism design.
\newblock In \emph{Proceedings of the 23rd ACM Conference on Economics and
  Computation}, pp.\  897--914, 2022.

\bibitem[Cai \& Huang(2013)Cai and Huang]{DBLP:journals/corr/abs-1210-3560}
Yang Cai and Zhiyi Huang.
\newblock Simple and nearly optimal multi-item auctions.
\newblock In \emph{Proceedings of the Twenty-Fourth Annual ACM-SIAM Symposium
  on Discrete Algorithms}, pp.\  564--577. SIAM, 2013.

\bibitem[Cai et~al.(2012{\natexlab{a}})Cai, Daskalakis, and
  Weinberg]{10.1145/2213977.2214021}
Yang Cai, Constantinos Daskalakis, and S~Matthew Weinberg.
\newblock An algorithmic characterization of multi-dimensional mechanisms.
\newblock In \emph{Proceedings of the forty-fourth annual ACM symposium on
  Theory of computing}, pp.\  459--478, 2012{\natexlab{a}}.

\bibitem[Cai et~al.(2012{\natexlab{b}})Cai, Daskalakis, and Weinberg]{6375290}
Yang Cai, Constantinos Daskalakis, and S~Matthew Weinberg.
\newblock Optimal multi-dimensional mechanism design: Reducing revenue to
  welfare maximization.
\newblock In \emph{2012 IEEE 53rd Annual Symposium on Foundations of Computer
  Science}, pp.\  130--139. IEEE, 2012{\natexlab{b}}.

\bibitem[Cai et~al.(2013)Cai, Daskalakis, and Weinberg]{cai2013understanding}
Yang Cai, Constantinos Daskalakis, and S~Matthew Weinberg.
\newblock Understanding incentives: Mechanism design becomes algorithm design.
\newblock In \emph{2013 IEEE 54th Annual Symposium on Foundations of Computer
  Science}, pp.\  618--627. IEEE, 2013.

\bibitem[Cai et~al.(2016)Cai, Devanur, and Weinberg]{cai2016duality}
Yang Cai, Nikhil~R Devanur, and S~Matthew Weinberg.
\newblock A duality based unified approach to bayesian mechanism design.
\newblock In \emph{Proceedings of the forty-eighth annual ACM symposium on
  Theory of Computing}, pp.\  926--939, 2016.

\bibitem[Casella et~al.(2004)Casella, Robert, and
  Wells]{casella2004generalized}
George Casella, Christian~P Robert, and Martin~T Wells.
\newblock Generalized accept-reject sampling schemes.
\newblock \emph{Lecture Notes-Monograph Series}, pp.\  342--347, 2004.

\bibitem[Chawla et~al.(2007)Chawla, Hartline, and
  Kleinberg]{chawla2007algorithmic}
Shuchi Chawla, Jason~D Hartline, and Robert Kleinberg.
\newblock Algorithmic pricing via virtual valuations.
\newblock In \emph{Proceedings of the 8th ACM Conference on Electronic
  Commerce}, pp.\  243--251, 2007.

\bibitem[Chawla et~al.(2010)Chawla, Hartline, Malec, and
  Sivan]{10.1145/1806689.1806733}
Shuchi Chawla, Jason~D Hartline, David~L Malec, and Balasubramanian Sivan.
\newblock Multi-parameter mechanism design and sequential posted pricing.
\newblock In \emph{42nd ACM Symposium on Theory of Computing, STOC 2010}, pp.\
  311--320, 2010.

\bibitem[Che \& Yoo(2001)Che and Yoo]{che2001optimal}
Yeon-Koo Che and Seung-Weon Yoo.
\newblock Optimal incentives for teams.
\newblock \emph{American Economic Review}, 91\penalty0 (3):\penalty0 525--541,
  2001.

\bibitem[Clarke(1971)]{clarke1971multipart}
Edward~H Clarke.
\newblock Multipart pricing of public goods.
\newblock \emph{Public choice}, pp.\  17--33, 1971.

\bibitem[Daskalakis(2015)]{10.1145/2845926.2845928}
Constantinos Daskalakis.
\newblock Multi-item auctions defying intuition?
\newblock \emph{ACM SIGecom Exchanges}, 14\penalty0 (1):\penalty0 41--75, 2015.

\bibitem[Dughmi et~al.(2014)Dughmi, Han, and
  Nisan]{DBLP:journals/corr/DughmiHN14}
Shaddin Dughmi, Li~Han, and Noam Nisan.
\newblock Sampling and representation complexity of revenue maximization.
\newblock In \emph{International Conference on Web and Internet Economics},
  pp.\  277--291. Springer, 2014.

\bibitem[Goldner \& Karlin(2016)Goldner and
  Karlin]{10.1007/978-3-662-54110-4_12}
Kira Goldner and Anna~R Karlin.
\newblock A prior-independent revenue-maximizing auction for multiple additive
  bidders.
\newblock In \emph{International Conference on Web and Internet Economics},
  pp.\  160--173. Springer, 2016.

\bibitem[Han \& Dai(2025)Han and Dai]{han2025online}
Jiale Han and Xiaowu Dai.
\newblock Online auction design using distribution-free uncertainty
  quantification with applications to e-commerce.
\newblock \emph{Journal of the American Statistical Association}, pp.\  1--12,
  2025.

\bibitem[Joseph et~al.(2016)Joseph, Kearns, Morgenstern, and
  Roth]{NIPS2016_eb163727}
Matthew Joseph, Michael Kearns, Jamie~H Morgenstern, and Aaron Roth.
\newblock Fairness in learning: Classic and contextual bandits.
\newblock \emph{Advances in neural information processing systems}, 29, 2016.

\bibitem[Morgenstern \& Roughgarden(2016)Morgenstern and
  Roughgarden]{DBLP:journals/corr/MorgensternR16}
Jamie Morgenstern and Tim Roughgarden.
\newblock Learning simple auctions.
\newblock In \emph{Conference on Learning Theory}, pp.\  1298--1318. PMLR,
  2016.

\bibitem[Myerson(1981)]{myerson1981optimal}
Roger~B Myerson.
\newblock Optimal auction design.
\newblock \emph{Mathematics of operations research}, 6\penalty0 (1):\penalty0
  58--73, 1981.

\bibitem[Scott(1992)]{scott1992multivariate}
DW~Scott.
\newblock Multivariate density estimation.
\newblock \emph{Multivariate Density Estimation}, 1992.

\bibitem[Strassen(1965)]{strassen1965existence}
Volker Strassen.
\newblock The existence of probability measures with given marginals.
\newblock \emph{The Annals of Mathematical Statistics}, 36\penalty0
  (2):\penalty0 423--439, 1965.

\bibitem[Vickrey(1961)]{vickrey1961counterspeculation}
William Vickrey.
\newblock Counterspeculation, auctions, and competitive sealed tenders.
\newblock \emph{The Journal of finance}, 16\penalty0 (1):\penalty0 8--37, 1961.

\bibitem[Yao(2014)]{yao2014n}
Andrew Chi-Chih Yao.
\newblock An n-to-1 bidder reduction for multi-item auctions and its
  applications.
\newblock In \emph{Proceedings of the twenty-sixth annual ACM-SIAM symposium on
  Discrete algorithms}, pp.\  92--109. SIAM, 2014.

\bibitem[Zambom \& Ronaldo(2013)Zambom and Ronaldo]{zambom2013review}
Adriano~Z Zambom and Dias Ronaldo.
\newblock A review of kernel density estimation with applications to
  econometrics.
\newblock \emph{International Econometric Review}, 5\penalty0 (1):\penalty0
  20--42, 2013.

\end{thebibliography}
\bibliographystyle{iclr2026_conference}

\appendix

\newpage
\setcounter{equation}{0}
\setcounter{section}{0}
\renewcommand{\theequation}{C.\arabic{equation}}

\section{Proofs}
\noindent 
In this section, we present the technical proofs for the main results outlined in our paper. We begin by introducing Lemma \ref{lemma1}, which serves as a foundational component for our subsequent proofs. Following that, in Section \ref{fair}, we demonstrate the proof of Proposition \ref{fairness}, which validates the fairness of our method in the process of winnowing down potential winners. Moving on to Section \ref{3.5}, we provide the proof of Theorem \ref{New1}, establishing the incentive compatibility of our proposed mechanism. Finally, we delve into Section \ref{Theorem} where we prove the lower bound of the revenue generated by the degenerated VCG mechanism as outlined in Theorem \ref{T1}.

\begin{lemma}[Bernoulli's Inequality]\label{lemma1}
For any real number $x\geq -1$ and any natural number $n$, $(1+x)^n\geq 1+nx$.
\end{lemma}
\begin{proof}[Proof of Lemma \ref{lemma1}]
    See \cite{brannan2006first}. A First Course in Mathematical Analysis. \emph{Cambridge University Press. p. 20.}
\end{proof}

\subsection{Proof of Proposition \ref{fairness}}\label{fair}
\begin{proof}~
\noindent 

For each item $j\in S \subset [N]$ and every bidder $B_k$, the revised mechanism ensures $\alpha$-fairness as follows:

1. If $B_k$ is not in $\mathcal{B}_j$ defined in Eq.~\eqref{Bb}, according to the definition of $\mathcal{B}_j$, we have ${t}_{k,j}^U\leq{t}_{i_{*}^j,j}^L$. Based on the definition of the credible intervals, with probability $1-\alpha/2$, $t_{k,j}$ is smaller than ${t}_{k,j}^U$. Similarly, with probability $1-\alpha/2$, the winner's bid $t_{i_{*}^j,j}$ is bigger than ${t}_{i_{*}^j,j}^L$. Since the bidders' type distributions are independent, we can write:
\begin{equation*}
\begin{aligned}
\mathbb{P}(t_{k,j}< t_{i_{*}^j,j})
&\geq \mathbb{P}(t_{k,j}<{t}_{k,j}^U\leq{t}_{i_{*}^j,j}^L<t_{i_{*}^j,j})\\ 
& =\mathbb{P}(t_{k,j}<{t}_{k,j}^U)\mathbb{P}({t}_{i_{*}^j,j}^L<t_{i_{*}^j,j})\\
& =(1-\alpha/2)^2\geq 1-\alpha.
\end{aligned}
\end{equation*}
The last inequality holds because $0\leq\alpha\leq 1$ and Lemma \ref{lemma1}.

2. If $B_k$ is in $\mathcal{B}_j$,  because the original mechanism is fair, which means it satisfies the fairness definition \ref{beta-fair} with $\delta=0$, so that: 
\[\mathbb{P}(t_{k,j}<t_{i_{*}^j,j})=1\geq 1-\alpha.\]

In conclusion, the revised mechanism guarantees $\alpha$-fairness.
\end{proof}

\subsection{Proof of Theorem \ref{New1}}\label{3.5}
\begin{proof}~
First, we prove the $\delta$-DSIR property of the mechanism.
For any bid $b$, the utility of a bidder $B_i$ with a private type $t_i$ in mechanism $\widehat{M}$, with allocations $(\hat{q}^j_i(b))$ and payments $(\hat{p}_1(b),\dots,\hat{p}_m(b))$, is formulated as:
\[u_i(t_i,\widehat{M}(b))=\sum_{j-1}^N t_{i,j}\hat{q}^j_i(b)-\hat{p}_i(b).\]

Because $\widehat{M}$ is DSIR with respect to $\widehat{D}$, it guarantees the following property for any private type $\hat{t}$ sampled from the distribution $\widehat{D}$:
\begin{equation}\label{C.1}
	\begin{split}
	u_i(\hat{t}_i,\widehat{M}(\hat{t}))&= \sum_{j\in Q_i} t_{i,j}\hat{q}^j_i(\hat{t})+\sum_{j\in L_i} \hat{t}_{i,j}\hat{q}^j_i(\hat{t})-\hat{p}_i(\hat{t})\\
	&=\sum_{j\in Q_i} t_{i,j}\hat{q}^j_i(\hat{t})+\sum_{j\in L_i} t^L_{i,j}\hat{q}^j_i(\hat{t})-\hat{p}_i(\hat{t})\\
        &\geq 0.
	\end{split}
\end{equation}
Since the mechanism $M$ only relies on query access to $\widehat{M}$, it ensures that if the bidder's total type is $t$, $\widehat{M}$ will solely make queries on types in set $Q$ and eventually operate on $\hat{t}$. Moreover, $M$ preserves the same allocation and payment functions as $\widehat{M}$ after the querying process, denoted as $M(t) = \widehat{M}(\hat{t})$.

Given that $\mathbb{P}(t_{i,j}\geq t^L_{i,j})=1-\alpha/2$, we can conclude that with a probability of $(1-\alpha/2)^{n_i}$, for every $j$ in $L_i$, $t_{i,j}\geq t^L_{i,j}$. Consequently, with a probability of $(1-\alpha/2)^{n_i}$, for all types $t$ sampled from the distribution $D$,
\begin{equation}\label{C.2}
	\begin{split}
	u_i({t}_i,{M}(t))&= u_i({t}_i,\widehat{M}(\hat{t}))\\
 &=\sum_{j\in Q_i} t_{i,j}\hat{q}^j_i(\hat{t})+\sum_{j\in L_i} {t}_{i,j}\hat{q}^j_i(\hat{t})-\hat{p}_i(\hat{t})\\
	&\geq \sum_{j\in Q_i} t_{i,j}\hat{q}^j_i(\hat{t})+\sum_{j\in L_i} t^L_{i,j}\hat{q}^j_i(\hat{t})-\hat{p}_i(\hat{t})\\
        &=u_i(\hat{t}_i,\widehat{M}(\hat{t}))\geq 0.
	\end{split}
\end{equation}
Then, for every $i$, we have
\[\mathbb{P}\left(u_i({t}_i,{M}(t))\geq 0\right)\geq \left(1-\frac{\alpha}{2}\right)^{\sum_{i=1}^m n_i}=\left(1-\frac{\alpha}{2}\right)^{n}\geq 1-\frac{\alpha n}{2}.\]
The last inequality holds because $0\leq\alpha\leq 1$ and Lemma \ref{lemma1}.
Therefore, our mechanism is ${M}$ is DSIR w.r.t. ${D}$ with $\delta = \frac{\alpha n}{2}$.

Next, we prove the DSIC property of the mechanism. 
%Because we only use query access to types in set $Q$, from the assumption of the theorem, the allocation and payment rule of ${M}$ will only be influenced by the change of the types in $Q$. 
%and the types in $Q$ has the same distribution in both $D$ and $\hat{D}$, 
Consider any bidder $B_i$ with type distribution $\widehat{D}$. Only the types for items in $Q_i$ can take different values due to the definition of the degenerated distribution $\widehat{D}$. Since the mechanism $\widehat{M}$ is DSIC with respect to $\widehat{D}$, untruthful reporting of types by bidder $B_i$ will result in a reduction in her utility in $\widehat{M}$. Thus, we have:
\begin{equation}\label{new1}
u_i(\hat{t}_i,\widehat{M}(\hat{t}_i,\hat{t}_{-i}))\geq u_i(\hat{t}_i,\widehat{M}(\hat{t}'_i,\hat{t}_{-i})),
\end{equation}
where the difference between $\hat{t}_i$ and $\hat{t}'_i$ occurs only in the items in $Q_i$.

Based on the assumption of the theorem, untruthful reporting of the types of items in $Q_i$ will only influence the allocation and payment for the items in $Q_i$. Let $\hat{p}_i(\hat{t})=\hat{p}_i^1(\hat{t})+ \dots+\hat{p}_i^N(\hat{t})$, where $\hat{p}_i^j(\hat{t})$ is the payment that bidder $B_i$ ultimately pays for item $j$ in the Mechanism $\widehat{M}$ given the type $\hat{t}\sim \widehat{D}$. For any item $j$ in $L_i$, we have $\hat{q}^j_i(\hat{t}_i,\hat{t}_{-i})=\hat{q}^j_i(\hat{t}'_i,\hat{t}_{-i})$ and $\hat{p}_i^j(\hat{t}_i,\hat{t}_{-i})=\hat{p}_i^j(\hat{t}'_i,\hat{t}_{-i})$.

Based on Eq.\eqref{C.1}, we can rewrite Eq.\eqref{new1} as follows:
\begin{equation*}
    \sum_{j\in Q_i} {t}_{i,j}\hat{q}^j_i(\hat{t}_i,\hat{t}_{-i})- \sum_{j\in Q_i}\hat{p}_i^j(\hat{t}_i,\hat{t}_{-i}) \geq     \sum_{j\in Q_i} {t}_{i,j}\hat{q}^j_i(\hat{t}'_i,\hat{t}_{-i})- \sum_{j\in Q_i}\hat{p}_i^j(\hat{t}'_i,\hat{t}_{-i}).
\end{equation*}

This equation is equivalent to:
\begin{equation*}
    u_i({t}_i,{M}({t}_i,{t}_{-i}))\geq u_i({t}_i,{M}({t}'_i,{t}_{-i})),
\end{equation*}
based on Eq.~\eqref{C.2}. Hence, we can conclude that ${M}$ is DSIC with respect to $D$.
\noindent 
\end{proof}

%Let the expected revenue of $M$ based on $D$ be $\text{REV}(M, D)$, and the expected revenue of $\widetilde{M}$ based on $D$ be $\text{REV}(\widetilde{M}, D)$. The difference between $M$ and $\widetilde{M}$ is that we degenerate the distribution of the types in $L$ as a one-point distribution, and that is the reason why the revenue is reduced. And for each bidder $B_i$, the payment will only change in the items in $L_i^*$. We have 

%\begin{equation}
%	\begin{split}
%	\text{REV}(\widetilde{M}, D)-\text{REV}(M,D)&=\mathbb{E}_{t\sim D}\left[\sum_{i\in [m]}\tilde{p}_i(t)\right]-\mathbb{E}_{t\sim D}\left[\sum_{i\in [m]}\hat{p}_i(\hat{t})\right]\\
%        &= \mathbb{E}_{t\sim D}\left[\sum_{i\in [m]}\sum_{j\in L_i^*}(\tilde{p}_i^j(t)-\hat{p}_i^j(\hat{t})) \right]
%	\end{split}
%\end{equation}
%$\hat{p}_i^j(\hat{t})\leq t^L_{i,j}$, with probability $(1-\alpha/2)$, $\tilde{p}_i^j(t)\leq t^L_{i,j}+d$

%with probability $(1-\alpha/2)$, $t_{i,j}\leq t^L_{i,j}+d$
\subsection{Proof of Theorem \ref{T1}}\label{Theorem}
\begin{proof}~
\noindent 
If the bidders have total types $t\sim D$, both $\widehat{M}$ and $M$ will generate the payment and allocation based on the revised types $\hat{t}\sim\widehat{D}$ as defined in Section \ref{Designate the value}, because $M$ uses only query access to $\widehat{M}$. On the other hand, $M_0$ directly generates the payment and allocation based on $t$. The difference between $t$ and $\hat{t}$ only occurs for the types in the set $L$ defined in Eq.~\eqref{L}.

Let $p_i(t) = p_i^1(t) + \dots + p_i^N(t)$, where $p_i^j(t)$ is the payment that bidder $B_i$ ultimately pays for item $j$ in Mechanism $M_0$ given the type $t\sim D$. Similarly, we can define $\hat{p}_i(\hat{t}) = \hat{p}_i^1(\hat{t}) + \dots + \hat{p}_i^N(\hat{t})$, where $\hat{p}_i^j(\hat{t})$ is the payment that bidder $B_i$ ultimately pays for item $j$ in Mechanism $\widehat{M}$ given the type $\hat{t}\sim \widehat{D}$.

In the original VCG mechanism $M_0$, the payment for item $j$ is defined as $P_j(t)\equiv\sum_{i=1}^{m}p_i^j(t)$. This payment is determined as the second highest bid among $t_{i,j}$ for all $i \in [m]$. In mechanism $M$ or $\widehat{M}$, the payment for item $j$ is defined as $P_j(\hat{t})\equiv\sum_{i=1}^{m}\hat{p}_i^j(\hat{t})=\sum_{i=1}^{m}{p}_i^j(\hat{t})$. This payment is determined as the second highest bid among $\hat{t}_{i,j}$ for all $i \in [m]$.

For every $i \in [m]$, $\mathbb{P}(\hat{t}_{i,j} \geq t_{i,j}-d)=(1-\alpha/2)$ holds for each item $j \in K$ as defined in Eq.~\eqref{K}. Let $n_{j,1}$ be the number of elements in the set $\{i|0<{t}_{i,j}^U-{t}_{i,j}^L\leq d,i\in[m]\}$, given by:
\begin{equation*}
n_{j,1} = \Bigl|\{i|0<{t}_{i,j}^U-{t}_{i,j}^L\leq d,i\in[m]\}\Bigr|.
\end{equation*}

1. If item $j \in K$, with probability $(1-\alpha/2)^{n_{j,1}}$, $\hat{t}_{i,j} \geq t_{i,j}-d$ holds for all $i\in [m]$. This implies:
\begin{equation*}
P_j(\hat{t}) > P_j(t)-d.
\end{equation*}

2. When $j \notin K$, for any $i\in[m]$, if ${t}_{i,j}^U-{t}_{i,j}^L=0$, then with probability $(1-\alpha)$, $\hat{t}_{i,j}>t_{i,j}$. If ${t}_{i,j}^U-{t}_{i,j}^L\neq0$, then $t_{i,j}$ is in $Q$, so $\hat{t}_{i,j}=t_{i,j}$. Let $n_{j,2}$ be the number of elements in the set $\{i|{t}_{i,j}^U-{t}_{i,j}^L=0,i\in[m]\}$, given by:
\begin{equation*}
n_{j,2} = \Bigl|\{i|{t}_{i,j}^U-{t}_{i,j}^L=0,i\in[m]\}\Bigr|.
\end{equation*}

Then with probability $(1-\alpha/2)^{n_{j,2}}$, we have:
\begin{equation*}
P_j(\hat{t}) \geq P_j(t).
\end{equation*}

Altogether, we have that with probability $(1-\alpha/2)^{\sum_{j=1}^N(n_{j,1}+n_{j,2})}=(1-\alpha/2)^{n}$, the following inequality holds:
  \begin{equation}\label{eq1}
	\begin{split}
	\sum_{j\in [N]} P_j(\hat{t}) &\geq \sum_{j\in [N]}P_j(t)-\sum_{j\in K}d\\
	&\geq \sum_{j\in [N]}P_j(t)-kd.
	\end{split}
    \end{equation}
    
By taking expectations on both sides of Eq.~\eqref{eq1}, we have:
    \begin{equation}\label{eq2}
   \mathbb{E}_{\hat{t}\sim \widehat{D}}\left[\sum_{j\in [N]}P_j(\hat{t})\right] \geq\mathbb{E}_{t\sim D}\left[\sum_{j\in [N]}P_j(t)\right]- kd,
    \end{equation}
where we used the fact that only the term $\sum_{j\in [N]}P_j(\hat{t})$ is random over the change of $\hat{t}$, and the term $\sum_{j\in [N]}P_j(t)$ is random over the change of $t$.

Because for any bidders' total bid $b$, we have:
\[\sum_{j\in [N]}P_j(b)=\sum_{j=1}^N\sum_{i=1}^{m}p_i^j(b)=\sum_{i=1}^{m}\sum_{j=1}^N p_i^j(b)=\sum_{i=1}^{m} p_i(b),\]
we can rewrite Eq.~\eqref{eq2} as:
    \begin{equation*}
   \mathbb{E}_{\hat{t}\sim \widehat{D}}\left[\sum_{i=1}^{m} {p}_i(\hat{t})\right] \geq\mathbb{E}_{t\sim D}\left[\sum_{i=1}^{m} p_i(t)\right]- kd,
    \end{equation*}
which can be further simplified to:
 \[\text{REV}({M},{D}) \geq \text{REV}({M}_0,{D})-kd.\]

\end{proof}

%The estimation of the type $\hat{t}$ is made possible by leveraging historical data and the confidence intervals derived from Section \ref{subsubsection3.2.1}. 
%e note that $\hat{t}$ is unique under these conditions. Meanwhile, the bids estimated in type I are determined by the actual values given by the bidders. The only source of uncertainty lies in type II estimated bids, specifically within the set $\{\hat{t}_{ij} | i\in[m], j\in L_i\}$. It is within this subset that the true value may differ from the estimated data.
%Given a fixed $t$, the following relationship holds for every $i \in [m]$ with a probability of $(1-\alpha/2)$,
%    \begin{equation}\label{ttt}
 %       \hat{t}_{ij} > t_{ij}-d,  \text{ where } j \in K \text{ as defined in Eq.~\eqref{K}}.   
 %   \end{equation}

\section{Allocation Complexity Analysis}\label{Complexity}
In this section, we establish the lower bound for the number of bundle selections in scenarios where each bound is constrained to a maximum of 2 items. This analysis serves to emphasize the inherent intractability of the allocation problem, particularly when confronted with a considerable number of items.

\begin{proposition}\label{prop1}
When each bundle has at most 2 items, the number of bundle selections grows exponentially as a function of the number of items $N$. Specifically, the number of potential bundle selections exceeds $(2N/3)^{\lfloor N/3\rfloor}$.
\end{proposition}

\begin{proof}~
    \noindent When each bundle just has one item, We only have 1 bundle selection rule: 
    \begin{equation*}
    \bigl\{\{1\},\{2\},\{3\},\dots,\{N\}\bigr\}.
    \end{equation*}

When at least a bundle has 2 items, We use $\lfloor N/2 \rfloor$ to denote the maximum integer which less than or equal to N/2. Then
if we have $i\ (i \leq \lfloor N/2 \rfloor)$ 2-item bundles in total, the number of bundle selections is 

\begin{equation*}
    \frac{\tbinom{N}{2i} \tbinom{2i}{2} \tbinom{2i-2}{2}\cdots\tbinom{4}{2}}{i!}=\frac{N!}{(N-2i)!i!2^i}.
\end{equation*}

When $N \geq 3i$, $N-i\geq 2i$, we have $(N-2i+1)(N-2i)\dots(N-i)\geq1\times2\times\dots\times 2i$, and $N-i\leq \frac{2N}{3}$, so 

\begin{equation}\label{!}
	\begin{split}
	\frac{N!}{(N-2i)!i!2^i} &= \frac{(N-2i+1)(N-2i) \cdots (N)}{1\times 2\times \cdots \times i \times 2^i}\\
	&= \frac{(N-2i+1)(N-2i) \cdots (N)}{2\times 4\times \dots 
\times 2i}\\
        &\geq (N-i+1)(N-i+2)\cdots N\\
        &\geq (N-i)^{i} \geq \big(\frac{2N}{3}\big)^i.
	\end{split}
\end{equation}

The total number of bundle selections is the sum of the number of bundle selections for $i\ (i \leq \lfloor N/2 \rfloor)$, which is
\begin{equation*}
	\begin{split}
	\sum_{i=1}^{\lfloor N/2 \rfloor}\frac{N!}{(N-2i)!i!2^i} &\geq \sum_{i=1}^{\lfloor N/3\rfloor}\frac{N!}{(N-2i)!i!2^i}\geq \sum_{i=1}^{\lfloor N/3\rfloor}\big(\frac{2N}{3}\big)^i\\
        &= \frac{2N}{2N-1}\Big(\big(\frac{2N}{3}\big)^{\lfloor N/3\rfloor}-1\Big)\geq \big(\frac{2N}{3}\big)^{\lfloor N/3\rfloor}-1,
	\end{split}
\end{equation*}
where the first inequality uses $N/2>N/3$, the second inequality uses inequality (\ref{!}), and the third inequality uses $\frac{2N}{2N-1}<1$.
Hence the total number of bundle selections is
\begin{equation*}
1+\sum_{i=1}^{\lfloor N/2 \rfloor}\frac{N!}{(N-2i)!i!2^i} \geq \left(\frac{2N}{3}\right)^{\lfloor N/3\rfloor}.
\end{equation*}
This completes the proof.
\end{proof}

\section{Algorithm}\label{Algorithm}

\begin{figure}[ht!]
        \centering
	\includegraphics[scale=0.4]{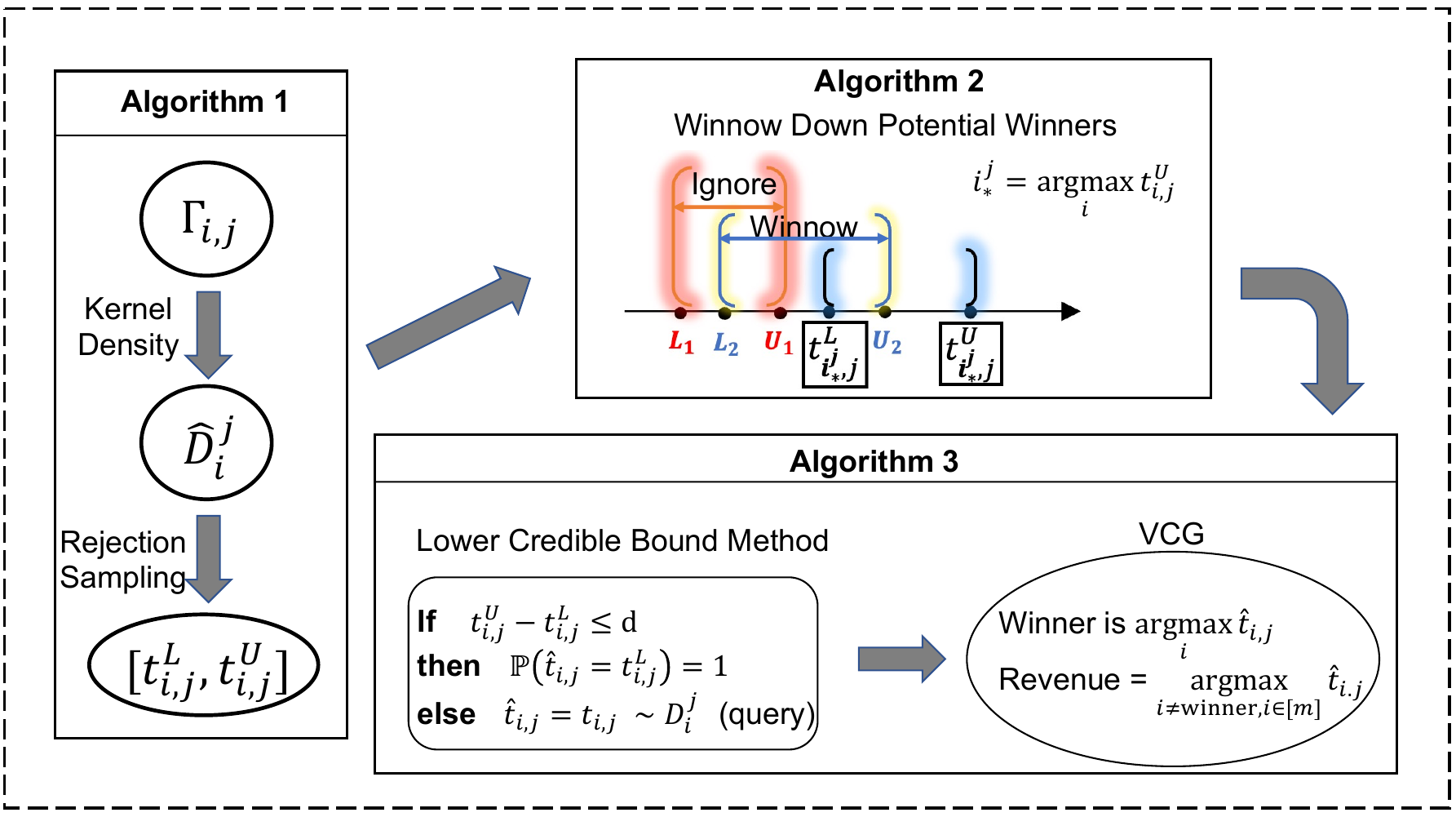}
        \vspace{-0.1in}
	\caption{Flow chart of the proposed mechanism.}
	\label{fig:flow chart}
\end{figure}

In this section, we give the three algorithms based on the methods proposed in Section \ref{sec:method}.
A flow chart of the proposed mechanism is shown in Figure \ref{fig:flow chart}. First, we employ kernel density estimation to estimate the distribution of the data set $\Gamma_{i,j}$ and obtain an estimated distribution $\widehat{D}_{i}^j$. Then, we use rejection sampling to calculate a $(1-\alpha)$ credible interval, denoted by $[{t}_{i,j}^L,{t}_{i,j}^U]$. This procedure is presented as Algorithm \ref{alg:one}. Second, we implement a winnow-down procedure for each item $j$ by considering only the intervals that intersect with the interval with the largest upper bound ${t}^U_{i^j_{*},j}$, while disregarding the others. For example, as shown in Figure \ref{fig:flow chart}, the interval $[L_2,U_2]$ is winnowed down, while $[L_1,U_1]$ is disregarded. 
 This procedure is presented as Algorithm \ref{alg:two}.
Finally, we employ the lower credible bound method on the VCG mechanism and obtain the implemented VCG mechanism using the degenerated types distribution.  This procedure is presented as Algorithm \ref{alg:three}.

\subsection{Algorithm of Distribution and Credible Intervals Estimation}\label{A1}
Algorithm \ref{alg:one} shows the details of estimating  bidder types  and constructing credible intervals, as outlined in Section  \ref{subsubsection3.2.1}. The inputs to the algorithm include the number of bidders, $m$, the number of items, $N$, a historical dataset, $\Gamma_{i,j}$, consisting of ${n_{i,j}}$ independently and identically distributed samples of the distribution $D^j_{i}$ for $i\in[m],j\in[N]$, the minimum and maximum values, $a_j$ and $b_j$, for item $j$, a critical value, $\alpha$, and the sampling number $\mathcal{N}$. The output of the algorithm is a $(1-\alpha)$ credible interval for each type, $t_{i,j}$, with lower bound, ${t}_{i,j}^L$, and upper bound, ${t}_{i,j}^U$.
The algorithm begins by utilizing a kernel density function to estimate the distribution, $\widehat{D}_{i}^j$. Then, a uniform distribution on $[a_j,b_j]$ is employed through rejection sampling. We sample $\mathcal{N}$ results from $\widehat{D}_{i}^j$, and take the $\alpha/2$th quantile as the lower bound, ${t}_{i,j}^L$, and take the $(1-\alpha/2)$th quantile as the upper bound, ${t}_{i,j}^U$. 
\begin{algorithm}[ht!]	
	\KwIn{$m$, $N$, historical dataset $\Gamma_{i,j}=\{x_{i,j}^1,x_{i,j}^2,\dots, x_{i,j}^{n_{i,j}}\}$, minimum value $a_{j}$, maximum value $b_{j}$, $a_{j}<b_{j}$, $i\in[m], j\in[N]$, $\alpha\in[0,1]$, $\mathcal{N}\in\mathbb{N}$}
	\KwOut{Lower bound ${t}_{i.j}^L$ and upper bound ${t}_{i,j}^U$ for $i\in[m],j\in[N]$}
        \For{$i\in[m], j\in[N]$}{
        t=0;\
        
        \Repeat{t = $\mathcal{N}$}{
         $K(t)=\frac{1}{\sqrt{2\pi}}e^{-t^2/2}$; $h=1.06\cdot \min{\bigl \{\hat{\sigma},\frac{IQR}{1.34}}\bigr\}\cdot {n_{i,j}^{-1/5}}$;
         $\widehat{D}_{i}^j(x) =\frac{1}{{n_{i,j}}h}\sum^{{n_{i.j}}}_{s=1}K(\frac{x-x_{i,j}^s}{h})$; $g(x)=\frac{1}{b_j-a_j}\mathbb{I}\{x\in[a_j,b_j]\}$\;
        %do simulation by rejection sampling from $f_{ij}$ to get $s$ results:
         sample a point $x\sim$Unif$[a_{j}, b_{j}]$\;
        choose $c$ big enough s.t. $\widehat{D}_{i}^j(x_0)/cg(x_0)<1$ for $\forall x_0\in [a_{j}, b_{j}]$\;
        sample a point $u\sim$Unif[0,1]\;
        \If{$u < \widehat{D}_{i}^j(x)/cg(x)$}{
            accept $x$\;
            t $\gets$ t+1
        }
		}
        ${t}_{i,j}^L$= $\alpha/2$th quantile of the accepted $\mathcal{N}$ results\;
        ${t}_{i,j}^U$= ($1-\alpha/2$)th quantile of the accepted $\mathcal{N}$ results
        }
	\caption{Estimation of distribution and credible intervals for bidder types }
	\label{alg:one}
\end{algorithm}

\subsection{Algorithm of Winnowing Down Potential Winners}\label{A2}
In Algorithm \ref{alg:two}, we present a method for identifying potential winners among bidders based on the linkage of their confidence intervals, as outlined in Section \ref{Winnow down the possible winners}.
The input to the algorithm includes the number of bidders, $m$, the number of items, $N$, and confidence intervals, $[{t}_{i,j}^L,{t}_{i,j}^U]$, for each $t_{i,j}$, where  $i\in [m], j \in [N]$. The output is a set of potential winners, $\mathcal{B}_j$, for each item $j$.
For each item $j$, we identify the bidder, $i_{*}^j$, with the highest upper bound, ${t}_{i_{*}^j,j}^U$, of the credible interval. We then consider only those bidders whose credible intervals for item $j$ are linked to the interval of bidder $i_{*}^j$. These bidders are then added to the set of potential winners, $\mathcal{B}_j$. The number of bidders that we will disregard for item $j$ can be determined by $N - |\mathcal{B}_j|$. Furthermore, the total number of bids that will be neglected in the multi-item auction is given by 
\begin{equation}\label{mstar}
    m^* \equiv mN - \sum_{j=1}^N |\mathcal{B}_j|.
\end{equation}

\begin{algorithm}[ht!]
	\KwIn{$m$, $N$, $[{t}_{i,j}^L,{t}_{i,j}^U] \text{ for } i\in[m], j\in[N]$}
	\KwOut{For each $j\in[N]$, output a bidders' set $\mathcal{B}_j$}
        \For{$j \in [N]$}{
        $\mathcal{B}_j=\emptyset$\;
        $i_{*}^j=\text{argmax}_i {t}_{i,j}^U$\;
            \For{$i \in [m]$}{
                \If{${t}_{i,j}^U>{t}_{i_{*}^j,j}^L$}{
                $\mathcal{B}_j\gets\mathcal{B}_j\cup B_i$
                }
            } 
        }
    
	\caption{Winnow down the potential winners}
	\label{alg:two}
\end{algorithm}

\subsection{Algorithm of The Implemented VCG Mechanism }\label{A3}
\noindent 
In this section, we begin by providing a detailed explanation of how to determine the suitable range of values for $d$. Subsequently, we present the algorithms for the implemented VCG mechanism.

\subsubsection{Tuning Parameters for The Degenerated VCG Mechanism}
\label{subsubsection3.3.5}

\noindent 
Theorems \ref{New1} and \ref{T1} provide a theoretical foundation for our mechanism. 
We consider two key tuning parameters, $n$ in Eq.~\eqref{eqn:defofn} and $d$ in Eq.~\eqref{eqn:defofli}.

First, we consider the tuning of $n$. In order to achieve the desired IR and expected revenue in high confidence, the value of $(1-\alpha/2)^n$ needs to be as large as possible. For any confidence rate $\eta\in[0,1]$,  let
\[(1-\alpha/2)^{n}\geq \eta.\]
This yields the inequality
$n \leq \log(\eta)/\log(1-\alpha/2)$.
Additionally, as defined in Section \ref{Designate the value}, $n$ must be less than $mN$. Therefore, we can set
\[n\leq n^{*}\equiv\min\left\{\frac{\log(\eta)}{\log(1-\alpha/2)},mN\right\}.\]

Second, we consider the tuning of $d$. We desire to control $d$ such that $n \leq n^{*}.$ Formally, this can be achieved by selecting $d$ as the maximum value that satisfies
\begin{equation}\label{md}
  n=\Bigl|\{(i,j)|{t}_{i,j}^U - {t}_{i,j}^L \leq d,i\in[m],j\in[N]\}\Bigr| \leq n^{*}.
\end{equation}
In practice, we begin by selecting $d$ as the theoretical maximum value given by Eq.~\eqref{md}. We then calculate the corresponding value of $kd$. If this value is too large, we reduce the value of $d$, whereas if the seller is willing to tolerate a reduced revenue and take on a small risk, we may increase the value of $d$ to degenerate more types' distributions, thereby reducing the number of queries required in the VCG mechanism. 
This process of adjusting $d$ can be performed by sorting the length of intervals in increasing order and selecting the $n^{*}$th length as the initial value of $d$, and iteratively adjusting until a satisfactory value is found.
However, when the product $mN$ is large, this method may encounter limitations. This can be observed when we consider the values $\eta=0.9$ and $\alpha=0.01$. In this scenario, we observe that $n^*=21$ when the product $mN$ exceeds 21. This implies that we can degenerate a maximum of 21 types' distributions.
In such cases, we can use the strategy outlined in Section \ref{Winnow down the possible winners} to filter out unnecessary intervals, thus reducing the number of intervals considered and allowing us to utilize the mechanism proposed in Section \ref{Designate the value} more effectively. In practice, we recommend starting with the value of $d$ given by Eq.~\eqref{md} for a specified value of $\eta$, and adjusting as necessary.

\subsubsection{Algorithm of the Implemented VCG Mechanism}

The proposed algorithm, outlined in Algorithm \ref{alg:three}, leverages the results of Algorithms \ref{alg:one} and \ref{alg:two} to provide a final implementation for determining the potential winners in an auction setting. Algorithm \ref{alg:one} yields the estimated confidence intervals $[{t}_{i,j}^L,{t}_{i,j}^U]$, while Algorithm \ref{alg:two} yields the set of potential bidders $\mathcal{B}_j$ for each item $j$ and the number of neglected bids $m^{*}$ defined in Eq.~\eqref{mstar}. Additionally, the algorithm introduces a hyper-parameter $q$, which represents the maximum loss the seller is willing to tolerate.
Algorithm \ref{alg:three} first filters the credible intervals based on the potential bidders $\mathcal{B}_j, j\in[N]$, by setting the lower and upper bounds of the interval to $0$ for bidders not in $\mathcal{B}_j$. The remaining intervals are sorted in increasing order according to the values of ${t}_{i,j}^U-{t}_{i,j}^L$. The value of $\hat{t}$ is then determined as described in Section \ref{Designate the value}, with the value of $d$ chosen to minimize the query number while ensuring that the maximum reduced revenue is less than $q$, as discussed in Section \ref{subsubsection3.3.5}. The algorithm begins by assuming an initial value of $d$, and repeatedly increases or decreases it until it fluctuates between two numbers $d{-}$ and $d{+}$, such that $kd_{-}\leq q< kd_{+}$, and the final value of $d$ is set to $d_{-}$. The complete procedure is presented in Algorithm \ref{alg:three}.

\begin{algorithm}[ht!]
	%\SetAlgoNoLine
    	\KwIn{$m$, $N$, $[{t}_{i,j}^L,{t}_{i,j}^U] \text{ for } i\in[m], j\in[N]$, $\alpha\in[0,1]$, accepted confidence rate $\eta\in[0,1]$, regret tolerance threshold $q\in(0,+\infty)$, $\mathcal{B}_j$ for $j\in[N]$, $m^{*}$}
	\KwOut{The winners set and the revenue}
        \For{j $\in$ $[N]$}{{\bf{if}} i $\notin$ $\mathcal{B}_j$ {\bf{then}} ${t}_{i,j}^L={t}_{i,j}^U=0$ \;}
        %{\bf{for}}  {\bf{do}}: 
         %{\bf{end}} \;

        $n^{*}=\min\left\{\frac{\log(\eta)}{\log(1-\alpha/2)},mN-m^{*}\right\}$\;
        sort the value of $({t}_{i,j}^U-{t}_{i,j}^L), i\in[m], j\in[N]$ in an increasing sequence $(\ell_1,\ell_2,\dots,\ell_{mN})$\;
        %$d = \arg\max\limits_{d}\{|\{i,j|\widehat{D}_{i,j}^U - \widehat{D}_{i,j}^L \leq d\}|<= n^{*}\} $\;
        $M = n^{*}+m^{*}$, $\ell_0=0$, $\ell_{mN+1}=+\infty$\;  
        
        \Repeat{\text{d fluctuates between the two values} $d_{-}$ \text{and} $d_{+}$\text{, where} $d_{-}<d_{+}$}{
        $d=\ell_{M}$, $K$ = \o\;
        \For{i $\in$ $[m]$}{
        {\bf{if}} $j$ $\in$ $L_i\cap L_i^*$ {\bf{then}} $K$ = $K\cup j$\;
        }
        $k$ = $|K|$\;
        {\bf{if}} $kd> q$ {\bf{then}} $M \gets M-1$ {\bf{else}} $M \gets M+1$\;
        %{\bf{if}} $M=0$ {\bf{then}} $d=0$; {\bf{if}} $M=mN+1$ {\bf{then}} $d=\ell_{mN}$;
        %\If{$kd$ \text{is so big}}{
        %$d = \ell_{M}$
        %}
        }
        $d = d_{-}$\;
        \For{i $\in$ $[m]$}{{\bf{if}} $j$ $\in$ $L_i$ {\bf{then}} $\hat{t}_{i,j}={t}_{i,j}^L$ {\bf{else}} $\hat{t}_{i,j}=t_{i,j}$ (make a query) \; }
       % {\bf{for}}  {\bf{do}}:    
       % {\bf{end}}
      revenue = 0\;
        \For{j $\in$ $[N]$}
        {winner for item $j$ = %winner$\cup
        $\arg\max\limits_{i}\hat{t}_{i,j}$;        revenue  $\gets$ revenue + $\arg\max\limits_{i\neq winner,i\in[m]}\hat{t}_{i,j}$\;
        }
        
	\caption{Degenerated VCG mechanism incorporates winnow down data and a regret tolerance threshold of $q$.}
	\label{alg:three}
\end{algorithm}

Algorithm \ref{alg:three} will eventually terminate, and it does not loop indefinitely. This is because the decrease in the value of $d$ ensures that $k$ becomes less than or equal to its original value, resulting in a strictly decreasing value of $kd$. When $M=0$, $kd$ becomes 0, and when $M=mN+1$, $kd$ becomes $+\infty$. Hence, the loop will eventually terminate since $q$ is within the range of $(0,+\infty)$.

\section{Experiments Results}\label{ER}
We consider a scenario where there are  $N$ items and $m$ bidders.
For each item $j \in [N]$, the minimum and maximum values are set at ${\color{black}{a_j=}}
10(j-1)$ and ${\color{black}{b_j=}}
10j$ respectively. The private valuations of each bidder $B_i, i \in [m]$, are modeled by a truncated Gaussian distribution $D_{i}^j$ with means chosen uniformly from the range $[10(j-1), 10j]$ and variances of $10^x$, where $x$ is chosen uniformly from the interval $[-1,1]$. The distribution is then normalized to $1$ over the range $[10(j-1),10j]$. Once the distribution $D$ is obtained, we can generate the types of bidders $t$ from $D$. Historical data sets, $\Gamma_{i,j}$, can then be simulated for each item $j$ by drawing $n_{i,j}$ samples from the corresponding true distribution $D_{i}^j$. In this simulation, we set $n_{i,j}=|\Gamma_{i,j}|=50$ for all $i$ and $j$. Additionally, we set $\mathcal{N}=1000$, $\alpha=0.01$, and $\eta=0.9$ for our experiments. The mechanism can be implemented once the values of $m$ and $N$ are determined. 

In our experiments, we do not specify a certain value for the hyper-parameter $q$ described in Algorithm \ref{alg:three} as we wish to analyze the revenue, regret, and number of queries for various values of $d$. To accomplish this, we start by initializing $d=0$ and increment it gradually in a continuous manner while simultaneously keeping track of the revenue generated and the corresponding value of $n$. In other words, we implement a modified version of Algorithm \ref{alg:three} by excluding the execution of lines 7, 13-15. Additionally, we expand the range of possible values for $d$ to include all values in the interval $[0,\ell_{mN}]$, instead of limiting it to $\ell_M$ alone.

We use three different revised mechanisms to verify our proposed strategies.

{\bf{Mechanism 1 (Kernel density \& Winnow down data)}}: This mechanism implements our entire proposed strategies, comprising of Algorithms \ref{alg:one}, Algorithm \ref{alg:two}, and Algorithm \ref{alg:three}, in sequence.
    
{\bf{Mechanism 2 (Kernel density \& Full data)}}: This mechanism is adopted to avoid potential compromise of auction fairness that may arise from winnowing down bidders and neglecting some bids. The revenue may be reduced if potential winners with higher types than the upper bound of their confidence intervals are neglected. In mechanism 2,  Algorithm \ref{alg:one} and Algorithm \ref{alg:three} are implemented, omitting the use of Algorithm \ref{alg:two} by setting $\mathcal{B}_j={\color{black}{[m]}}
,\forall j\in[N]$, and $m^*=0$ in the input of Algorithm \ref{alg:three}.

{\bf{Mechanism 3} (Ordinary method \& Winnow down data)}:  This mechanism does not estimate the distribution of $D$, and instead rely on the historical data $\Gamma_{i,j}=\{x_{i,j}^1,x_{i,j}^2,\dots,x_{i,j}^{n_{i,j}}\}, i\in[m], j\in[N]$. In this scenario, the minimum type of bidder $B_i$ for item $j$ is determined by the smallest value of $\Gamma_{i,j}$, while the maximum type of bidder $B_i$ for item $j$ is determined by the largest value of $\Gamma_{i,j}.$ Specifically, let $a_{i,j}=\min\{\Gamma_{i,j}\}$ and $b_{i,j}=\max\{\Gamma_{i,j}\}$. In mechanism 3, we just implement Algorithm \ref{alg:two} and   Algorithm \ref{alg:three}, without using Algorithm \ref{alg:one}. To do this, let ${t}_{i,j}^L=a_{i,j}$ and ${t}_{i,j}^U=b_{i,j}$, then we implement the Algorithm \ref{alg:two} and   Algorithm \ref{alg:three}.

\subsection{Small-Scale Data}
In order to evaluate the efficacy of our proposed methods, we conduct an initial experiment using a small dataset. Specifically, we consider a scenario in which there are $m=30$ bidders and $N=10$ items to be sold. Using the Vickrey-Clarke-Groves (VCG) mechanism to sell items individually, we calculate the revenue of the original VCG mechanism based on type $t$, represented by the green dashed line in the bottom-left plot of Figure \ref{fig:one}. This value, Revenue$=543$, serves as a benchmark against which we compare the performance of our proposed methods. Our goal is to obtain revenue values as close as possible to this benchmark through the application of our methods. In other words, we want to minimize the regret.

\subsubsection{Benefits of Density Estimation and Sampling Rejection}

\begin{figure}[htbp]
\centering
\subfigure[Comparison of Revenue]{
\begin{minipage}[t]{0.3\linewidth}
\centering
\label{4a}
\includegraphics[width=1.8in]{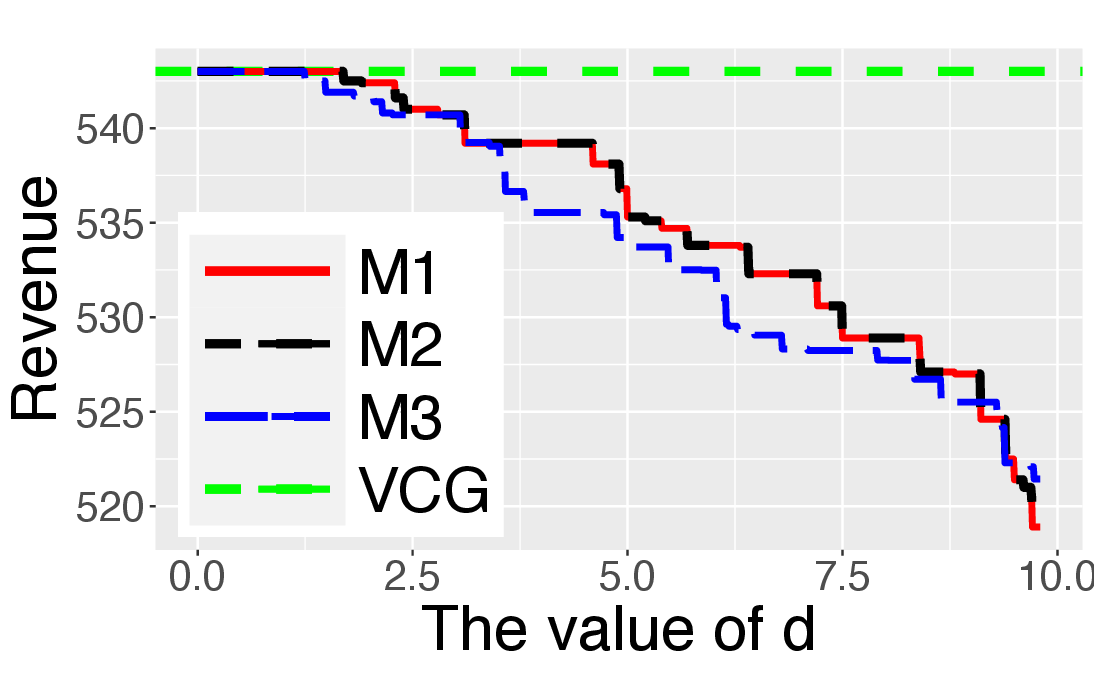}
%\caption{fig1}
\end{minipage}%
}%
\subfigure[Comparison of Regret]{
\begin{minipage}[t]{0.3\linewidth}
\centering
\label{4b}
\includegraphics[width=1.8in]{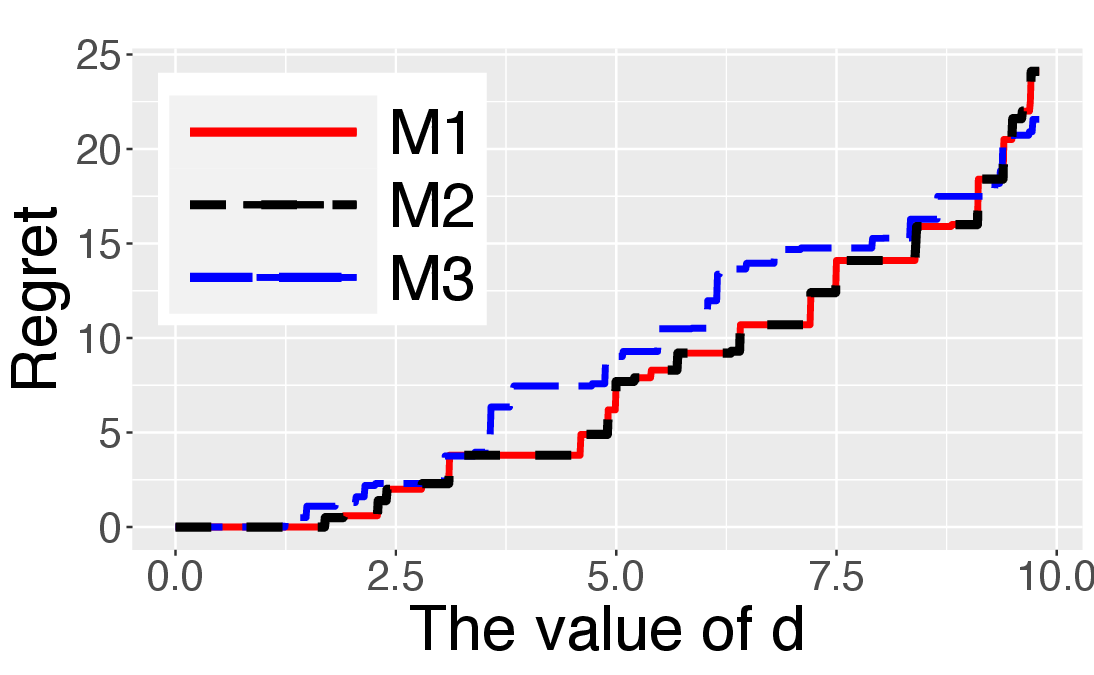}
%\caption{fig2}
\end{minipage}%
}%
\subfigure[Comparison of $kd$]{
\begin{minipage}[t]{0.3\linewidth}
\centering
\label{4c}
\includegraphics[width=1.8in]{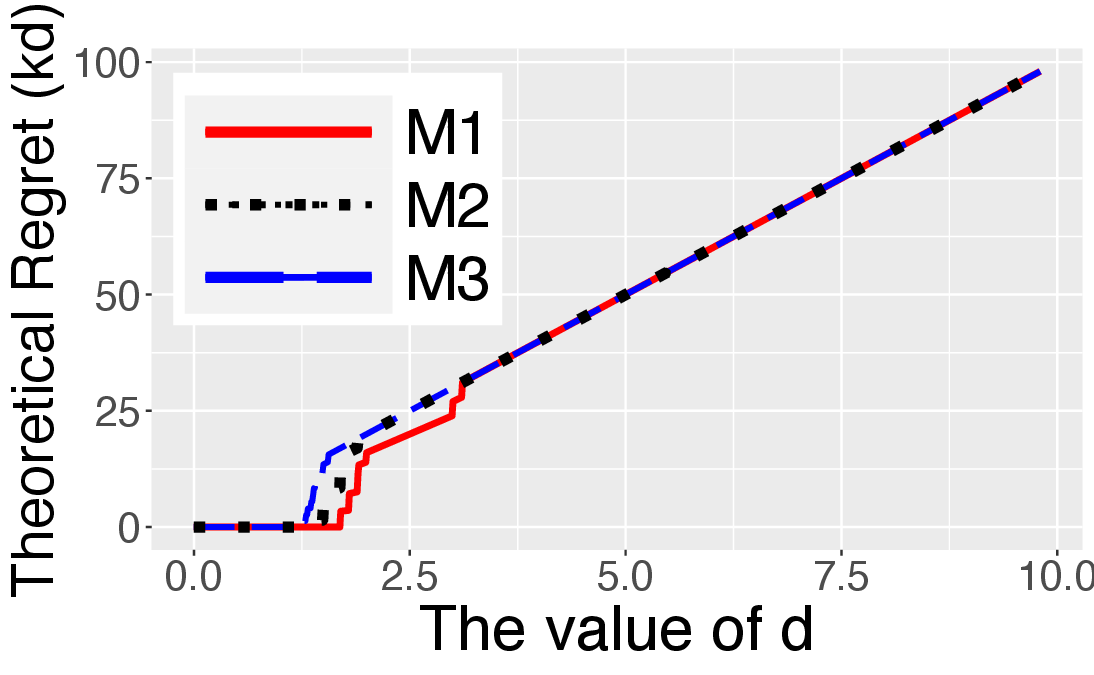}
%\caption{fig2}
\end{minipage}
}%
\centering
\caption{Comparison of the revenue and regret when $m=30$, $N=10$.}
\label{fig:one}
\end{figure}

In Figure \ref{fig:one}, we present a comparison of revenue and regret across the three mechanisms as the value of $d$ varies. Mechanism 1 is represented by a red solid line, Mechanism 2 by a black dashed line, and Mechanism 3 by a blue long dashed line. As shown in Figure \ref{4a} and \ref{4b}, which represent revenue and regret comparison respectively, all three mechanisms exhibit revenue exceeding 515 and regret less than 25, indicating overall good performance. However, we note that Mechanism 3 performs worse than Mechanism 1 and 2 in most cases. This is due to the absence of kernel density function in Mechanism 3, resulting in intervals of lower confidence compared to those obtained from Mechanism 1 and 2. Additionally, Mechanisms 1 and 2 overlap, indicating that the neglected data does not significantly impact the final outcome. This is in line with the $\alpha$-fair mechanism as proposed in Proposition \ref{fairness}, where $\alpha=0.01$, a value small enough to ensure fairness of the game.

In Figure \ref{4c}, Mechanism 1 is represented by a red solid line, Mechanism 2 by a black dotted line, and Mechanism 3 by a blue long dashed line. Figure \ref{4c} presents the comparison of theoretical regret and reveals that the lines of the three mechanisms overlap in most cases. This is because when $d$ is too small, the set $K$, as defined in Eq.~\eqref{K}, is empty, resulting in $k=0$. On the other hand, when $d$ is sufficiently large, $K$ tends towards $[N]$, resulting in $k=N$, and subsequently $kd=Nd$, a line with slope $N$. However, within a small interval, we find that for a fixed value of $d$, $kd$ of Mechanism 3 is the largest, and $kd$ of Mechanism 1 is the smallest. This is because, in the ordinary method, the length of the types' confidence intervals tends to be smaller, as it does not incorporate density function estimation, thus being limited to the historical data. As a result, for a fixed $d$, Mechanism 3 will have more intervals with a bandwidth smaller than $d$, resulting in a larger value of $k$. Conversely, Mechanism 1, by virtue of its sampling rejection method, already neglects some intervals, leading to a smaller number of intervals with a bandwidth smaller than $d$, and thus a smaller value of $k$.

In conclusion, our analysis of Figure \ref{fig:one} suggests that Mechanism 1 and Mechanism 2 perform better than Mechanism 3 in practice, highlighting the benefits of utilizing kernel density estimation and sampling rejection in Algorithm \ref{alg:one}.

\begin{figure}[htbp]
\centering
\subfigure[Comparison of Type's Proportion Without Query]{
\begin{minipage}[t]{0.5\linewidth}
\centering
\label{5a}
\includegraphics[width=2.2in]{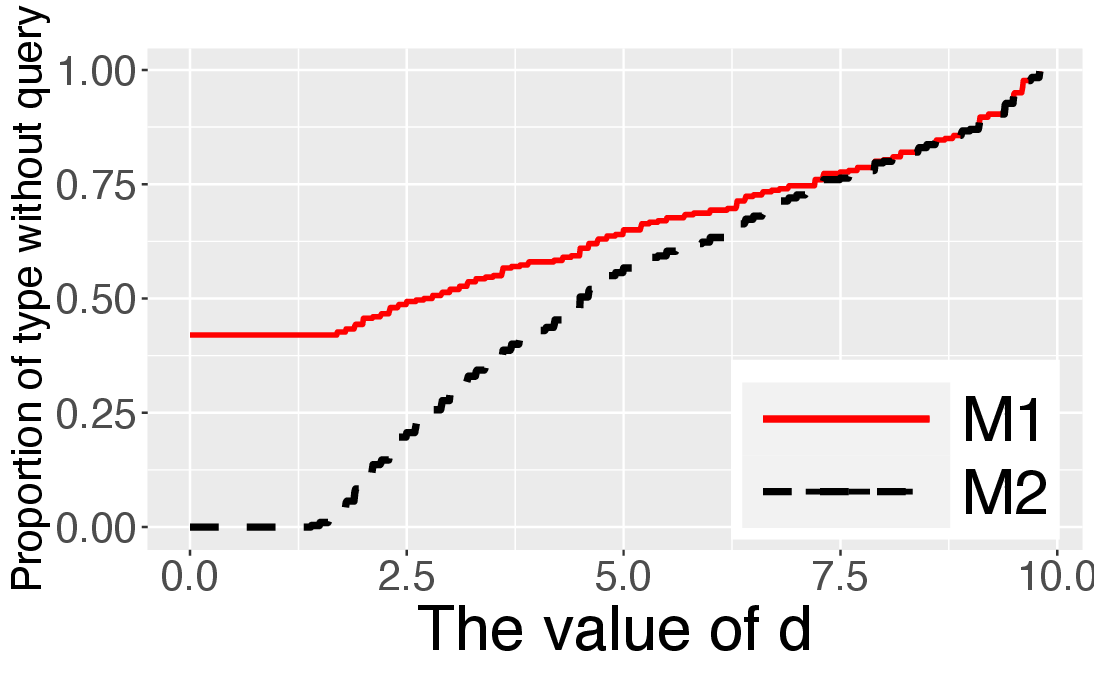}
%\caption{fig2}
\end{minipage}%
}%
\subfigure[Comparison of Confidence Rate]{
\begin{minipage}[t]{0.5\linewidth}
\centering
\label{5b}
\includegraphics[width=2.2in]{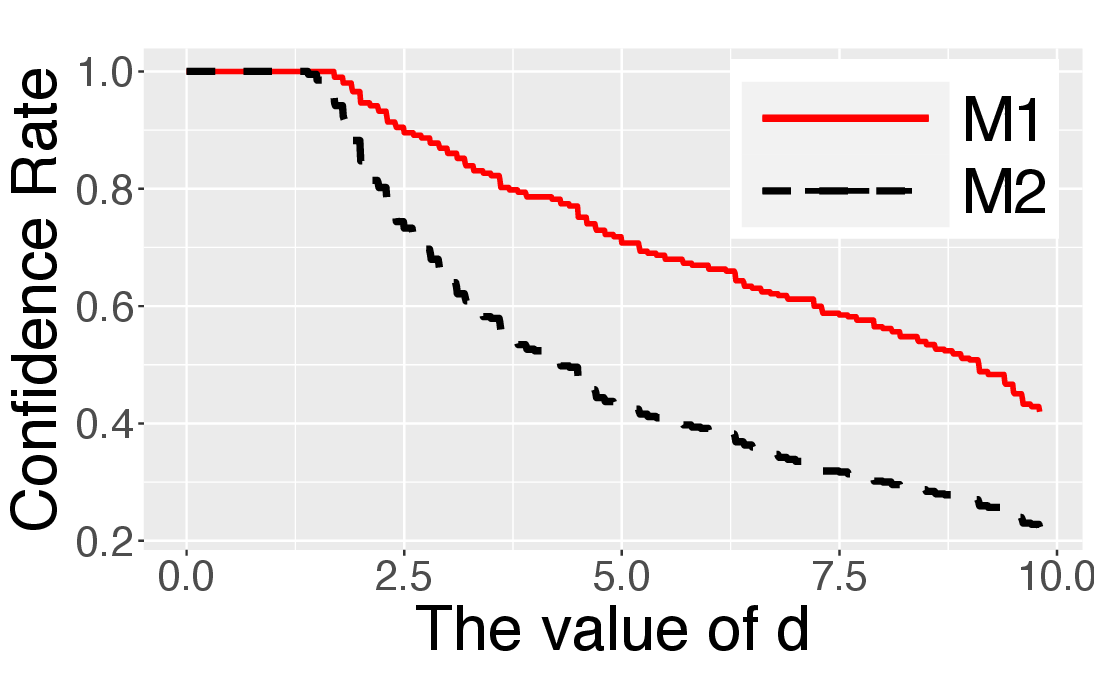}
%\caption{fig2}
\end{minipage}
}%
\centering
\caption{Comparison of the proportion of bidder types  without queries and comparison of the confidence rate when $m=30$, $N=10$.}
\label{fig:two}
\end{figure}

\subsubsection{Performance of Mechanism 1}

In Figure \ref{fig:two}, we compare the proportion of bidder types  without queries and the confidence rate of Mechanism 1 and Mechanism 2 as the value of $d$ varies. Mechanism 1 is represented by a red solid line, while Mechanism 2 is represented by a black dotted line.
Figure \ref{5a} illustrates that the proportion of types without query in Mechanism 1 is consistently higher or equal to that of Mechanism 2. This is because all types with ignored credible intervals in Mechanism 1 are dropped, and the seller does not make a query to those types. In our experiment, there were $m^*=126$ such types, resulting in an initial proportion of $126/300=0.42$ in Mechanism 1, while Mechanism 2 had a proportion of $0$ as $L$, defined in Eq.~\eqref{L}, was an empty set at the beginning. As $d$ increases, the number of elements in $L$ also increases, but since Mechanism 1 has dropped some credible intervals, the number of intervals whose bandwidths are smaller than a certain $d$ increases more slowly, resulting in a line with a smaller slope for Mechanism 1 compared to Mechanism 2.
Figure \ref{5b} illustrates the confidence rate, which is $(1-\alpha)^{n-m^*}$ in Mechanism 1 and $(1-\alpha)^{n}$ in Mechanism 2. The confidence rate of Mechanism 1 is consistently higher than that of Mechanism 2. 

The results presented in Figure \ref{fig:two} clearly demonstrate that Mechanism 1 outperforms Mechanism 2. As observed, to attain a confidence rate of 0.9, Mechanism 1 requires ${\color{black}{51}}
\%=1-(m^*+n^*)/mN=1-(126+21)/300$ of the total queries, while Mechanism 2 still requires $93\%=1-n^*/mN=1-21/300$ of the total queries. Additionally, both mechanisms have a practical regret of {\color{black}{no larger than}}
 2, which is relatively small when compared to the revenue of the original VCG mechanism, which is 543. 

Overall, the combination of Figures \ref{fig:one} and \ref{fig:two} reveal that Mechanism 1 has both the lowest practical and theoretical regret, requiring a mere $58\%=1-m^*/mN=1-126/300$ of total queries. Furthermore, when a confidence rate of $0.9$ is desired, Mechanism 1 only needs ${\color{black}{51}}
\%$ of the total queries, with a minimal practical revenue reduction of 2. These results highlight the effectiveness and efficiency of the proposed algorithm, making it an ideal choice in practical applications.

\subsection{Large-Scale Data}

\subsubsection{Advantage When the Number of Bidders is Large}

\begin{figure}[htbp]
\centering
\subfigure[Comparison of Revenue]{
\begin{minipage}[t]{0.3\linewidth}
\centering
\includegraphics[width=1.8in]{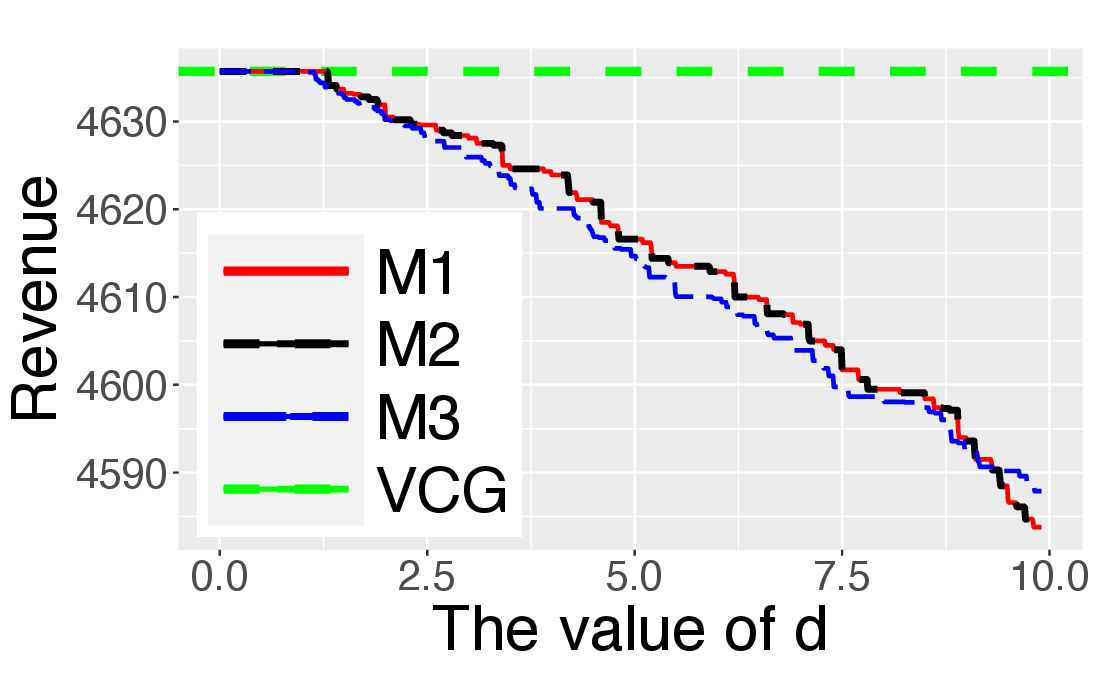}
%\caption{fig1}
\end{minipage}%
}%
\subfigure[Comparison of Regret]{
\begin{minipage}[t]{0.3\linewidth}
\centering
\includegraphics[width=1.8in]{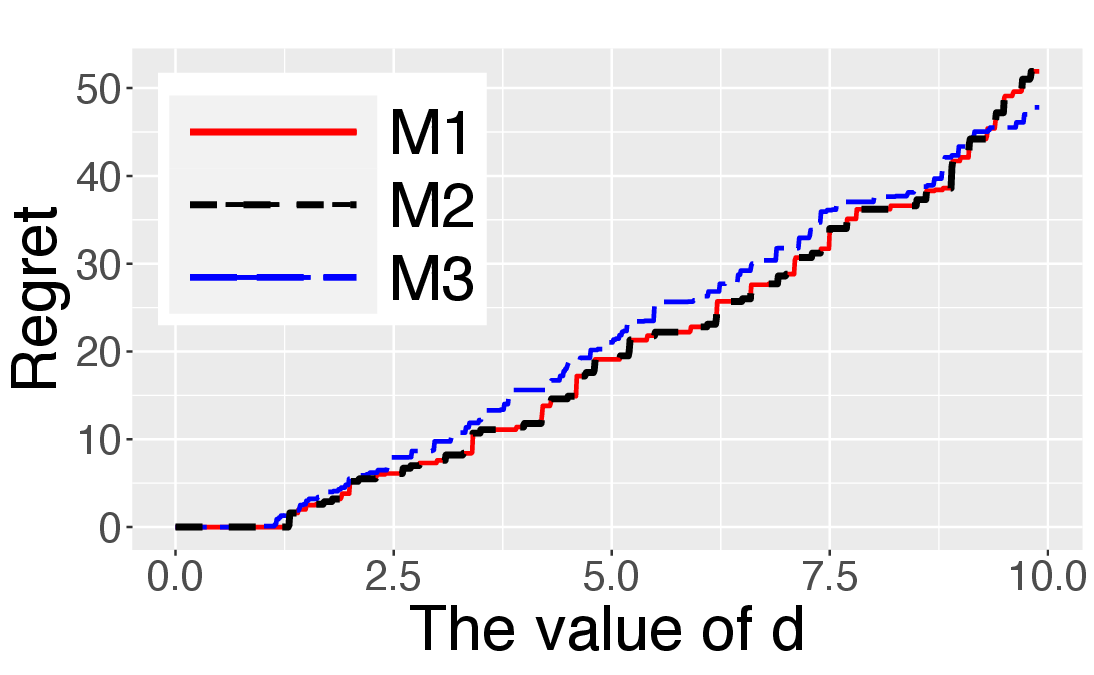}
%\caption{fig2}
\end{minipage}%
}%
\subfigure[Comparison of $kd$]{
\begin{minipage}[t]{0.3\linewidth}
\centering
\includegraphics[width=1.8in]{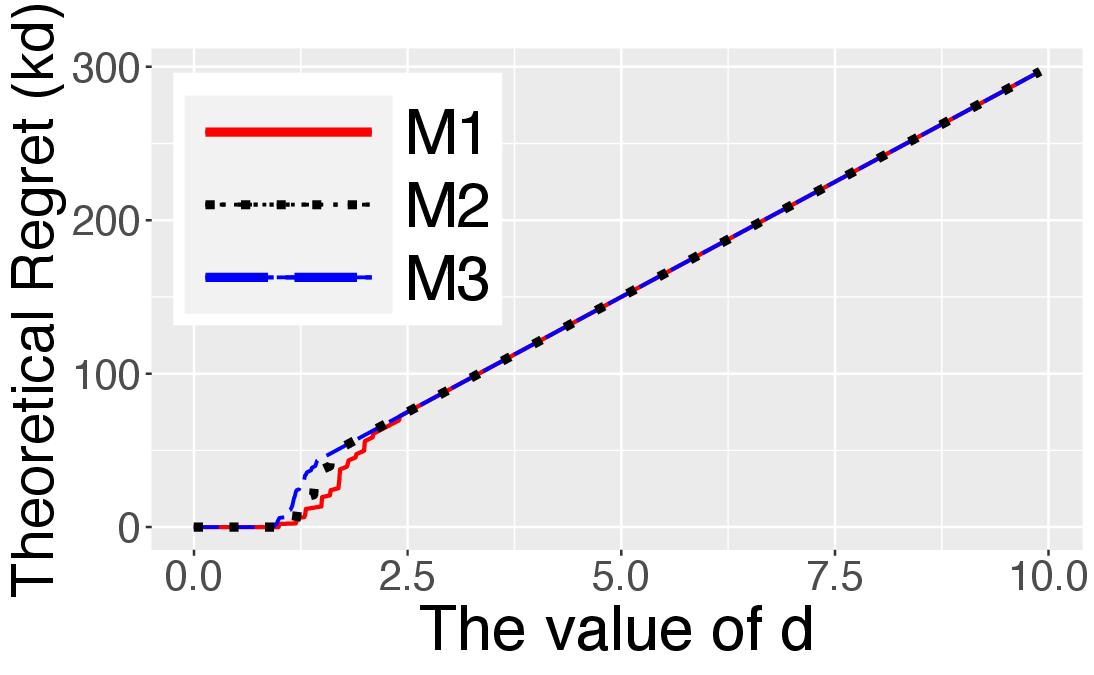}
%\caption{fig2}
\end{minipage}
}%
\centering
\caption{Comparison of the revenue and regret when $m=50$, $N=30$.}
\label{fig:three}
\end{figure}

In our experiments utilizing large-scale data, we first set the number of bidders to $m=50$ and the number of items to $N=30$, creating a scenario in which the bidders must compete for the items as $m > N$. 

The notations used in Figure \ref{fig:three} are consistent with those in Figure \ref{fig:one}. The revenue of the original VCG mechanism in this case is 4635.7. The results in Figure \ref{fig:three} reveal that the revenue generated by Mechanism 3 is consistently lower than that of Mechanism 1 and Mechanism 2 in most cases.  By comparing the results in Figure \ref{fig:one} and Figure \ref{fig:three}, we know that the use of kernel density estimation yields superior performance when the data scale is increased. In this scenario, the maximum theoretical regret $kd$ is 300, which is relatively small in comparison to the revenue of the original VCG mechanism, which is 4635.7. These results demonstrate the significant advantage of our mechanism in scenarios where the number of items is limited but the number of bidders is large, as previously analyzed in Section \ref{subsubsection3.2.5}.

Additionally, we also evaluate our method under a scenario where the number of bidders is less than the number of items. Specifically, we set $m=50$ and $N=100$. 
The results, shown in Figure \ref{fig:five}, exhibit a similarity to those presented in Figure \ref{fig:three}, resulting in a revenue of 50457.2 for the original VCG mechanism. However, as the number of items $m$ increases, a substantial increase in both practical and theoretical regret is observed, with the maximum regret reaching approximately 175 and 1000, respectively.
In our scenario, the basic value of items increases as the number of items increases, making a maximum regret of 1000 acceptable due to the high true revenue. Nonetheless, if the basic value of items remains constant, this outcome may not be optimal. Hence our method is most appropriate in scenarios where the number of items is limited, or the values of items are high but exhibit limited variance.

\begin{figure}[htbp]
\centering
\subfigure[Comparison of Revenue]{
\begin{minipage}[t]{0.3\linewidth}
\centering
\includegraphics[width=1.8in]{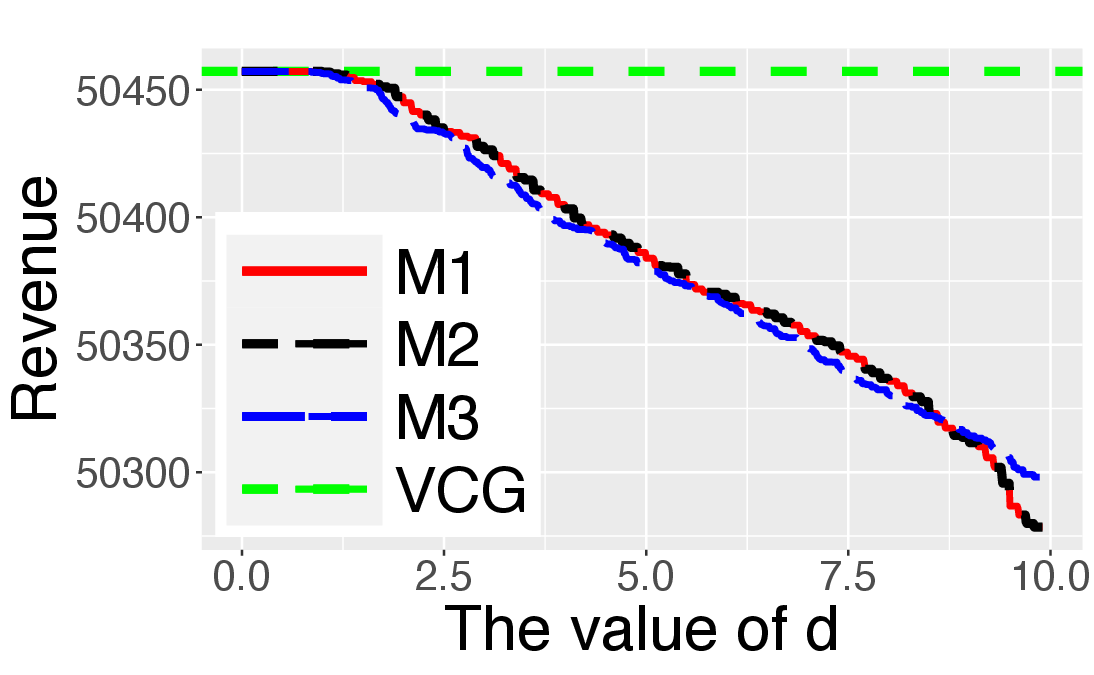}
%\caption{fig1}
\end{minipage}%
}%
\subfigure[Comparison of Regret]{
\begin{minipage}[t]{0.3\linewidth}
\centering
\includegraphics[width=1.8in]{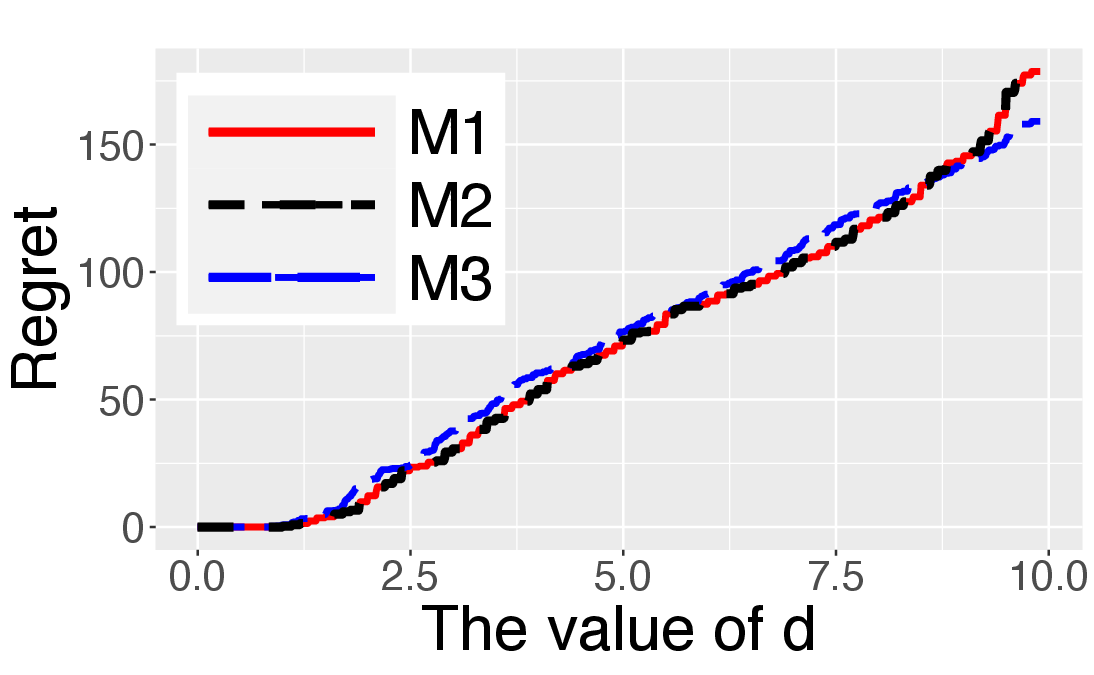}
%\caption{fig2}
\end{minipage}%
}%
\subfigure[Comparison of $kd$]{
\begin{minipage}[t]{0.3\linewidth}
\centering
\includegraphics[width=1.8in]{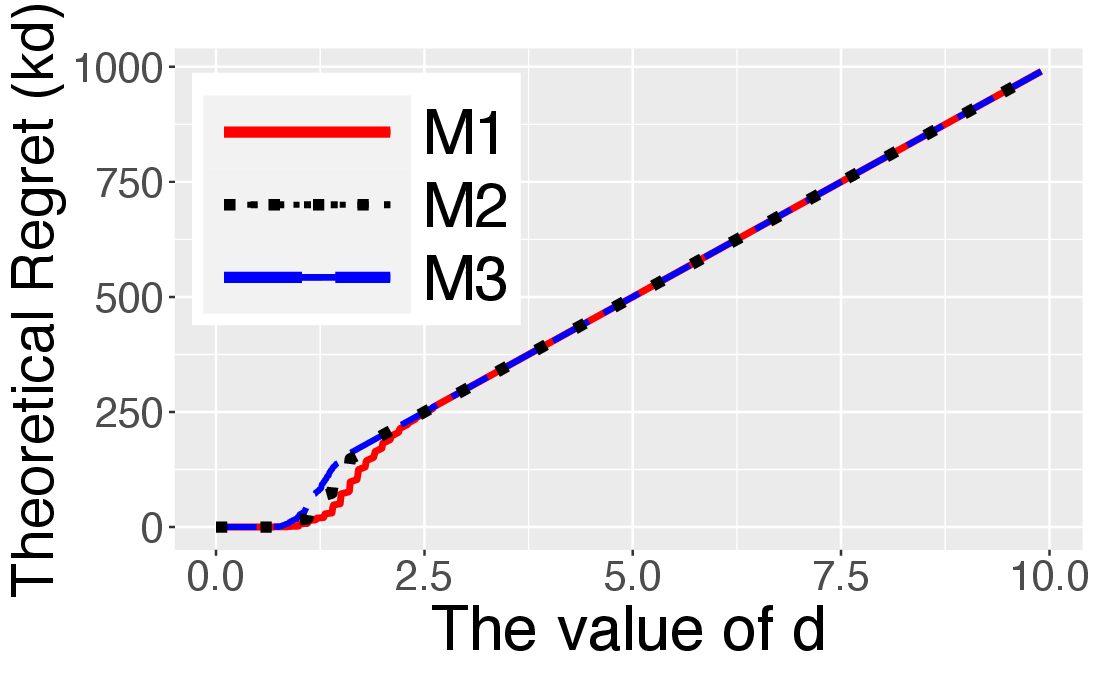}
%\caption{fig2}
\end{minipage}
}%
\centering
\caption{Comparison of the revenue and regret when $m=50$, $N=100$.}
\label{fig:five}
\end{figure}

\subsubsection{Performance of Mechanism 1}

\begin{figure}[htbp]
\centering
\subfigure[Comparison of Type's Proportion Without Query]{
\begin{minipage}[t]{0.5\linewidth}
\centering
\includegraphics[width=2.2in]{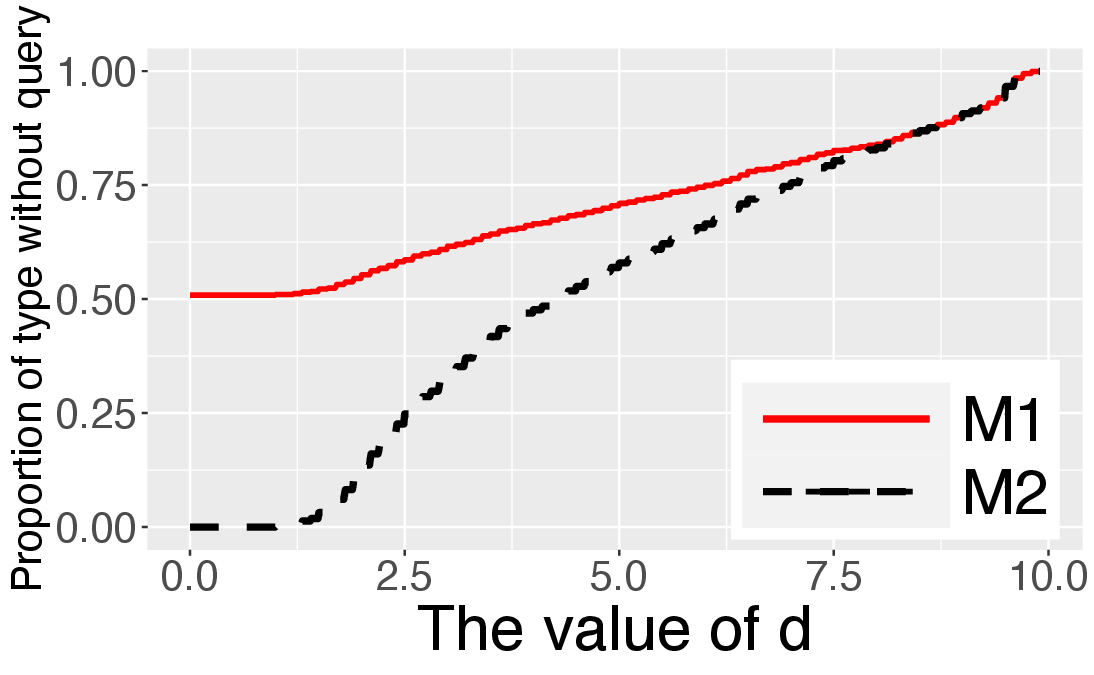}
%\caption{fig2}
\end{minipage}%
}%
\subfigure[Comparison of Confidence Rate]{
\begin{minipage}[t]{0.5\linewidth}
\centering
\includegraphics[width=2.2in]{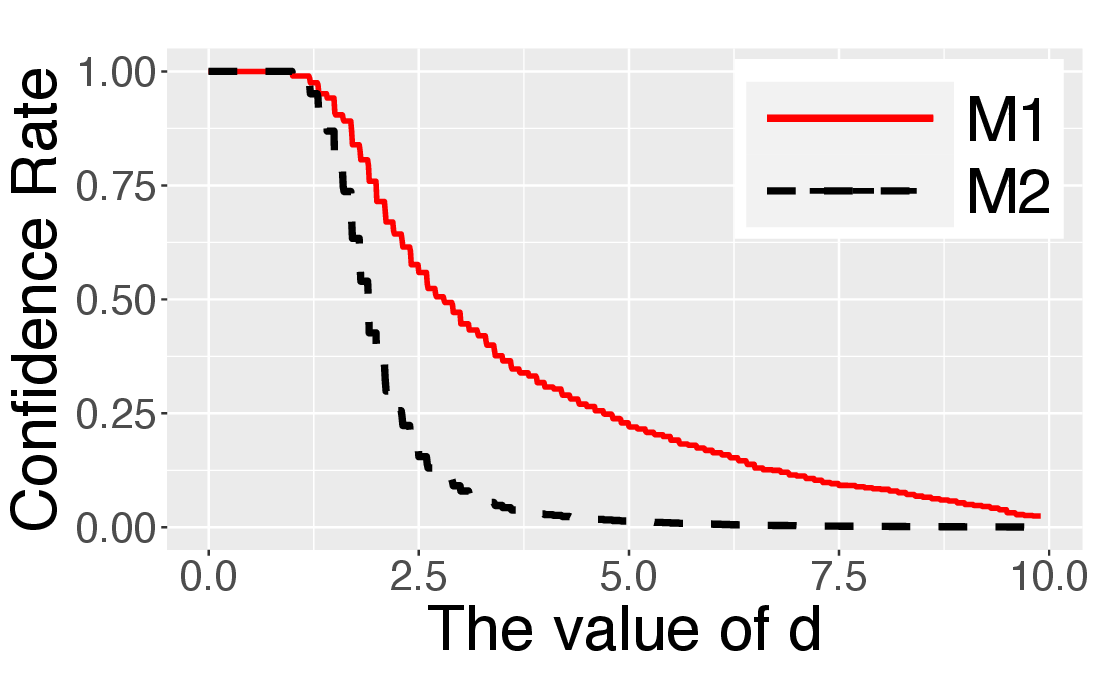}
%\caption{fig2}
\end{minipage}
}%
\centering
\caption{Comparison of the proportion of bidder types  without queries and comparison of the confidence rate when $m=50$, $N=30$.}
\label{fig:four}
\end{figure}

In Figure \ref{fig:four}, we present the results of our method under the scenario of large-scale data when $m=50, N=30$. Specifically, we set $m^*=763$ and $mN=1500$, which results in an initial query rate of less than $49.2\%$, as approximated by $(mN-m^*)/mN$. With a confidence rate $\eta$ of 0.9, Mechanism 1 further reduces the query rate to less than $(mN-m^*-n^*)/mN=47.74\%$, resulting in just 2.5 practical regrets and 24 theoretical regrets. These results demonstrate that the proposed Mechanism 1 works well, even when dealing with large datasets.

In Figure \ref{fig:six}, we also obtain results for an even larger dataset with $m=50$ and $N=100$.  In this scenario, we set $m^*=2456$ and  $mN=5000$, which results in an initial query rate of  less than $50.88\%$, as approximated by $(mN-m^*)/mN$. With a confidence rate of $0.9$, Mechanism 1 further reduces the query rate to $50.46\%$ as approximated by $(mN-m^*-n^*)/mN$, resulting in just 1.6 practical regrets and 72 theoretical regrets.
We note that  as the value of $mN$ increases, the confidence rate decreases at an accelerated rate. In this scenario, the seller must make a trade-off between classifying the intervals and making all queries to the winnowed-down data ($d=0$), or reducing the number of queries by choosing a larger value of $d$, at the risk of sacrificing some revenue. Our results show that when $d=3.75$, although the confidence rate is low (less than 0.06), we can still reduce the number of queries to less than $37.5\%=1-62.5\%$. This reduction results in a reduction of more than $mN*(50-37.5)\% = 625$ in the number of queries, with only a negligible reduction in the revenue of the original VCG mechanism (50 out of a total of 50457.2). This analysis supports the discussion in Section \ref{subsubsection3.3.5} and highlights the importance of carefully considering the choice of $d$ in practical applications.

\begin{figure}[htbp]
\centering
\subfigure[Comparison of Type's Proportion Without Query]{
\begin{minipage}[t]{0.5\linewidth}
\centering
\includegraphics[width=2.2in]{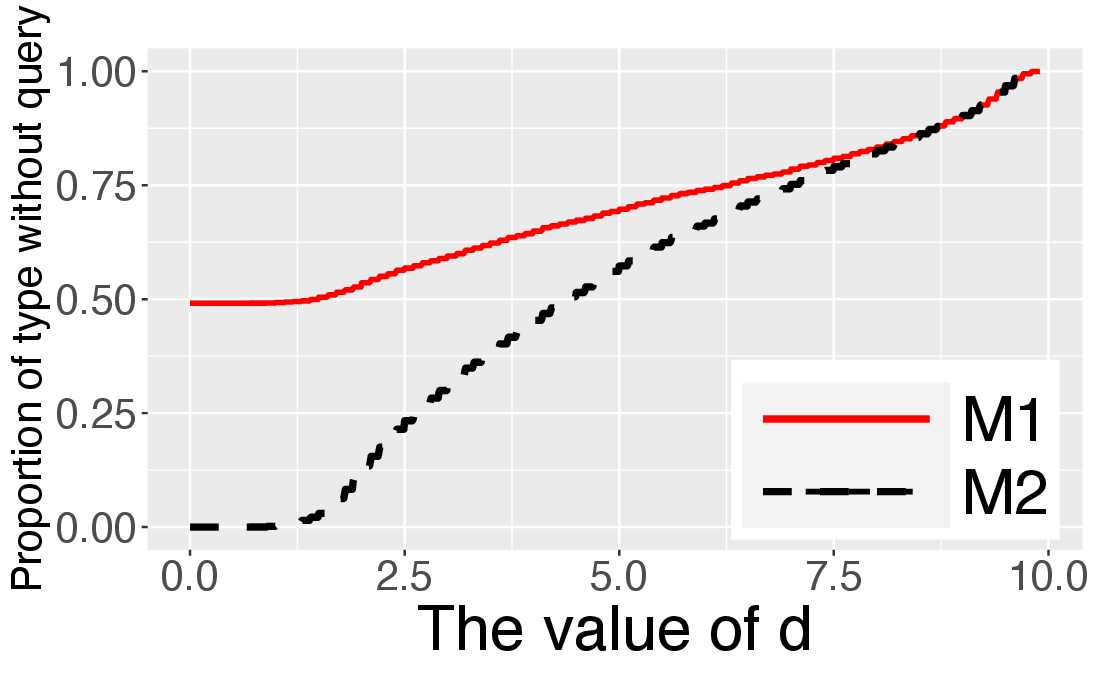}
%\caption{fig2}
\end{minipage}%
}%
\subfigure[Comparison of Confidence Rate]{
\begin{minipage}[t]{0.5\linewidth}
\centering
\includegraphics[width=2.2in]{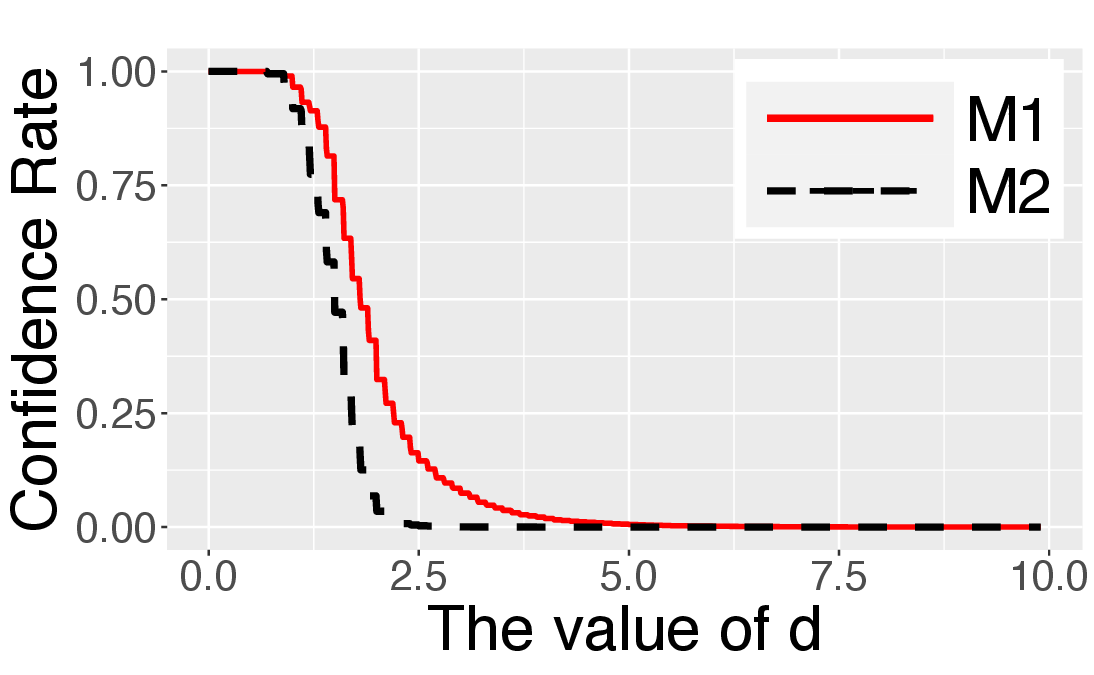}
%\caption{fig2}
\end{minipage}
}%
\centering
\caption{Comparison of the proportion of bidder types  without queries and comparison of the confidence rate when $m=50$, $N=100$.}
\label{fig:six}
\end{figure}

\subsection{Analysis of Error Bars}
In the bidding process, we make the assumption that the distributions of the bidder types remain fixed, as mentioned in the initial paragraph of the main results. Consequently, once the distributions are fixed, their credible intervals should also be fixed. In other words, $t^L_{i,j}$ and $t^U_{i,j}$ represent fixed constants for all $i\in[m]$ and $j\in[N]$. Therefore, the intervals obtained from the estimation based on simulated data may contain some errors when compared to the true intervals. In this section, we will analyze these errors.

First, we determine the true intervals by giving the following proposition.
\begin{proposition}\label{error}
    Suppose the truncated Gaussian distribution $D_i^j$ has the mean $\mu_{i,j}\in[10(j-1),10j]$ and the variance $\sigma_{i,j}^2>0$, and it is re-normalized within the interval $[10(j-1),10j]$. Let $\phi(x;\mu_{i,j},\sigma_{i,j}^2)$ and $\Phi(x;\mu_{i,j},\sigma_{i,j}^2)$ represent the probability density function and cumulative distribution function of $\mathcal{N}(\mu_{i,j}, \sigma_{i,j}^2)$, respectively. Then, the $\alpha/2$th quantile $t_{i,j}^L$ and the $(1-\alpha/2)$th quantile $t_{i,j}^U$ of the $D_i^j$ are calculated as follows: 
    \[t_{i,j}^L=\Phi^{-1}\biggl(\Phi\Bigl(10j;\mu_{i,j},\sigma_{i,j}^2\Bigr)-(1-\alpha/2)\Phi\Bigl(10(j-1);\mu_{i,j},\sigma_{i,j}^2\Bigr); \mu_{i,j},\sigma_{i,j}^2\biggr),\]
    \[t_{i,j}^U=\Phi^{-1}\biggl(\Phi\Bigl(10j;\mu_{i,j},\sigma_{i,j}^2\Bigr)-(\alpha/2)\Phi\Bigl(10(j-1);\mu_{i,j},\sigma_{i,j}^2\Bigr); \mu_{i,j},\sigma_{i,j}^2\biggr).\]
    for any $\alpha\in[0,1].$
\end{proposition}
\begin{proof}
    \begin{equation*}
    \begin{aligned}
    \phi\Bigl(t_{i,j};\mu_{i,j},\sigma_{i,j}^2|10(j-1)<t_{i,j}<10j\Bigr) & =\frac{\phi(t_{i,j};\mu_{i,j},\sigma_{i,j}^2)}{\mathbb{P}(10(j-1)<t_{i,j}<10j)}\\
    &=\frac{\phi(t_{i,j};\mu_{i,j},\sigma_{i,j}^2)}{\Phi(10j;\mu_{i,j},\sigma_{i,j}^2)-\Phi(10(j-1);\mu_{i,j},\sigma_{i,j}^2)}\\
    &\equiv\frac{\phi(t_{i,j};\mu_{i,j},\sigma_{i,j}^2)}{C},
    \end{aligned}
    \end{equation*}
    where we define $C=\Phi(10j;\mu_{i,j},\sigma_{i,j}^2)-\Phi(10(j-1);\mu_{i,j},\sigma_{i,j}^2)$.

    With reference to the definition of $t_{i,j}^L$, we can derive the following expression: 
    \begin{equation*}
    \begin{aligned}
    &\quad \quad \quad \int_{10(j-1)}^{t_{i,j}^L}\frac{\phi(t_{i,j};\mu_{i,j},\sigma_{i,j}^2)}{C}dt_{i,j}=(\alpha/2) \\
    &\iff \int_{10(j-1)}^{t_{i,j}^L}\phi(t_{i,j};\mu_{i,j},\sigma_{i,j}^2)dt_{i,j}=(\alpha/2)C\\
    &\iff \Phi(t_{i,j}^L;\mu_{i,j},\sigma_{i,j}^2)-\Phi\Bigl(10(j-1);\mu_{i,j},\sigma_{i,j}^2\Bigr)=(\alpha/2)C\\
    &\iff  \Phi(t_{i,j}^L;\mu_{i,j},\sigma_{i,j}^2)=(\alpha/2)C+\Phi\Bigl(10(j-1);\mu_{i,j},\sigma_{i,j}^2\Bigr)\\
    &\iff  t_{i,j}^L=\Phi^{-1}\biggl((\alpha/2)C+\Phi\Bigl(10(j-1);\mu_{i,j},\sigma_{i,j}^2\Bigr);\mu_{i,j},\sigma_{i,j}^2\Bigr); \mu_{i,j},\sigma_{i,j}^2\biggr)\\
     &\iff t_{i,j}^L=\Phi^{-1}\biggl(\Phi\Bigl(10j;\mu_{i,j},\sigma_{i,j}^2\Bigr)-(1-\alpha/2)\Phi\Bigl(10(j-1);\mu_{i,j},\sigma_{i,j}^2\Bigr); \mu_{i,j},\sigma_{i,j}^2\biggr).
    \end{aligned}
    \end{equation*}
    
    Likewise, by replacing $t_{i,j}^L$ with $t_{i,j}^U$ and $\alpha/2$ with $1-\alpha/2$, we obtain:
    \[t_{i,j}^U=\Phi^{-1}\biggl(\Phi\Bigl(10j;\mu_{i,j},\sigma_{i,j}^2\Bigr)-(\alpha/2)\Phi\Bigl(10(j-1);\mu_{i,j},\sigma_{i,j}^2\Bigr); \mu_{i,j},\sigma_{i,j}^2\biggr).\]
\end{proof}

Based on the true intervals, we can introduce two additional mechanisms to visualize the error of the random simulation points.

{\bf{Mechanism 4}} (True intervals \& Winnow down data): This mechanism does not utilize simulated historical data but directly employs the true intervals. To implement this, we apply Algorithms \ref{alg:two} and \ref{alg:three}, considering the true values for $t_{i,j}^L$ and $t_{i,j}^U$ as described in Proposition \ref{error}.

{\bf{Mechanism 5}} (True intervals \& Full data): In comparison to Mechanism 4, this mechanism solely employs Algorithm \ref{alg:three} and sets $\mathcal{B}_j=[m], \forall j\in[N]$, and $m^*=0$ as inputs to Algorithm \ref{alg:three}.

\begin{figure}[htbp]
\centering
\subfigure[Comparison of Revenue.]{
\includegraphics[width=4.2cm]{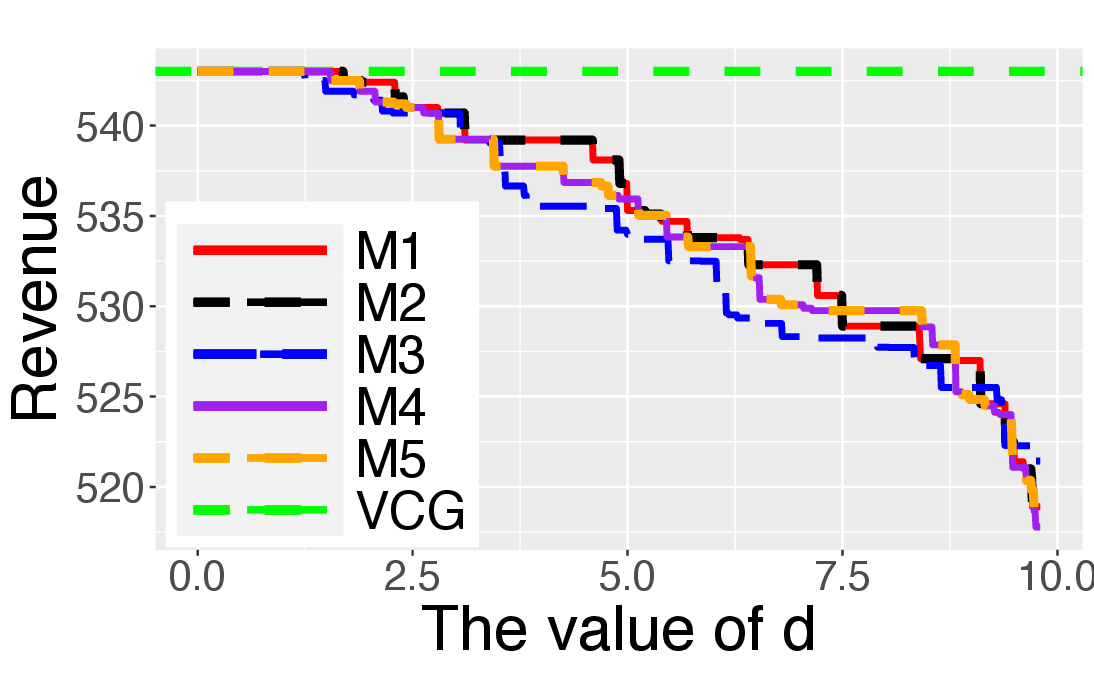}
\label{11a}
%\caption{fig1}
}
\quad
\subfigure[Comparison of Regret.]{
\includegraphics[width=4.1cm]{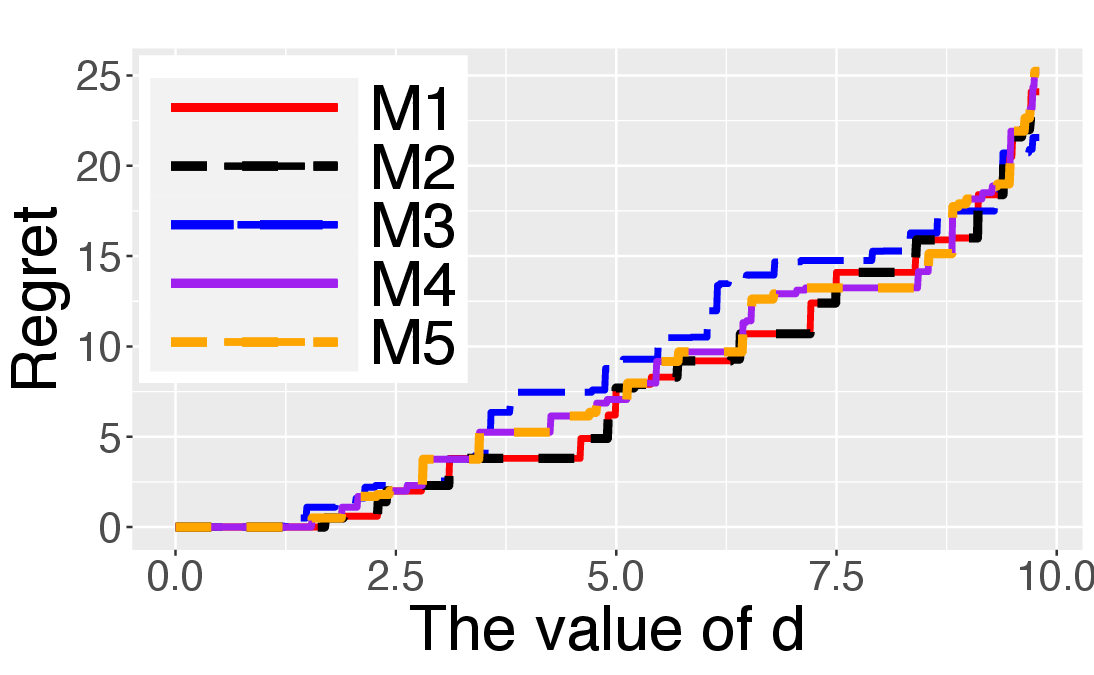}
\label{11b}
}
\quad
\subfigure[Error in Revenue of the Simulated Data.]{
\includegraphics[width=4.1cm]{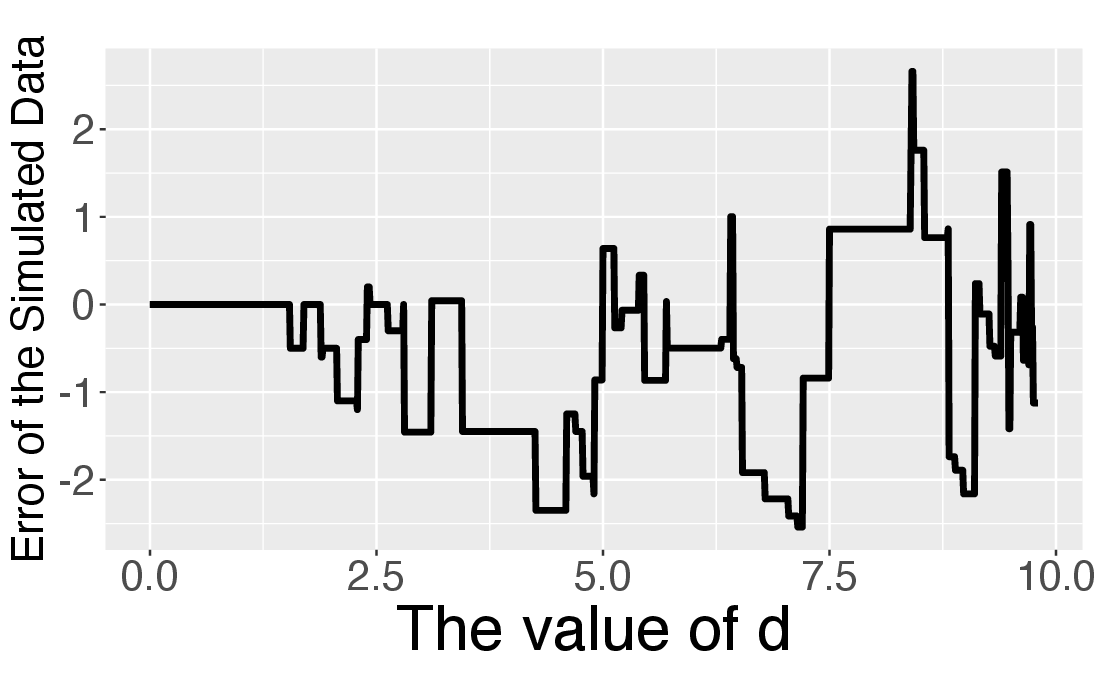}
\label{11c}
}
\quad
\subfigure[Comparison of $kd$.]{
\includegraphics[width=4.1cm]{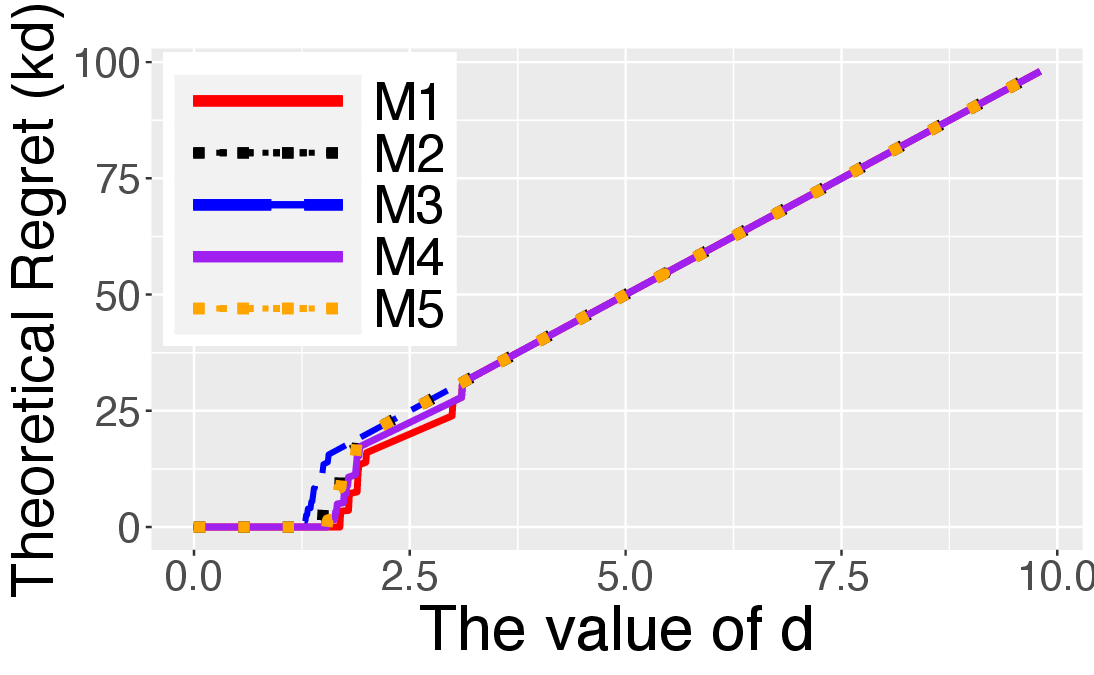}
\label{11d}
}
\quad
\subfigure[Comparison of Type's Proportion Without Query.]{
\includegraphics[width=4.1cm]{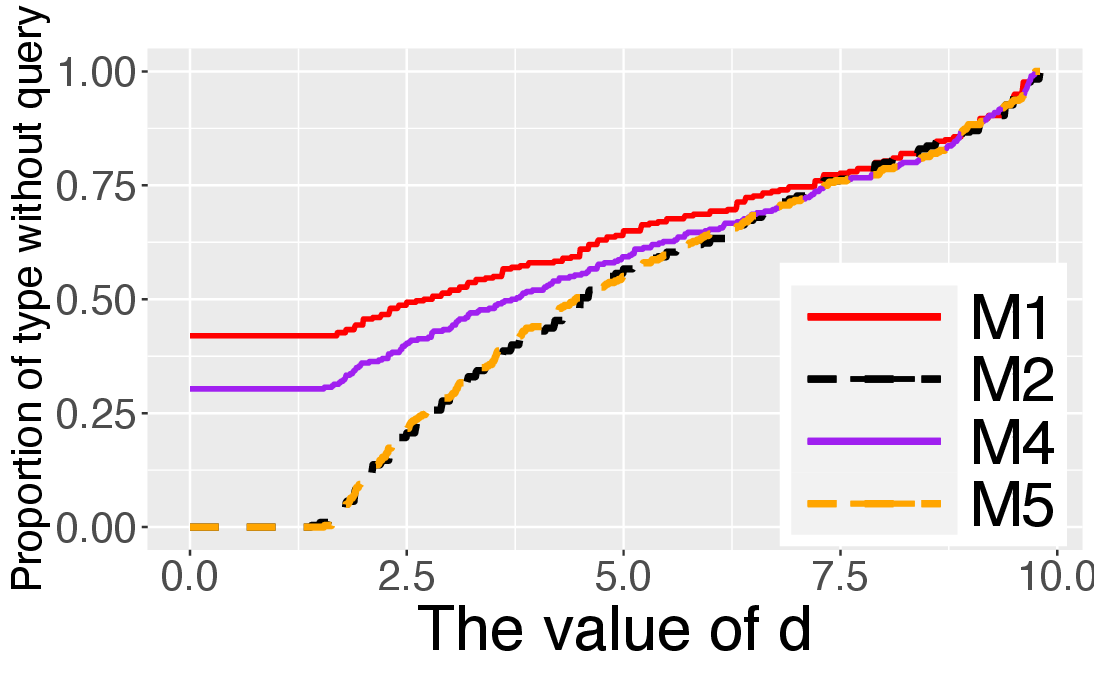}
\label{11e}
}
\quad
\subfigure[Comparison of the Confidence Rate.]{
\includegraphics[width=4.1cm]{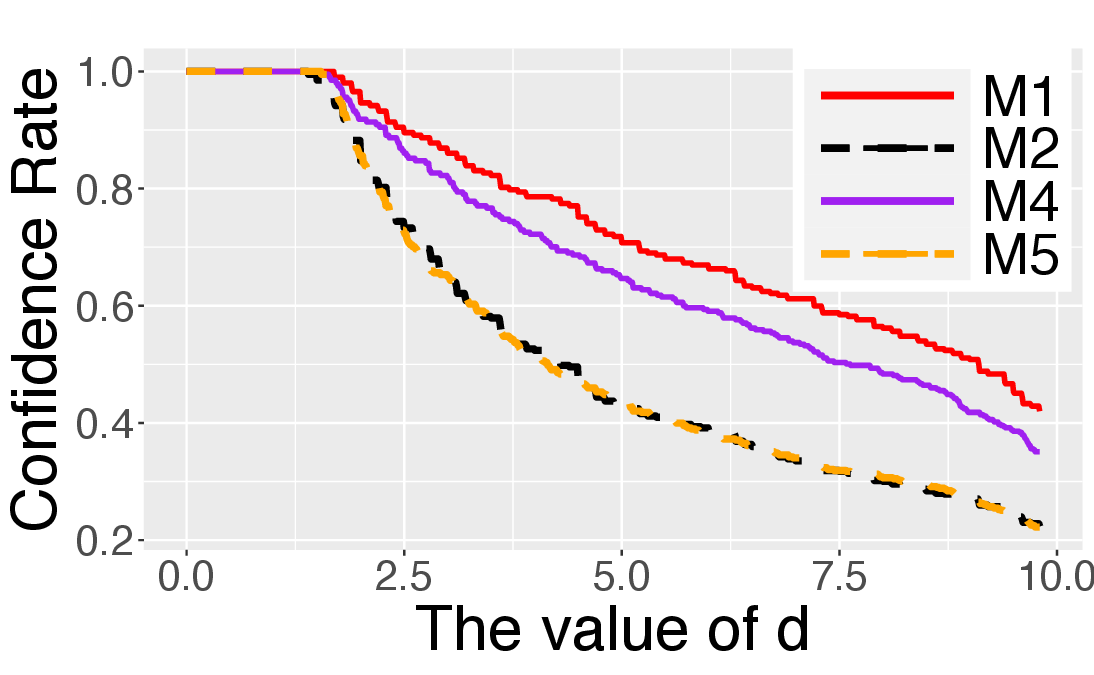}
\label{11f}
}
\caption{The visualization of error when $m=30$ and $N=10$.}
\label{fig:ten}
\end{figure}

\begin{figure}[ht!]
\centering
\subfigure[Comparison of Revenue.]{
\includegraphics[width=4.2cm]{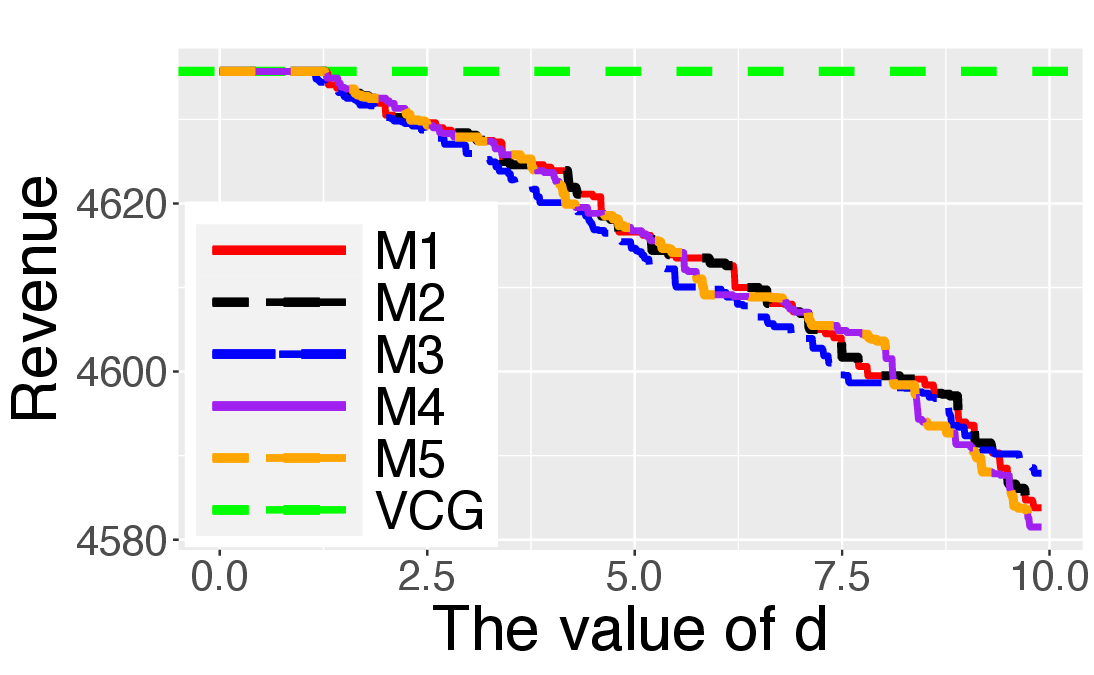}
\label{12a}
%\caption{fig1}
}
\quad
\subfigure[Comparison of Regret.]{
\includegraphics[width=4.1cm]{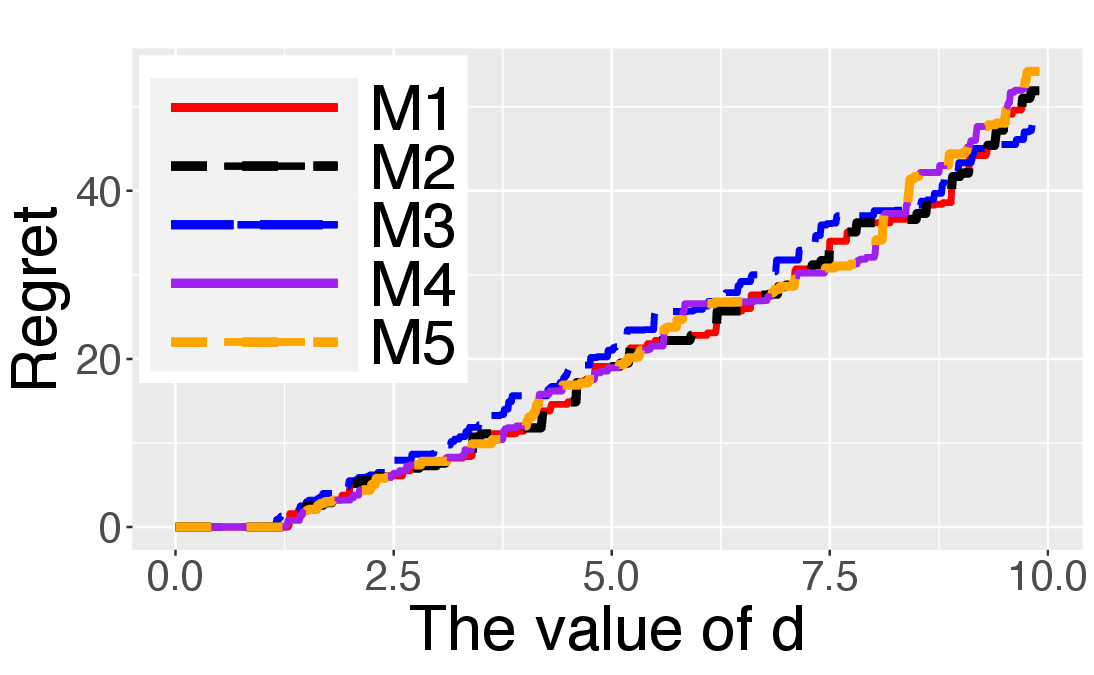}
\label{12b}
}
\quad
\subfigure[Error in Revenue of the Simulated Data.]{
\includegraphics[width=4.1cm]{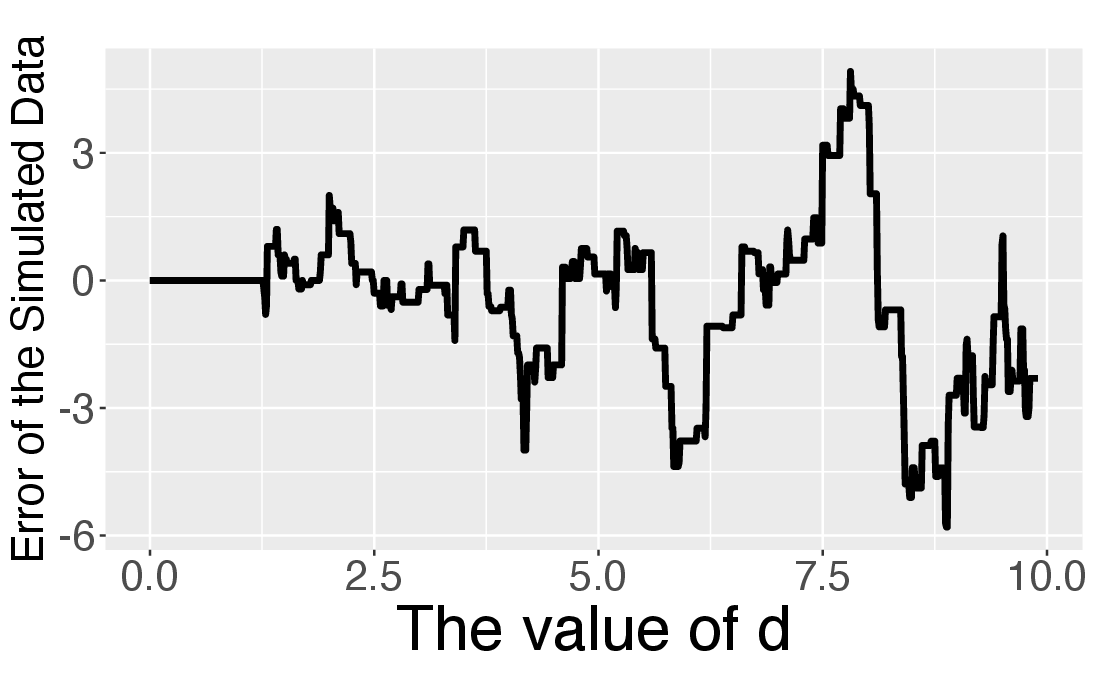}
\label{12c}
}
\quad
\subfigure[Comparison of $kd$.]{
\includegraphics[width=4.1cm]{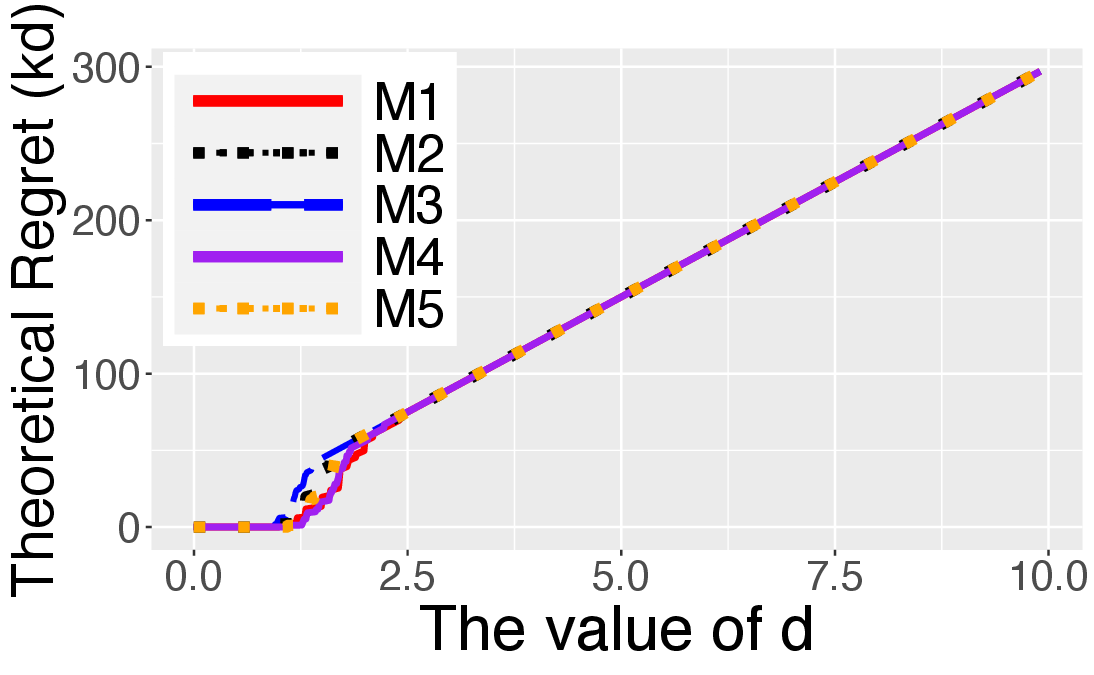}
\label{12d}
}
\quad
\subfigure[Comparison of Type's Proportion Without Query.]{
\includegraphics[width=4.1cm]{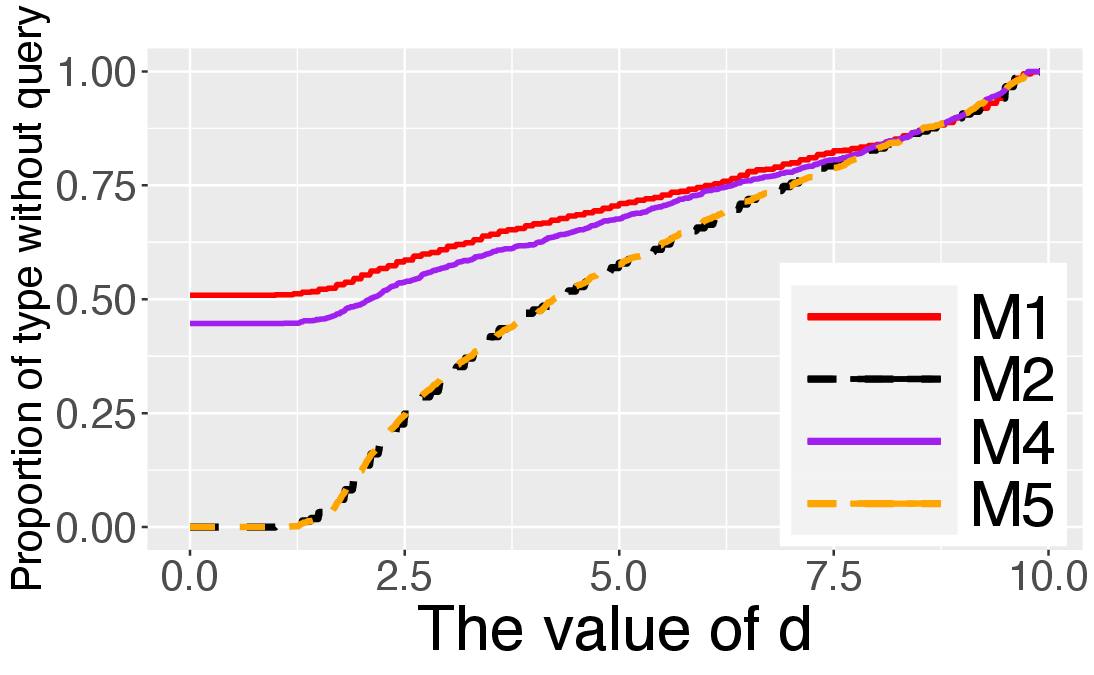}
\label{12e}
}
\quad
\subfigure[Comparison of the Confidence Rate.]{
\includegraphics[width=4.1cm]{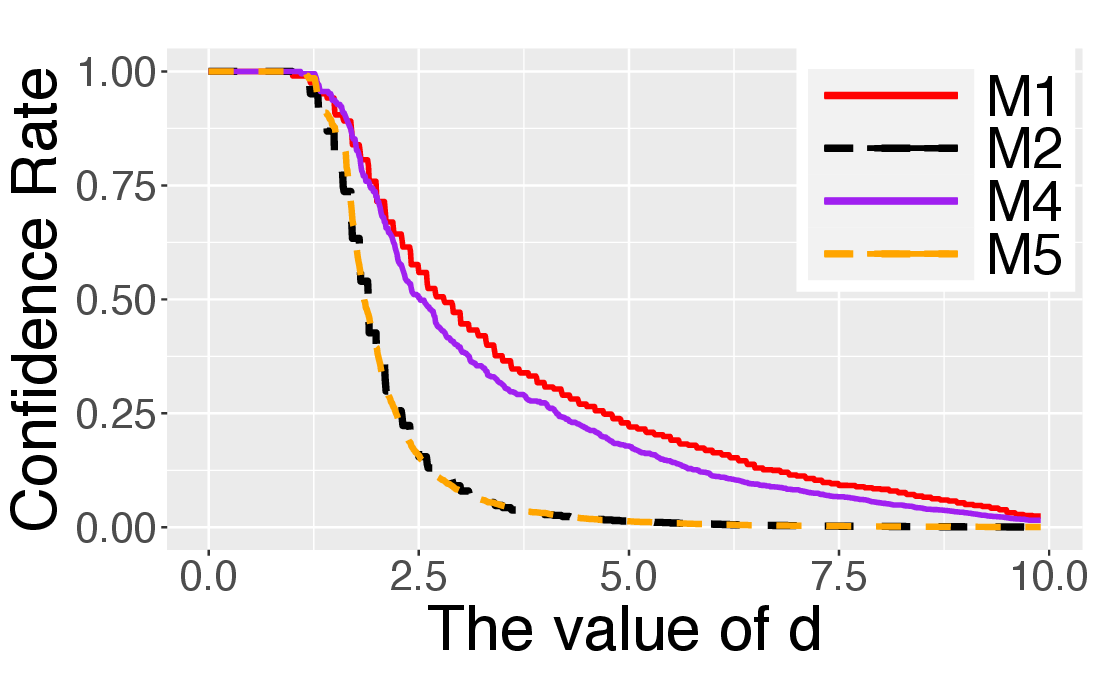}
\label{12f}
}
\caption{The visualization of error when $m=50$ and $N=30$.}
\label{fig:eleven}
\end{figure}

\begin{figure}[ht!]
\centering
\subfigure[Comparison of Revenue.]{
\includegraphics[width=4.2cm]{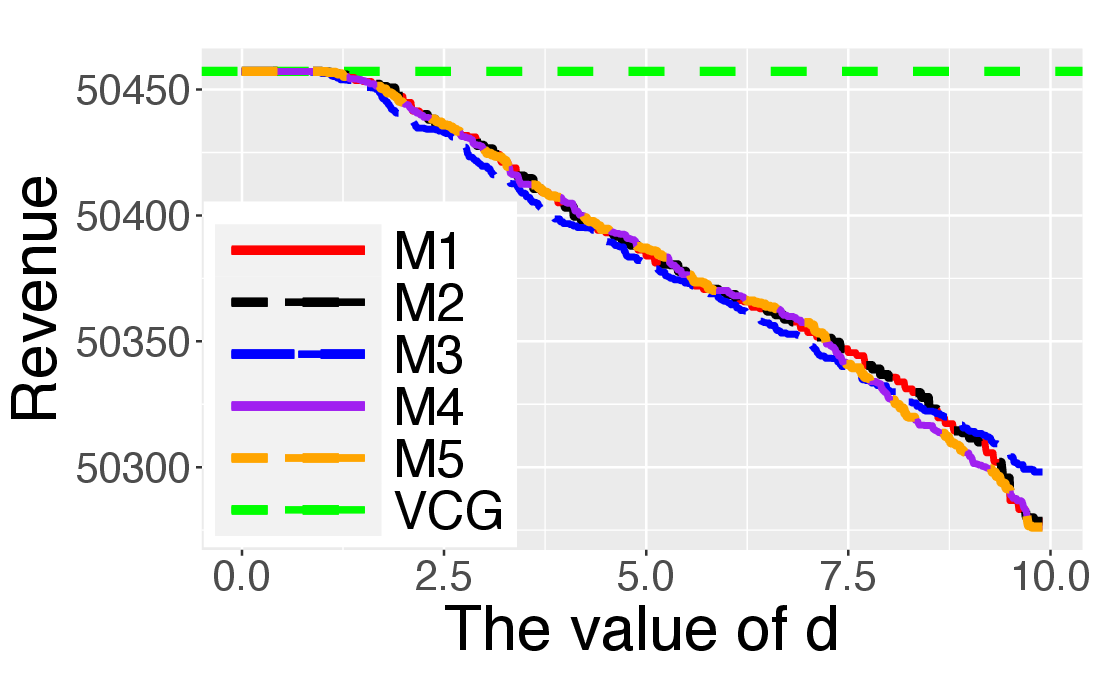}
\label{13a}
%\caption{fig1}
}
\quad
\subfigure[Comparison of Regret.]{
\includegraphics[width=4.1cm]{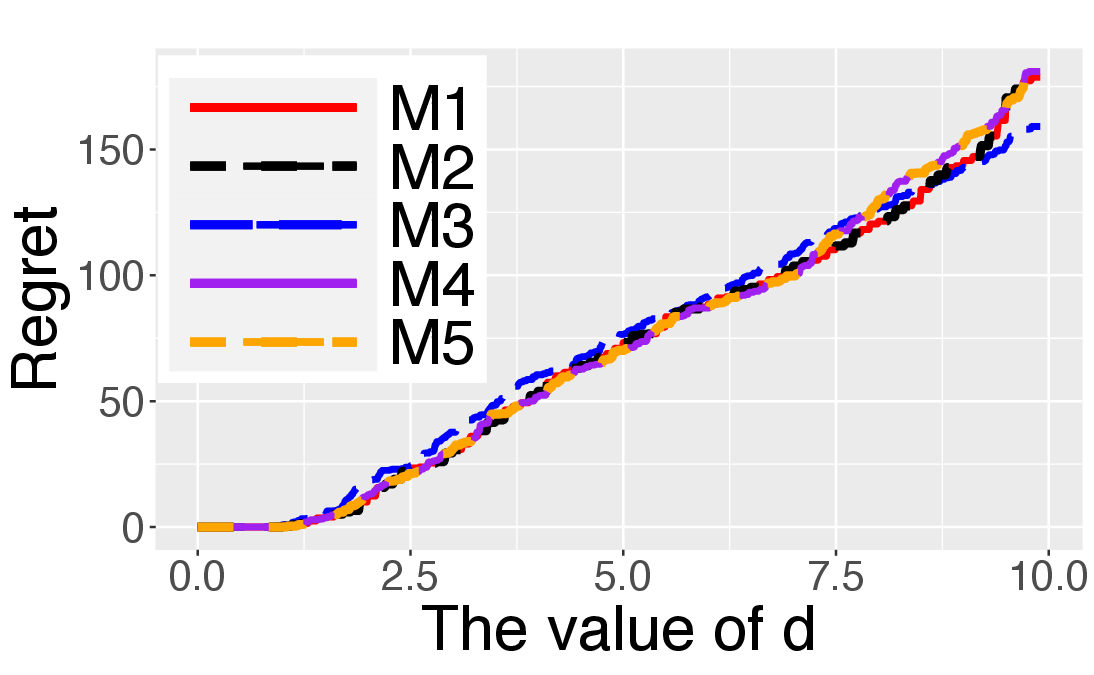}
\label{13b}
}
\quad
\subfigure[Error in Revenue of the Simulated Data.]{
\includegraphics[width=4.1cm]{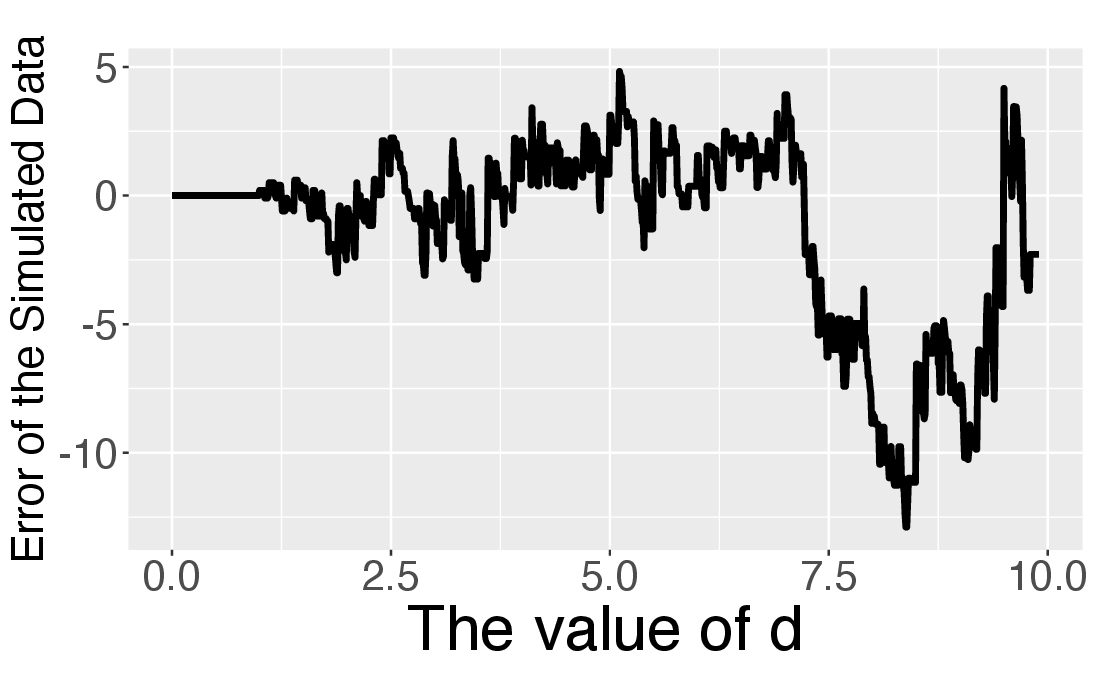}
\label{13c}
}
\quad
\subfigure[Comparison of $kd$.]{
\includegraphics[width=4.1cm]{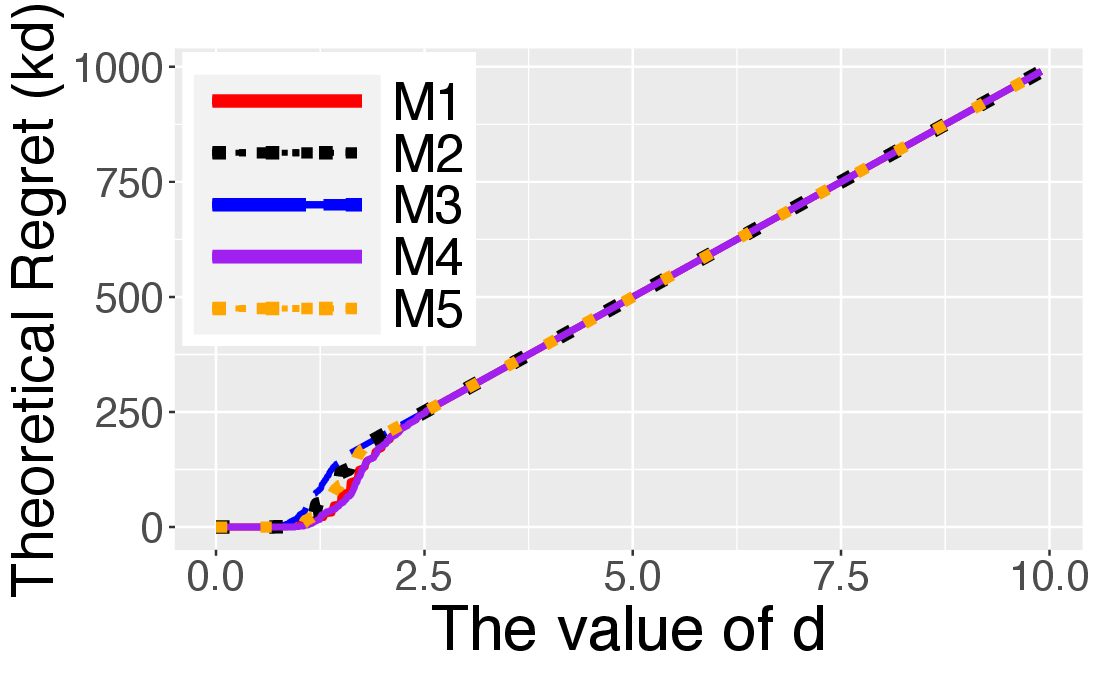}
\label{13d}
}
\quad
\subfigure[Comparison of Type's Proportion Without Query.]{
\includegraphics[width=4.1cm]{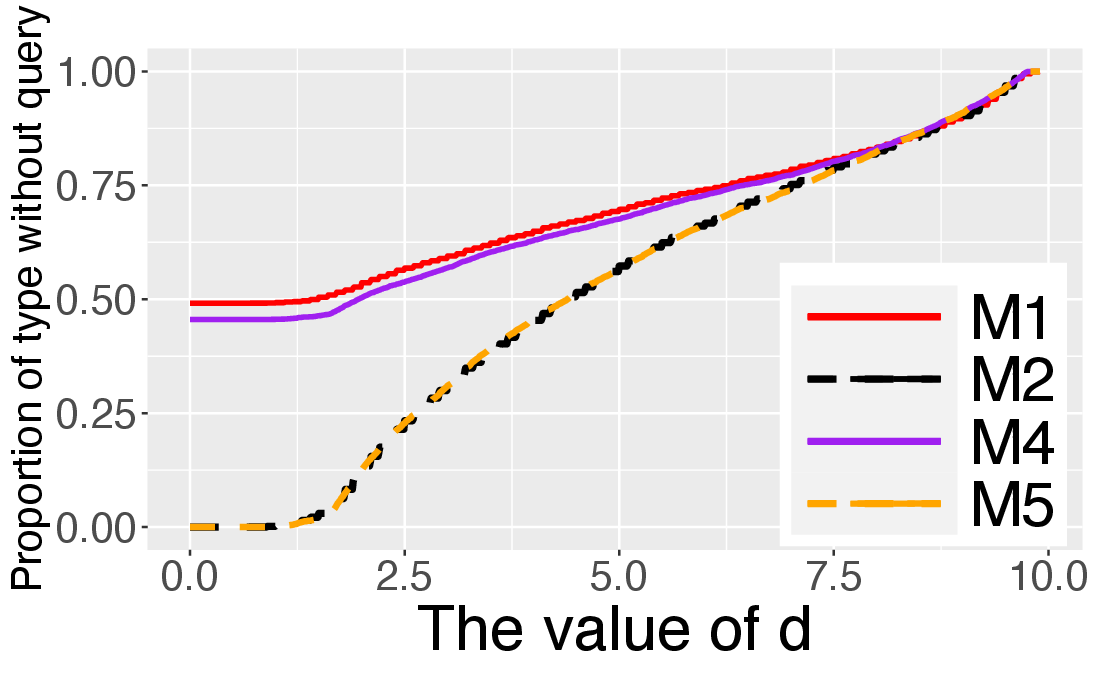}
\label{13e}
}
\quad
\subfigure[Comparison of the Confidence Rate.]{
\includegraphics[width=4.1cm]{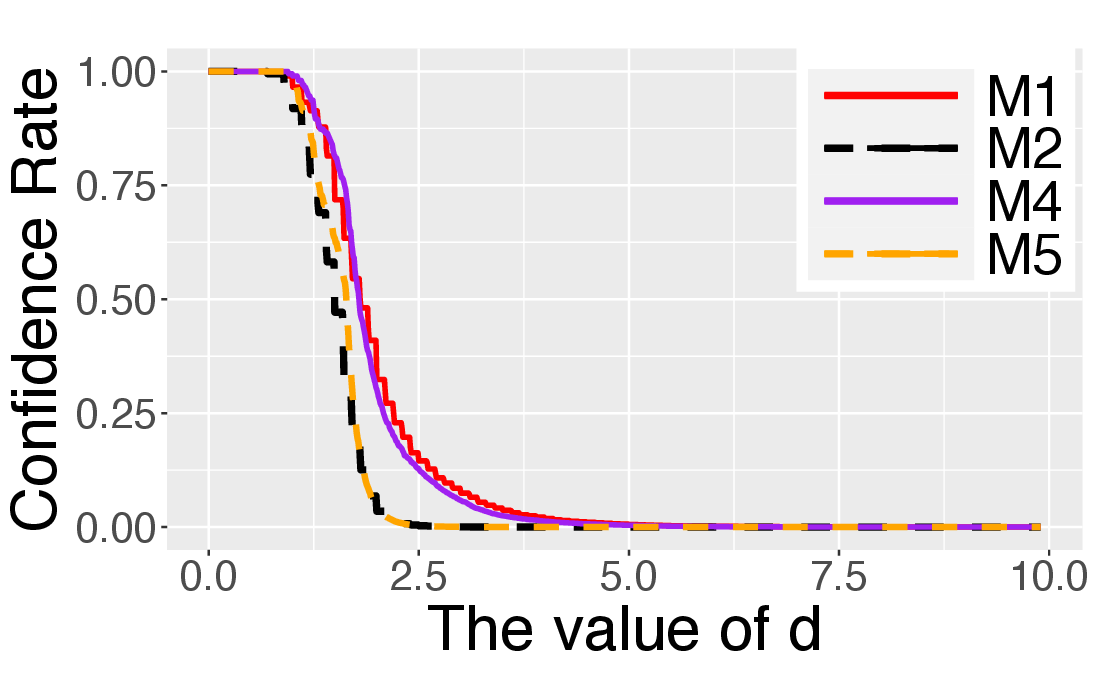}
\label{13f}
}
\caption{The visualization of error when $m=50$ and $N=100$.}
\label{fig:twelve}
\end{figure}

Figure \ref{fig:ten}, \ref{fig:eleven}, and \ref{fig:twelve} depict the visualization of errors in each auction size. 
Mechanism 4 (M4) is represented by the solid purple line, while Mechanism 5 (M5) is indicated by the dashed or dotted orange line. Mechanism 4 and Mechanism 5 can be considered as benchmarks for Mechanism 1 and Mechanism 2, respectively, with fixed true intervals. We can apply the same analysis discussed earlier by replacing Mechanism 1 with Mechanism 4 and Mechanism 2 with Mechanism 5, yielding similar results.

The relative error decreases as the number of bidders and items increases, highlighting the advantages of our strategies in large-size auctions. Moreover, in Figure \ref{11c}, \ref{12c}, and \ref{13c}, the errors in revenue are depicted, which are calculated by subtracting the revenue of Mechanism 1 (Mechanism 2) from the revenue of Mechanism 4 (Mechanism 5) for each scenario. It is evident that the error in revenue, obtained through our strategies utilizing credible intervals derived from the simulated data, is remarkably small compared to the revenue generated by the original VCG mechanism across all scenarios. This demonstrates that our proposed methods are not sensitive to changes in the credible intervals, indicating the robustness of our multi-item auction design.

\end{document}